\documentclass[11pt]{article}
\usepackage[margin=1in]{geometry}
\usepackage{booktabs}
\usepackage{array}
\usepackage{multirow}
\usepackage{multicol}

\usepackage{graphicx} 

\usepackage{multirow,amssymb,amsmath,tabularx, amsthm, amsfonts,color}
\usepackage{import}
\usepackage{placeins}
\usepackage[ruled,linesnumbered]{algorithm2e}
\usepackage{chngcntr, apptools}
\usepackage{mathtools}

\usepackage{tikz}
\usetikzlibrary{positioning, arrows.meta, fit}

\let\oldnl\nl
\newcommand{\nonl}{\renewcommand{\nl}{\let\nl\oldnl}}

\newlength\mylen
\newcommand\myinput[1]{\nonl
  \settowidth\mylen{\KwIn{}}%
  \setlength\hangindent{\mylen}%
  \hspace*{\mylen}#1\\}

\renewcommand{\P}[1]{\mathbb{P}\left[#1\right]}
\newcommand{\Ps}[1]{\mathbb{P}[#1]}
\newcommand{\EV}[1]{\mathbb{E}\left[#1\right]}
\newcommand{\I}[1]{\mathbb{I}\left[#1\right]}
\newcommand{\Is}[1]{\mathbb{I}[#1]}

\newcommand{\tY}{\Tilde{Y}}
\newcommand{\tP}{\Tilde{P}}
\newcommand{\tF}{\Tilde{F}}
\newcommand{\tf}{\Tilde{f}}

\newcommand{\htau}{{\widehat{\tau}}}

\def\independenT#1#2{\mathrel{\rlap{$#1#2$}\mkern2mu{#1#2}}}

\newtheorem{theorem}{Theorem}

\newtheorem{assumption}{Assumption}

\newtheorem{lemma}{Lemma}
\newtheorem{proposition}{Proposition}
\newtheorem{corollary}{Corollary}
\newtheorem{definition}{Definition}

\newcommand\numberthis{\addtocounter{equation}{1}\tag{\theequation}}

\newcommand\independent{\protect\mathpalette{\protect\independenT}{\perp}}
\def\independenT#1#2{\mathrel{\rlap{$#1#2$}\mkern2mu{#1#2}}}

\usepackage{natbib}
\usepackage[hidelinks]{hyperref}
\usepackage{url}

\title{Noise-Adaptive Conformal Classification with Marginal Coverage}
\date{}
\author{Teresa Bortolotti\textsuperscript{1}, Y. X. Rachel Wang\textsuperscript{2}, Xin Tong\textsuperscript{3},\\ Alessandra Menafoglio\textsuperscript{1}, Simone Vantini\textsuperscript{1}, and Matteo Sesia\textsuperscript{4,5}
}

\begin{document}

\maketitle
\let\thefootnote\relax
\footnotetext{\hspace{-0.57cm}\textsuperscript{1}MOX, Department of Mathematics, Politecnico di Milano, Piazza Leonardo da Vinci 32, 20133, Milan, Italy.\\
\textsuperscript{2}School of Mathematics and Statistics, University of Sydney, Sydney, Australia.\\
\textsuperscript{3}Business School, University of Hong Kong, Hong Kong, China. \\
\textsuperscript{4}Department of Data Sciences and Operations, University of Southern California, Los Angeles, CA, USA.\\
\textsuperscript{5}Department of Computer Science, University of Southern California, Los Angeles, CA, USA.}


\begin{abstract}
Conformal inference provides a rigorous statistical framework for uncertainty quantification in machine learning, enabling well-calibrated prediction sets with precise coverage guarantees for any classification model. However, its reliance on the idealized assumption of perfect data exchangeability limits its effectiveness in the presence of real-world complications, such as low-quality labels---a widespread issue in modern large-scale data sets. This work tackles this open problem by introducing an adaptive conformal inference method capable of efficiently handling deviations from exchangeability caused by random label noise, leading to informative prediction sets with tight marginal coverage guarantees even in those challenging scenarios. We validate our method through extensive numerical experiments demonstrating its effectiveness on synthetic and real data sets, including BigEarthNet and CIFAR-10H.
\\

\end{abstract} 

\bigskip

\section{Introduction}
\label{sec:introduction}

\subsection{Background and Motivation}
Conformal prediction provides rigorous uncertainty quantification for any ``black-box'' machine learning model, avoiding parametric assumptions \citep{Vovk2005}. In classification, it constructs a prediction set for the label of a new test point by analyzing the empirical distribution of suitable residuals (or {\em non-conformity scores}) computed by the model on a held-out calibration data set assumed to be {\em exchangeable} with the test data.
Perfect exchangeability, however, is not always realistic. This work tackles scenarios where the calibration labels may be randomly contaminated, breaking exchangeability, invalidating classical conformal methods, and often leading to overly conservative predictions. Label noise is pervasive, as modern data sets are frequently affected by privacy-preserving mechanisms, human annotation errors (e.g., in crowdsourcing platforms like Amazon Mechanical Turk), or other measurement inaccuracies.

Our main contribution is an adaptive method for constructing informative prediction sets that provably achieve tight \emph{marginal coverage} under label noise; that is, sets containing the correct label for a predefined proportion of future test points.
Stronger guarantees are possible but may come at the expense of very large and, thus, uninformative sets. For example, label-conditional coverage \citep{vovk2003mondrian}, which requires the correct label to be covered within each subgroup defined by the true labels, can be overly conservative in applications with many labels \citep{ding2024class} or significant class imbalance, due to the statistical ``bottleneck'' created by rarer labels. Marginal coverage instead strikes a practical balance between informativeness and rigor, although it is relatively challenging to attain under label noise without resorting to unnecessarily large prediction sets.

\subsection{Main Contributions and Related Work} \label{sec:contributions}

Prior works have studied conformal prediction beyond exchangeability \citep{barber2023conformal}, typically by re-weighting the calibration data.
While re-weighting is effective at addressing covariate shift \citep{tibshirani2019conformal} and label shift \citep{podkopaev2021distribution}, it does not solve the problem studied in this paper.
As shown in Appendix~\ref{appendix:weighted-cp-comparison}, under label noise, weighted conformal prediction would require knowledge of the true conditional label distribution, and assuming that to be known would be unrealistic and would give away the problem.
Therefore, a different line of work has emerged to handle label noise.
Under general label contamination, \citet{einbinder2024label} identify conditions under which standard conformal prediction becomes conservative, while \citet{penso2024conformal} propose computing more robust scores. Assuming a known or estimable label noise model, \citet{sesia2025adaptive} introduce an adaptive method providing \emph{label-conditional} coverage.

Building on \citet{sesia2025adaptive}, we target \emph{marginal} rather than
label-conditional coverage. While marginal coverage is weaker, in this setting it is more
difficult to achieve efficiently. \citet{sesia2024adaptive_arxiv} show that it can be
attained through a direct extension of the label-conditional approach, but that over-covers by $\mathcal{O}(1/\sqrt{n_{\min}})$, where
$n_{\min}$ is the sample size of the minority class. In contrast, we achieve
$\mathcal{O}(\sqrt{\log K/n})$ efficiency, where $K$ is the number of classes and $n$ is the
total calibration sample size. Thus, our prediction sets are more informative in applications with
many labels or strong class imbalance, which are precisely the ones that motivate marginal
coverage.

Our main innovation is to analyze the empirical process describing
marginal coverage directly, without conditioning on the labels. This analysis is more delicate 
than a conditional one because it does not reduce to applying the Dvoretzky–Kiefer–Wolfowitz (DKW) inequality,
requiring instead different tools from empirical process theory. We address this challenge in two complementary ways,
by deriving a \emph{finite-sample} correction through
chaining arguments, which leads to the $\mathcal{O}(\sqrt{\log K/n})$ scaling described above, and by obtaining an even tighter \emph{asymptotic}
approximation, which is our recommended approach in practice.

\citet{clarkson2024split} propose a related method that also builds on \citet{sesia2025adaptive} to target marginal coverage, but under different assumptions and with a different analysis. (i) They model the conditional distribution of the clean labels given the contaminated ones, whereas we model the conditional distribution of the contaminated labels given the clean ones, which facilitates the estimation of our contamination model from a small auxiliary sample. (ii) They assume the clean and contaminated label frequencies are known, whereas we estimate them from the contaminated data and account for the resulting uncertainty. (iii) They control only the error in estimating the coverage inflation factor, by applying the DKW inequality label by label at a cost of $\mathcal{O}(\sqrt{K/n})$, without accounting for the data-dependence of the resulting calibrated threshold; as a result, their method does not come with a rigorous finite-sample coverage guarantee. Our method does, and at a cost of only $\mathcal{O}(\sqrt{\log K/n})$; we achieve this by applying a chaining argument on the marginal empirical process that controls both sources of error jointly, and we complement this finite-sample analysis with an even more efficient asymptotic approximation that has no counterpart in their work. Appendix~\ref{appendix:clarkson-comparison} presents a more detailed comparison and discussion.

Beyond conformal prediction, prior research on classification with label noise developed robust loss functions \citep{cannings2020classification}, pre-training methods to filter out likely-corrupted samples \citep{northcutt2021confident}, and classifiers that model the contamination explicitly by incorporating a label transition matrix into the training loss \citep{patrini2017making}; see \citet{frenay2013classification} for a review.
Our contribution is mostly complementary, as we quantify the uncertainty of any model. We do borrow from the third area, however: our method takes a label transition matrix as an input, and in Section~\ref{sec:contamination_estimation} we estimate this matrix using an approach similar to those employed to fit robust classifiers.

\section{Methods} \label{sec:methods}

\subsection{Problem Statement and Notation} \label{sec:methods-def}

Consider $n+1$ data points $(X_i,Y_i,\tilde{Y}_i)$, for $i \in [n+1] = \{1,\ldots,n+1\}$, where $X_i \in \mathbb{R}^d$ is a feature vector, $Y_i \in [K]$ is a latent categorical label with $K \geq 2$, and $\tilde{Y}_i \in [K]$ is an observable label that we interpret as a contaminated version of $Y_i$.
Assume the data are i.i.d.~random samples from an unknown distribution $\mathcal{P} = \mathcal{P}_X \cdot \mathcal{P}_{Y \mid X} \cdot \mathcal{P}_{\tilde{Y} \mid X, Y}$.
We seek a prediction set $\hat{C}(X_{n+1}) \subseteq [K]$ for $Y_{n+1}$ given $X_{n+1}$, leveraging the observations $(X_i, \tilde{Y}_i)$ for $i \in [n]$, which we collectively denote as $\mathcal{D}$. 
We want this set to guarantee \textit{marginal coverage}, $\P{ Y_{n+1} \in \hat{C}(X_{n+1}) } \geq 1-\alpha$,
at the desired miscoverage level $\alpha \in (0,1)$, while being as tight as possible; that is, the miscoverage probability should be close to $\alpha$, especially if $n$ is large.
Note that this {\em marginal} probability is taken with respect to $(X_{n+1},Y_{n+1})$ and the data in $\mathcal{D}$, both of which are random.
To achieve this, some assumptions are needed.

\begin{assumption}\label{assumption:linear-contam}
$\tilde{Y}$ is independent of $X$ given $Y$; i.e., $\tilde{Y} \independent X \mid Y$, or $\mathcal{P}_{\tilde{Y} \mid X, Y} = \mathcal{P}_{\tilde{Y} \mid Y}$.
\end{assumption}\noindent
While we rely on Assumption~\ref{assumption:linear-contam} to design our method and analyze it theoretically, we will also evaluate it on real data where the assumption may not hold exactly.

Additionally, we assume that the contamination model is known, with $T \in [0,1]^{K \times K}$ denoting the \emph{invertible} stochastic transition matrix such that $T_{kl} = \Ps{\tilde{Y} = k \mid Y = l}$.

\begin{assumption}\label{assumption:known-contam}
The matrix $T$ is known and invertible, with inverse $W := T^{-1}$.
\end{assumption}

Assumption~\ref{assumption:known-contam} bundles two reasonable requirements.
The assumption that $T$ is known may hold exactly in some applications, including when data are intentionally contaminated by a known privacy algorithm \citep{dwork2006calibrating}. More generally, the method developed in this paper can often be successfully applied using a suitable estimate of $T$ in place of the true matrix, as discussed in Section~\ref{sec:contamination_estimation} and demonstrated empirically in Sections~\ref{sec:numerical_experiments} and~\ref{sec:case_study}.

The assumption that $T$ be invertible is satisfied by a broad class of contamination mechanisms. For example, $T$ is invertible if strictly diagonally dominant by columns, i.e., $T_{ll} > \sum_{k \neq l} T_{kl} = 1-T_{ll}$ for every $l \in [K]$, or equivalently if the true label is preserved with probability greater than $1/2$ for every class. On the other hand, invertibility fails if two classes become indistinguishable after contamination; i.e., if $\P{\tY=k \mid Y=l} = \P{\tY = k \mid Y = l'}$, for all $k \in [K]$ and for any $l \neq l'$, then columns $l$ and $l'$ of $T$ coincide and $T$ is singular. Such extreme label contamination cannot be handled by our method.

\subsection{Relevant Technical Background}
\label{sec:relevant_technical_background}

Consider a fixed classification model $\hat{\pi}$, trained on data $\mathcal{D}^{\text{train}}$ independent of the test point and of the $n$ calibration points in $\mathcal{D}$, that gives an arbitrary estimate $\hat{\pi}(x,k) \in [0,1]$ of $\P{\tY = k \mid X=x}$ for any $k \in [K]$ and $x \in \mathbb{R}^d$.
Without loss of generality, assume that $\hat{\pi}$ is normalized: $\sum_{k=1}^{K} \hat{\pi}(x,k) = 1$ for any $x$.
This model output is converted into a conformal prediction set by calibrating a \emph{nested} family of prediction sets \citep{gupta2022nested}.

\begin{definition}[Nested family] \label{def:pred-function}
A nested family is a collection $\{\mathcal{C}(\cdot, \tau)\}_{\tau \in [0,1]}$ of set-valued functions, whose form may depend on the model $\hat{\pi}$, each mapping $x \in \mathbb{R}^d$ to a subset of $[K]$, such that $\mathcal{C}(x, \tau_1) \subseteq \mathcal{C}(x, \tau_2)$ whenever $\tau_1 < \tau_2$, and $\mathcal{C}(x, 1) = [K]$ for any $x$.
\end{definition}

The dependence of a nested family $\mathcal{C}$ on $\hat{\pi}$ is typically kept implicit; i.e., $\mathcal{C}(x, \tau) = \mathcal{C}(x, \tau; \hat{\pi}) \subseteq [K]$. For any $\mathcal{C}$, the corresponding {\em non-conformity score} function $\hat{s}: \mathbb{R}^d \times [K] \mapsto [0,1]$, also implicitly depending on $\hat{\pi}$, outputs the smallest $\tau$ for which label $k$ is included:
\begin{align}  \label{eq:conf-scores}
  \hat{s}(x, k)
  = \inf \left\{ \tau \in [0,1] : k \in \mathcal{C}(x, \tau) \right\}.
\end{align}
One example is $\mathcal{C}(x, \tau; \hat{\pi}) := \{ k \in [K] : \hat{\pi}(x,k) \geq 1 - \tau \}$, with scores $\hat{s}(x,k) = 1-\hat{\pi}(x,k)$.
All our results also extend to other nested families, including that of \citet{romano2020classification}.

The standard conformal approach computes $\hat{s}(X_i, \tilde{Y}_i)$ for all $i \in [n]$, calibrates $\htau^{\mathrm{std}}$ as the $\lceil (1+n)(1-\alpha) \rceil$-th smallest value in $\{\hat{s}(X_i,\tilde{Y}_i)\}_{i=1}^{n} \cup \{1 \}$, and outputs $\hat{C}^{\mathrm{std}}(X_{n+1})=\mathcal{C}(X_{n+1}, \htau^{\mathrm{std}})$. To facilitate the comparison with the adaptive method proposed below, it is helpful to 
introduce $S_{(1)} \leq \ldots \leq S_{(n)}$ as the order statistics of $\{\hat{s}(X_i,\tilde{Y}_i)\}_{i=1}^n$ and equivalently rewrite $\htau^{\mathrm{std}} := S_{(\hat{i})}$ where $\hat{i} := \min \hat{\mathcal{I}}^{\mathrm{std}}$ and 
\begin{align} \label{eq:I-set-tau-marg}
  & \hat{\mathcal{I}}^{\mathrm{std}} := \left\{ i \in [n] : \hat{F}(S_{(i)}) \geq 1 - \alpha + \frac{1-\alpha}{n} \right\},
& \hat{F}(t) &:= \frac{1}{n} \sum_{i=1}^{n} \I{\hat{s}(X_i,\tilde{Y}_i) \leq t},
\end{align}
with the proviso that $\min \emptyset = \infty$ and $S_{(\infty)} := 1$, so that $\htau^{\mathrm{std}} = 1$ if $\hat{\mathcal{I}}^{\mathrm{std}}$ is empty.

This approach, summarized by Algorithm~\ref{alg:standard-marg} in Appendix~\ref{appendix:standard-marg}, guarantees marginal coverage above $1-\alpha$ if $Y = \tilde{Y}$ almost surely, with an almost-matching coverage upper bound (Proposition~\ref{prop:standard-coverage-marginal}).
In the next section, we study what happens when $\tilde{Y} \neq Y$.

\subsection{The Marginal Coverage Inflation Factor}

For any $t \in [0,1]$, define $F(t) := \Ps{\hat{s}(X,Y) \leq t}$ and $\tilde{F}(t) := \Ps{\hat{s}(X,\tilde{Y}) \leq t}$---the marginal CDFs of $\hat{s}(X,Y)$ and $\hat{s}(X,\tilde{Y})$, respectively, implicitly conditioning on the model $\hat{\pi}$ trained using $\mathcal{D}^{\text{train}}$.
Then, for any $t \in [0,1]$, we define the {\em marginal coverage inflation factor} as:
\begin{align} \label{eq:Delta-marg}
    \Delta(t) & := F(t) - \tF(t).
\end{align}
As made precise by the following theorem, the function $\Delta$ controls the effect of label noise on the marginal coverage guarantee for standard conformal prediction sets.

\begin{theorem}\label{thm:coverage-marginal}
Suppose $(X_i,Y_i,\tilde{Y}_i)$ are i.i.d.~for $i \in [n+1]$.
Fix any nested family $\mathcal{C}$ (Definition~\ref{def:pred-function}) and let $\hat{C}^{\mathrm{std}}(X_{n+1})$ indicate the standard conformal prediction set output by Algorithm~\ref{alg:standard-marg} applied using the corrupted labels $\tilde{Y}_i$ instead of the clean labels $Y_i$, for all $i \in [n]$.
Then, $ \P{Y_{n+1} \in \hat{C}^{\mathrm{std}}(X_{n+1})} \geq 1 - \alpha + \EV{\Delta(\htau^{\mathrm{std}})}$.
Further, if the scores $\hat{s}(X_i,\tilde{Y}_i)$ used by Algorithm~\ref{alg:standard-marg} are almost-surely distinct,
$\P{Y_{n+1} \in \hat{C}^{\mathrm{std}}(X_{n+1})} \leq 1 - \alpha + \frac{1}{n+1} + \EV{\Delta(\htau^{\mathrm{std}})}$.
\end{theorem}

Theorem~\ref{thm:coverage-marginal} applies to any contamination model and does not depend on Assumptions~\ref{assumption:linear-contam} or~\ref{assumption:known-contam}. However, under Assumption~\ref{assumption:linear-contam} and an additional condition on the distributions of the non-conformity scores---interpreted as the model being sufficiently accurate to typically identify the correct label as the most likely point prediction---the standard prediction sets generated by Algorithm~\ref{alg:standard-marg} are marginally conservative because $\Delta(\htau^{\mathrm{std}}) \geq 0$ almost-surely; see Corollary~\ref{cor:coverage-marg} in Appendix~\ref{appendix:conservativeness-standard}. This robustness aligns with similar results by \citet{barber2023conformal}, \citet{einbinder2024label}, and \citet{sesia2025adaptive}.

Next, we build on Theorem~\ref{thm:coverage-marginal} to develop an adaptive method that leverages an empirical estimate of the function $\Delta$ to achieve tight marginal coverage under label noise.

\subsection{Adaptive Prediction Sets with Marginal Coverage} \label{sec:adaptive-marginal}

For any $k,l \in [K]$ and $t \in [0,1]$, define
  \begin{align} \label{eq:cdf-scores}
   F_l^k(t) := \P{\hat{s}(X,k) \leq t \mid Y=l}, \qquad
   & \tilde{F}_l^k(t) := \P{\hat{s}(X,k) \leq t \mid \tY = l}.
 \end{align}
Above, $F_l^k(t)$ is the CDF of $\hat{s}(X,k)$, based on a fixed function $\hat{s}$ applied to a random $X$ from the distribution of $X \mid Y=l$, while $\tF_l^k(t)$ is the corresponding CDF of $\hat{s}(X,k)$, with $X$ conditioned on $\tilde{Y}=l$.
For any $k \in [K]$, define also $\rho_k := \Ps{Y=k}$ and $\tilde{\rho}_k := \Ps{\tY=k}$,
and assume $\rho_k > 0$ for all $k \in [K]$.

Under Assumptions~\ref{assumption:linear-contam} and~\ref{assumption:known-contam}, the coverage inflation factor~\eqref{eq:Delta-marg} can be expressed in terms of quantities that are either known or estimable from the contaminated data.

\begin{proposition} \label{prop:Delta-marg-linear}
Under Assumptions~\ref{assumption:linear-contam} and~\ref{assumption:known-contam}, with $W = T^{-1}$, for any $t \in [0,1]$,
\begin{align} \label{eq:Delta-marg-linear}
  \Delta(t)
  & = \sum_{k=1}^{K} \sum_{l=1}^{K} W_{kl} \tilde{\rho}_l \tilde{F}_l^{k}(t) - \tilde{F}(t).
\end{align}
\end{proposition}

Therefore, a natural empirical estimator of $\Delta(t)$ is $\hat{\Delta}(t) := \hat{F}_0(t) - \hat{F}(t)$, where
\begin{align} \label{eq:Delta-hat-marg-linear}
  & \hat{F}_0(t) := \sum_{k,l=1}^K W_{kl}\, \hat{\rho}_l\, \hat{F}^k_l(t),
  & \hat{F}_l^{k}(t) := \frac{1}{n_l} \sum_{i \in \mathcal{D}_l} \I{\hat{s}(X_i,k) \leq t},
\end{align}
and $\hat{F}(t)$ is the same empirical CDF of $\hat{s}(X_i,\tilde{Y}_i)$ from~\eqref{eq:I-set-tau-marg}, which is unbiased for $\tilde{F}(t)$.
Here, each $\hat{F}_l^{k}(t)$ is the empirical CDF of $\hat{s}(X_i,k)$ over $i \in \mathcal{D}_l := \{ i \in [n] : \tilde{Y}_i = l\}$ and is unbiased for $\tilde{F}_l^{k}(t)$, while $\hat{F}_0(t)$ is unbiased for $F(t)$, the CDF of the scores evaluated at the unobserved clean labels.
This mirrors how unbiased estimates of the loss function are obtained when training classifiers robust to label noise \citep{patrini2017making}.
It is instructive to write $\hat{F}_0$ compactly as $\hat{F}_0(t) = \frac{1}{n} \sum_{i=1}^{n} \sum_{k=1}^{K} W_{k \tilde{Y}_i} \, \I{\hat{s}(X_i,k) \leq t}$, which shows that $\hat{F}_0$ is the distribution function of a random signed measure placing mass $W_{k \tilde{Y}_i}/n$ at each score $\hat{s}(X_i,k)$.
Since these weights may be negative, this measure is not a probability measure, and $\hat{F}_0$ is therefore not an empirical CDF: it need not be non-decreasing and may take values outside $[0,1]$. Consequently, $\hat{\mathcal{I}}$ need not be an upper interval, and $\htau$ should not be understood as an exact empirical quantile.
This is a key difference between our method and standard (weighted) conformal prediction \citep{tibshirani2019conformal}.

We use $\hat{\Delta}$ within a calibration approach similar to~\eqref{eq:I-set-tau-marg}, computing
\begin{align} \label{eq:I-set-marg}
\begin{gathered}
  \htau := S_{(\hat{i})}, \qquad \hat{i} := \min \hat{\mathcal{I}}, \\
  \hat{\mathcal{I}} := \left\{ i \in [n] : \frac{i}{n} \geq 1 - \alpha - \hat{\Delta}(S_{(i)}) + \delta(n) \right\}
  = \left\{ i \in [n] : \hat{F}_0(S_{(i)}) \geq 1 - \alpha + \delta(n) \right\},
\end{gathered}
\end{align}
where $\delta(n) \geq 0$ is a finite-sample correction factor accounting for estimation errors, which we study in detail below; the second expression for $\hat{\mathcal{I}}$ follows because $\hat{F}(S_{(i)}) = i/n$.
Then, our adaptive prediction set is $\hat{C}(X_{n+1}) = \mathcal{C}(X_{n+1}, \htau)$, and the overall procedure is outlined by Algorithm~\ref{alg:adaptive-conformal-marg}.
We will show in the following that, when $\delta(n)$ is properly set, Algorithm~\ref{alg:adaptive-conformal-marg} achieves marginal coverage efficiently.
The additional difficulty of adapting to label noise will require us to apply a correction term $\delta(n)$ that vanishes with $n$ only as $\mathcal{O}(n^{-1/2})$, instead of the $(1-\alpha)/n$ term used in standard conformal prediction~\eqref{eq:I-set-tau-marg}.

In Appendix~\ref{sec:optimistic-calibration}, we study an ``optimistic" variation of Algorithm~\ref{alg:adaptive-conformal-marg} that selects the smaller of the prediction sets output by Algorithm~\ref{alg:adaptive-conformal-marg} and the standard conformal approach (Algorithm~\ref{alg:standard-marg}).
We show that this preserves marginal coverage (up to a small gap of negligible order) as long as the standard conformal prediction sets are sufficiently conservative in the presence of label noise---a relatively mild assumption as demonstrated in Appendix~\ref{appendix:conservativeness-standard}.

\begin{algorithm}[!htb]
\DontPrintSemicolon

\KwIn{Contaminated data set $\{(X_i, \tilde{Y}_i)\}_{i=1}^{n}$ with noisy labels $\tilde{Y}_i \in [K]$.}
\myinput{The inverse $W$ of the matrix $T$ describing the contamination model.}
\myinput{Pre-trained $K$-class classification model $\hat{\pi}$. Nested family $\mathcal{C}$ (Definition~\ref{def:pred-function}).}
\myinput{Desired miscoverage probability $\alpha \in (0,1)$. Correction factor $\delta(n) \geq 0$.}
\myinput{Unlabeled test point with features $X_{n+1}$.}

Compute the scores $\hat{s}(X_i,k)$ for all $i \in [n]$ and $k \in [K]$, based on $\mathcal{C}$ and $\hat{\pi}$ using~\eqref{eq:conf-scores}.\;
Compute the empirical contaminated label frequency $\hat{\rho}_k = n_k/n$, for all $k \in [K]$. \;
Compute the empirical CDFs $\hat{F}$ and $\hat{F}_l^{k}$, for all $k,l \in [K]$. \;
  Sort $\{\hat{s}(X_i,\tilde{Y}_i) : i \in [n] \}$ into $(S_{(1)}, S_{(2)}, \dots, S_{(n)})$, in ascending order. \;
  Compute $\hat{\Delta}(S_{(i)})$ for all $i \in [n]$, as in \eqref{eq:Delta-hat-marg-linear}. \;
  Construct $\hat{\mathcal{I}} \subseteq [n]$ and evaluate $\htau$ as in \eqref{eq:I-set-marg}.\;
\nonl
\textbf{Output: } Conformal prediction set $\hat{C}(X_{n+1}) = \mathcal{C}(X_{n+1}, \htau; \hat{\pi}) \subseteq [K]$ for $Y_{n+1}$.

\caption{Contamination-adaptive classification with marginal coverage}
\label{alg:adaptive-conformal-marg}
\end{algorithm}

\subsection{Lower Bound on Coverage of Adaptive Prediction Sets} \label{sec:coverage-guarantees-lower-bound}

Define the following zero-mean empirical process, along with its expected supremum,
\begin{align}\label{eq:hat-psi}
  &  \hat{\psi}(t) \coloneqq \sum_{k=1}^K \sum_{l=1}^K W_{kl} \left(\hat{\rho}_l \hat{F}_l^k(t) - \tilde{\rho}_l \tilde{F}_l^k(t) \right),
  &  \delta^{*}(n) = \mathbb{E} \left[\underset{t \in [0,1]}{\sup} \; \hat{\psi}(t) \right].
\end{align}

The following result states that, when applied with a correction factor $\delta(n) \geq \delta^*(n)$, Algorithm~\ref{alg:adaptive-conformal-marg} outputs prediction sets that are marginally valid in finite samples.

\begin{theorem} \label{thm:algorithm-lower-bound}
Suppose $\{(X_i,Y_i,\tilde{Y}_i)\}_{i=1}^{n+1}$ are i.i.d.~and that Assumptions~\ref{assumption:linear-contam} and~\ref{assumption:known-contam} hold.
For any nested family $\mathcal{C}$, let $\hat{C}(X_{n+1})$ be the prediction set output by Algorithm~\ref{alg:adaptive-conformal-marg} applied with a correction factor $\delta(n)$. Then, $\Ps{Y_{n+1} \in \hat{C}(X_{n+1})} \geq 1-\alpha$ if $\delta(n) \geq \delta^*(n)$.
\end{theorem}

Further, the closer $\delta(n)$ is to $\delta^*(n)$, the tighter the prediction sets output by Algorithm~\ref{alg:adaptive-conformal-marg} are; this is studied in more detail in Section~\ref{sec:coverage-guarantees-upper-bound}, where Theorem~\ref{thm:algorithm-lower-bound} is complemented by a coverage upper bound for our adaptive method. 
Therefore, $\delta^{*}(n)$ represents the ideal correction factor for Algorithm~\ref{alg:adaptive-conformal-marg}.
Unfortunately, this quantity is not readily available because it depends on the distribution of $\hat{\psi}(t)$, which is unknown and potentially complicated.

The following sections present two alternative solutions for calculating a valid correction factor $\delta(n)$ without being too conservative.
Section~\ref{sec:adaptive_pred_finitesample} gives a relatively tight finite-sample upper bound for $\delta^*(n)$, 
while Section~\ref{sec:adaptive_pred_asymptotic} provides an even closer asymptotic approximation.

\section{Implementation using Finite-Sample Techniques} \label{sec:adaptive_pred_finitesample}

In this section, we explain how to replace the ideal correction factor $\delta^*(n)$ required by Algorithm~\ref{alg:adaptive-conformal-marg} with an upper bound that is guaranteed to be conservative in finite samples.

\subsection{Correction Factor for the Randomized Response Model} \label{sec:adaptive_pred_finitesample_simplified}

We begin by focusing on the classical randomized response model \citep{warner1965randomized}.
In this special case, the transition matrix $T$ is characterized by a scalar noise parameter $\epsilon \in [0,1)$ and takes the form
$T = (1-\epsilon) \cdot I + \epsilon/K \cdot J$,
where $I$ is the $(K \times K)$ dimensional identity matrix and $J$ is the $(K \times K)$ matrix of all ones, and thus  $W = T^{-1} = 1/(1-\epsilon) \cdot I - \epsilon/[K(1-\epsilon)] \cdot J$.

In this case, it is not difficult to derive a finite-sample upper bound for the ideal correction factor $\delta^{*}(n)$, ensuring $1/\sqrt{n}$ scaling.
Let $U_1,\dots,U_{n}$ be i.i.d.~uniform random variables on $[0,1]$, and denote their order statistics as $U_{(1)} \leq \dots \leq U_{(n)}$.
Then, define the constant
\begin{align} \label{eq:c-definition}
c(n) := \EV{\sup_{i \in [n]} \left\{ \frac{i}{n} - U_{(i)}\right\}},
\end{align}
which satisfies $c(n) \leq \sqrt{\pi/(8n)}$  (Lemma~\ref{lemma:bound-on-c}) and can be easily computed via Monte Carlo.

\begin{assumption}\label{assumption:score-cdf-continuous}
For any $k \in [K]$, $\tilde{F}^k(t) := \mathbb{P}[\hat{s}(X,k) \leq t]$ is continuous on $[0,1]$.
\end{assumption}
This assumption is mild and easy to satisfy for the non-conformity scores typically used in practice. For the classical scores $\hat{s}(x,k) = 1 - \hat{\pi}(x,k)$, it can always be enforced by adding a small amount of independent random noise to the model output. 
It also holds for the randomized version of the adaptive scores of~\citet{romano2020classification}; see Appendix~\ref{app:adaptive-scores}.
It is easy to see that, when $\tilde{\rho}_l > 0$ for all $l \in [K]$,
Assumption~\ref{assumption:score-cdf-continuous} implies that $\tF_l^k$ is continuous for all $k,l \in [K]$, and hence $F$ and $\tF$ are also continuous. 
The condition $\tilde{\rho}_l > 0$ for all $l \in [K]$ is implied by the invertibility of $T$: since $\tilde{\rho} = T \rho$ and $\rho_k > 0$ for all $k$, a vanishing $\tilde{\rho}_k$ would require the $k$-th row of $T$ to be zero, making $T$ singular.

\begin{theorem} \label{thm:finite-sample-factor-simplified}
Under the assumptions of Theorem~\ref{thm:algorithm-lower-bound}, assume also that Assumption~\ref{assumption:score-cdf-continuous} holds. If $T = (1-\epsilon) \cdot I + \epsilon/K \cdot J$ for some $\epsilon \in [0,1)$,
then $\delta^{*}(n) \leq c(n) \cdot (1+\epsilon)/(1-\epsilon)$.
\end{theorem}

Therefore, applying Algorithm~\ref{alg:adaptive-conformal-marg} with $\delta(n) = \frac{1 + \epsilon}{1-\epsilon} c(n)$ achieves finite-sample coverage efficiently, regardless of the number of classes $K$. While this solution serves as a useful starting point, extending it beyond the randomized response model is more challenging. 

\subsection{Correction Factor for General Contamination Models}
\label{sec:adaptive_pred_finitesample_generic}

The proof of Theorem~\ref{thm:finite-sample-factor-simplified} exploits the structure of $W$ under the randomized response model to derive a bound for $\delta^{*}(n)$ that is independent of $K$ and scales well with $n$. This motivates decomposing the empirical process~\eqref{eq:hat-psi} for the general case into a well-behaved component, analogous to the randomized response case, and an often smaller remainder.

Consider a matrix $\bar{W} \in \mathbb{R}^{K \times K}$, parameterized by $K+1$ coefficients $\beta = (\beta_0, \beta_1, \dots, \beta_K)$, with entries $\bar{W}_{kl} = \beta_0 \I{k=l} + \beta_k/K$, and let $\Omega$ denote the matrix difference between $W$ and $\bar{W}$, such that $\Omega_{kl}=W_{kl}-\bar{W}_{kl}$.
This parametrization is motivated by the fact that, if $W = \bar{W}$, the expected supremum of the empirical process $\hat{\psi}(t)$ defined in~\eqref{eq:hat-psi} can be analyzed using the same approach as in Theorem~\ref{thm:finite-sample-factor-simplified}. Notably, the matrix $W$ under the randomized response model is a special case of $\bar{W}$, corresponding to a specific choice of $\beta$.

With this setup, we can introduce our general finite-sample upper bound for $\delta^{*}(n)$, derived using chaining arguments \citep{massart2000some,wainwright2019high}, which takes the form:
\begin{equation}
    \delta^{\mathrm{FS}}(n) := \inf_{\beta \in \mathbb{R}^{K+1}} \left\{ c(n) \left(\lvert \beta_0 \rvert+\frac{\sum_{k=1}^{K} \lvert \beta_k \rvert}{K} \right) + \frac{B(K,n,\beta)}{\sqrt{n}}  \right\},
    \label{eq:delta_n_finitesample}
\end{equation}
where $B(K,n, \beta)$ is given by the following expression, with $\|\Omega\|_1 := \max_{l \in [K]} \sum_{k=1}^{K} |\Omega_{kl}|$:
\begin{align} \label{eq:B-definition}
\begin{split}
    B(K,n, \beta) :=
    2 \|\Omega\|_1 \min \Bigg\{ &
    \sqrt{2 \log(2K n + 2)},  24 \sqrt{\log (K + 2)} + \frac{24}{\sqrt{\log (K + 2)}}  \Bigg\}.
  \end{split}
\end{align}
The definition of $\delta^{\mathrm{FS}}(n)$ in~\eqref{eq:delta_n_finitesample} involves an optimization over $\beta$, which can be solved as a convex optimization problem; see Appendix~\ref{app:optimization}. This makes our upper bound computable for any contamination model.
Note that $\beta$ in~\eqref{eq:delta_n_finitesample} is optimized over the unconstrained space $\mathbb{R}^{K+1}$ because $\bar{W}$ is an auxiliary matrix used to control $\hat{\psi}(t)$ through the decomposition $\Omega = W - \bar{W}$, and it need not itself be the inverse of a stochastic transition matrix.



\begin{theorem} \label{thm:finite-sample-factor}
Under the assumptions of Theorem~\ref{thm:finite-sample-factor-simplified}, $\delta^*(n) \leq \delta^{\mathrm{FS}}(n)$, with $\delta^{\mathrm{FS}}(n)$ in~\eqref{eq:delta_n_finitesample}.
\end{theorem}

Having established the practicality and validity of $\delta^{\mathrm{FS}}(n)$ as a general upper bound for $\delta^{*}(n)$, we turn to discussing its scaling properties with respect to $n$ and $K$.
We begin by noting that the behavior of $c(n)$, described by Lemma~\ref{lemma:bound-on-c}, implies
\begin{align*}
  \delta^{\mathrm{FS}}(n)
  & \leq \sqrt{\frac{\pi}{8n}} \cdot \inf_{\beta \in \mathbb{R}^{K+1}} \left( \lvert \beta_0 \rvert + \frac{\sum_{k=1}^{K}  \lvert \beta_k \rvert }{K} + \sqrt{\frac{8}{\pi}}B(K,n,\beta) \right).
\end{align*}

Therefore, $\delta^{\mathrm{FS}}(n)$ decays as $1/\sqrt{n}$ with $K$ fixed, as desired, since
$B(K,n,\beta)$ does not grow with $n$. By contrast, its scaling with respect to $K$ depends on
the contamination model. Under the randomized response model, a suitable $\beta$ can be found that
gives  $B(K,n,\beta) = 0$, so $\delta^{\mathrm{FS}}(n) \leq \frac{1+\epsilon}{1-\epsilon} c(n)$ uniformly in $K$, as in
Section~\ref{sec:adaptive_pred_finitesample_simplified}. In general, evaluating~\eqref{eq:delta_n_finitesample} at $\beta = 0$, where $\Omega = W$, 
gives $\delta^{\mathrm{FS}}(n) = \mathcal{O}(\|W\|_1 \sqrt{\log K /n})$, with $\|W\|_1 := \max_{l} \sum_{k} |W_{kl}|$.
This compares favorably with the $\mathcal{O}(\|W\|_1 \sqrt{K/n})$ that one would obtain from a DKW-based analysis in the style of \citet{sesia2025adaptive}.
Moreover, optimizing over $\beta$ can bring additional improvements.
For well-behaved models, such
as the {\em two-level} model of Appendix~\ref{appendix:Special-cases}, $\|W\|_1$ does not grow with $K$ and thus $\delta^{\mathrm{FS}}(n) = \mathcal{O}(\sqrt{\log K /n})$.

We demonstrate the practical utility of our method in
Sections~\ref{sec:numerical_experiments} and~\ref{sec:case_study}, where we show numerically
that Algorithm~\ref{alg:adaptive-conformal-marg}, applied with $\delta^{\mathrm{FS}}(n)$, produces informative prediction sets and significantly outperforms
standard conformal prediction under label noise.
That being said, the performance of Algorithm~\ref{alg:adaptive-conformal-marg} can be further enhanced by replacing $\delta^{\mathrm{FS}}(n)$ with a tighter asymptotic approximation of $\delta^{*}(n)$, as discussed next.

\section{Implementation using Asymptotic Techniques}
\label{sec:adaptive_pred_asymptotic}

\subsection{Convergence in Distribution and High-Level Strategy}

To obtain an asymptotic approximation of the ideal correction factor $\delta^*(n)$, we impose an additional mild condition on the smoothness of the non-conformity score distributions.
\begin{assumption}\label{ass:score-cdf-strictly-increasing}
For any $k \in [K]$, the marginal score distribution function
$\tilde{F}^k(t) := \mathbb{P}[\hat{s}(X,k) \leq t]$ is strictly increasing on $[0,1]$.
\end{assumption}
This assumption is often reasonable in practice and can always be satisfied by adding to the non-conformity scores random noise with positive probability density on $[0,1]$.

\begin{theorem} \label{thm:asymptotic-factor}
Under the assumptions of Theorem~\ref{thm:finite-sample-factor} and Assumption~\ref{ass:score-cdf-strictly-increasing}, consider the empirical process $\hat{\psi}$ in~\eqref{eq:hat-psi}.
Let $\mathrm{GBB}(t)$ denote a zero-mean Gaussian process, indexed by $t \in [0,1]$, with covariance function $G : [0,1]^2 \mapsto \mathbb{R}$ such that $G(t_1, t_2) = \EV{f_{t_1}f_{t_2}} - \EV{f_{t_1}} \EV{f_{t_2}}$, for any $t_1,t_2 \in [0,1]$,
where $f_{t} \coloneqq \sum_{k,l=1}^K W_{kl} \Is{\hat{s}(X, k) \leq t, \tY = l}$ for any $t \in [0,1]$.
Then, $\sqrt{n} \hat{\psi} \rightsquigarrow   \mathrm{GBB}$ as $n \to \infty$, with $\rightsquigarrow$ representing weak convergence of the process.
\end{theorem}

This result suggests approximating $\delta^*(n)$ with
\begin{align}\label{eq:delta_n_asymptotic}
    \delta^{\mathrm{asy}}(n) := \frac{1}{\sqrt{n}} \EV{\underset{t \in [0,1]}{\sup} \mathrm{GBB}(t)}.
\end{align}

The Generalized Brownian Bridge (GBB) in Theorem~\ref{thm:asymptotic-factor} generalizes the F-Brownian bridge (see \citet{van2000asymptotic}, Chapter 19.1). While the distribution of the supremum for the F-Brownian bridge matches that of the classical Brownian bridge and its expected value is available in closed form, computing~\eqref{eq:delta_n_asymptotic} for the GBB is more difficult. This is because its covariance function depends on the distribution of the non-conformity scores, and its expected supremum does not appear to be analytically tractable.

To address these challenges, we propose a discretization strategy for evaluating~\eqref{eq:delta_n_asymptotic}. First, the interval $[0,1]$ is divided into a finite grid, replacing the continuous process $\text{GBB}(t)$ with a Gaussian random vector of length $N$. Second, the covariance matrix of this random vector is approximated using an estimate of the covariance function in Theorem~\ref{thm:asymptotic-factor} based on the observed contaminated data. Third, the expected maximum element of the random vector is estimated with a Monte Carlo approach. Finally, numerical extrapolation is used to reduce the discretization bias. We now describe the key steps in more detail.

\subsection{Empirical Covariance Estimate}

According to  Theorem~\ref{thm:asymptotic-factor}, for any $t_1,t_2 \in [0,1]$, the covariance between $\mathrm{GBB}(t_1)$ and $\mathrm{GBB}(t_2)$ can be written as $G(t_1,t_2) = \EV{f_{t_1}f_{t_2}} - \EV{f_{t_1}} \EV{f_{t_2}}$, where
\begin{align*}
    \EV{f_{t}} &=
    \sum_{l,k=1}^K W_{kl} \P{\hat{s}(X_i, k) \leq t, \tY_i = l} = \sum_{l,k=1}^K W_{kl} \tilde{\rho}_l \tilde{F}_l^k(t),\\
    \EV{f_{t_1}f_{t_2}} &=
    \sum_{l,k,k'=1}^K  W_{kl} W_{k'l} \tilde{\rho}_l \P{\hat{s}(X_i, k) \leq t_1, \hat{s}(X_i, k') \leq t_2 \mid \tY_i = l}.
\end{align*}
An intuitive plug-in estimate of $G(t_1,t_2)$ is  $\hat{G}(t_1, t_2) = \hat{\mathbb{E}}[f_{t_1}f_{t_2}] - \hat{\mathbb{E}}[f_{t_1}] \hat{\mathbb{E}}[f_{t_2}]$, where
\begin{align*}
    \hat{\mathbb{E}}[f_t] &=
    \sum_{l,k=1}^K W_{kl} \hat{\rho}_l \frac{1}{n_l}
    \sum_{i \in \mathcal{D}_l} \I{\hat{s}(X_i, k) \leq t},\\
    \hat{\mathbb{E}}[f_{t_1}f_{t_2}] &=
    \sum_{l,k,k'=1}^K W_{kl} W_{k'l} \hat{\rho}_l \frac{1}{n_l}
    \sum_{i \in \mathcal{D}_l} \I{\hat{s}(X_i, k) \leq t_1, \,
    \hat{s}(X_i, k') \leq t_2},
\end{align*}
with $\mathcal{D}_l$ denoting the subset of indices $i \in [n]$ such that $\tY_i = l$, and $n_l = |\mathcal{D}_l|$.

\subsection{Monte Carlo Simulation of the Generalized Brownian Bridge}

For a given grid resolution parameter $h \in (0,1)$, define a sequence $0=t_1 < t_2 = h < t_3 = 2h < \dots < t_N =1$ discretizing the interval $[0,1]$ at $N = 1/h+1$ equal-spaced points.
Define also the covariance matrix $\hat{\Sigma} \in \mathbb{R}^{N \times N}$ such that $\hat{\Sigma}_{ij} = \hat{G}(t_i, t_j)$ for all $i,j \in [N]$, and let $\xi^{(1)}, \ldots, \xi^{(M)} \in \mathbb{R}^N$ denote $M$ i.i.d.~realizations of an $N$-dimensional Gaussian random vector $\xi \sim \mathcal{N}(0, \hat{\Sigma})$, with $\xi^{(m)} = (\xi^{(m)}_1,\ldots,\xi^{(m)}_N)$ for each $m \in [M]$.
Then, as a preliminary Monte Carlo estimate of the factor $\delta^{\mathrm{asy}}(n)$ defined in~\eqref{eq:delta_n_asymptotic}, we can compute
\begin{equation}
    \hat{\delta}^{\mathrm{asy}}(n, h) := \frac{1}{\sqrt{n}} \cdot \frac{1}{M} \sum_{m=1}^{M} \max_{i \in [N]} \xi^{(m)}_i,
    \label{eq:delta_n_h}
\end{equation}
highlighting explicitly the dependence of $\hat{\delta}^{\mathrm{asy}}(n, h)$ on the grid resolution parameter $h$.

To minimize the negative bias of $\hat{\delta}^{\mathrm{asy}}(n, h)$, the resolution $h$ should be as small as possible. However, the computational cost increases rapidly with $1/h$. To balance this trade-off, we employ Richardson extrapolation \citep{Richardson1911}, as detailed in Appendix~\ref{appendix:Richardson-extrapolation}.
This leads to a principled solution for estimating $\delta^{\mathrm{asy}}(n)$, which significantly improves the performance of Algorithm~\ref{alg:adaptive-conformal-marg} relative to the finite-sample correction.

\section{Theoretical Upper Bound on Coverage}
\label{sec:coverage-guarantees-upper-bound}

In order to prove that the prediction sets output by Algorithm~\ref{alg:adaptive-conformal-marg} are not overly conservative, we need to assume
that the marginal distributions of the scores with respect to the true and the noisy labels are continuous with strictly positive and bounded densities (Assumption~\ref{assumption:regularity-dist}), and that the calibration sample size is sufficiently large (Assumption~\ref{assumption:regularity-dist-delta-marg}).

\begin{assumption} \label{assumption:regularity-dist}
  The marginal CDFs $F$ and $\tF$ are differentiable on $(0,1)$, and the corresponding densities $f$ and $\tf$ are uniformly bounded with $\| f \|_{\infty} \leq f_{\max}$ and $\| \tf \|_{\infty} \leq \tf_{\max}$, for some $f_{\max}, \tf_{\max} > 0$. Further, $f_{\min} :=  \inf_{t\in(0,1)} f(t) > 0$ and $\tf_{\min} :=  \inf_{t\in(0,1)} \tf(t) > 0$.
\end{assumption}

\begin{assumption}\label{assumption:regularity-dist-delta-marg}
The calibration sample size $n$ is large enough that
$\delta(n) + \sqrt{\frac{\log (2 n)}{2 n}} + d(n) + L \frac{\log n}{n} \leq \alpha$,
where
$d(n) := \frac{1}{\sqrt{n}} \cdot \inf_{\beta} \{ \sqrt{\pi/2} [|\beta_0|+ (\sum_{k=1}^{K} |\beta_k|)/K ] + B(K,n,\beta) \} + \|W\|_1 \sqrt{2 \log(n) / n}$
and $L := (f_{\max} + \tf_{\max})/\tf_{\min}$.
\end{assumption}

\begin{theorem} \label{thm:algorithm-upper-bound}
Under the setup of Theorem~\ref{thm:algorithm-lower-bound}, let $\hat{C}(X_{n+1})$ be the prediction set output by Algorithm~\ref{alg:adaptive-conformal-marg} based on the inverse $W$ of the model matrix $T$. Suppose that Assumptions~\ref{assumption:regularity-dist} and~\ref{assumption:regularity-dist-delta-marg} hold.
Then,
$\Ps{Y_{n+1} \in \hat{C}(X_{n+1})} \leq  1 - \alpha + \delta(n) + \varphi(n)$,
where
\begin{align}
    \varphi(n) =  3 \delta^{**} (n) + \sqrt{\frac{2 \pi}{n}} + \frac{4}{n}
    + \frac{1 + L \left[2 + \log(n+1)\right]}{n+1} = \mathcal{O}\left(\frac{1}{\sqrt{n}}\right),
\end{align}
with $\delta^{**}(n) := \mathbb{E}[ \sup_{t \in [0,1]} | \hat{\psi}(t) | ]$ and $L$ as in Assumption~\ref{assumption:regularity-dist-delta-marg}.
\end{theorem}

Combined with Theorem~\ref{thm:algorithm-lower-bound}, this shows that Algorithm~\ref{alg:adaptive-conformal-marg} produces asymptotically tight prediction sets, whose coverage converges to $1-\alpha$ from above at rate $\mathcal{O}(1/\sqrt{n})$.

\section{Estimating the Label Contamination Process}
\label{sec:contamination_estimation}

Within our distribution-free framework, $T$ is not identifiable without additional assumptions or data with clean labels.
A common approach in the broader literature is to leverage {\em anchor points}, whose clean label can be deduced exactly from $X$ alone (due to a degenerate $\mathcal{P}_{Y \mid X}$) and which can be detected from contaminated data by existing algorithms \citep[e.g.,][]{patrini2017making}.
We adopt the closely related but more transparent assumption of having access to a small auxiliary set $\mathcal{D}^{\mathrm{cl}}$ of $n_{\mathrm{cl}}$ observations $(X_i^{\mathrm{cl}},Y_i^{\mathrm{cl}}, \tY_i^{\mathrm{cl}})$ with clean labels, independent of $\mathcal{D}$ and $\mathcal{D}^{\mathrm{train}}$, which may or may not include the matching contaminated labels $\tY_i^{\mathrm{cl}}$.
This set need not be large, nor representative of the target population: we only assume its observations are i.i.d.~from a distribution $\mathcal{Q} = \mathcal{Q}_X \cdot \mathcal{P}_{Y \mid X} \cdot \mathcal{P}_{\tilde{Y} \mid Y}$ departing from the target $\mathcal{P} = \mathcal{P}_X \cdot \mathcal{P}_{Y \mid X} \cdot \mathcal{P}_{\tilde{Y} \mid Y}$ by a covariate shift.
Such a shift should often be expected in practice, as clean labels may be available only for anchor points or other easier-to-classify observations.
It is also what makes estimating $T$ worthwhile: without it, Algorithm~\ref{alg:adaptive-conformal-marg} could be bypassed by calibrating directly on $\mathcal{D}^{\mathrm{cl}}$, while with it such direct calibration would be anti-conservative; see Sections~\ref{sec:simulations-estimated-contamination} and~\ref{sec:case_study}.
Weighted conformal prediction \citep{tibshirani2019conformal} can correct for covariate shift in general, but only if the density ratio $\mathrm{d}\mathcal{P}_X/\mathrm{d}\mathcal{Q}_X$ is both well-defined and known.
In our setting, that ratio is infinite-dimensional and thus much harder to estimate than $T$, which has at most $K(K-1)$ parameters and a single one under the randomized response model.
We thus consider the estimation of $T$ under two scenarios, depending on whether the contaminated labels $\tY_i^{\mathrm{cl}}$ are observed.

\subsection{Paired Clean and Noisy Labels}
\label{sec:contamination_estimation-paired}
Suppose each observation in $\mathcal{D}^{\mathrm{cl}}$ is a triplet $(X_i^{\mathrm{cl}}, Y_i^{\mathrm{cl}}, \tY_i^{\mathrm{cl}})$, as when a small subset of the contaminated data has been manually verified.
Then, $T$ can be estimated by $\hat{T}_{kl} = (n_{kl}^{\mathrm{cl}} + \lambda) / (n_{l}^{\mathrm{cl}} + \lambda K)$, where $n_{kl}^{\mathrm{cl}} = \sum_{i \in \mathcal{D}^{\mathrm{cl}}} \I{Y_i^{\mathrm{cl}} = l, \tY_i^{\mathrm{cl}} = k}$, $n_{l}^{\mathrm{cl}} = \sum_{i \in \mathcal{D}^{\mathrm{cl}}} \I{Y_i^{\mathrm{cl}} = l}$, and $\lambda \geq 0$ is a Laplace smoothing parameter useful if some classes are rarely observed.
This estimator is consistent under Assumption~\ref{assumption:linear-contam} alone, without further conditions on $\mathcal{Q}_X$: since $T_{kl} = \Ps{\tY = k \mid Y = l}$ does not depend on $X$, it remains valid even if $\mathcal{D}^{\mathrm{cl}}$ over-represents easier observations.
If the parametric form of $T$ is known, its $K(K-1)$ free entries reduce to a lower-dimensional parameter, further relaxing the need for every class to be observed. Under the randomized response model $T(\epsilon) = (1-\epsilon) I_K + (\epsilon/K) J_K$, for example, $T_{kk} = 1-\epsilon+\epsilon/K$ for all $k$, so one can take $\hat{\epsilon} = \frac{K}{K-1} ( 1 - \frac{1}{K} \sum_{k=1}^{K} \hat{T}_{kk} )$.

\subsection{Unpaired Clean and Noisy Labels}
\label{sec:contamination_estimation-unpaired}
A more challenging setting arises when the contaminated labels in $\mathcal{D}^{\mathrm{cl}}$ are not observed.
The estimator of Section~\ref{sec:contamination_estimation-paired} is then unavailable, and $T$ must be recovered by comparing the clean and contaminated label distributions across $\mathcal{D}^{\mathrm{cl}}$ and $\mathcal{D}^{\mathrm{train}}$.
Our approach is to fit a model in which those two ingredients appear as separate components, so that the estimate of $T$ can be read off after training.
Let $\pi(x,k;\theta)$ be a {\em backbone} classifier with parameters $\theta$, e.g., a neural network, estimating $\Ps{Y = k \mid X = x}$, and let $T(\Phi)$ be a {\em contamination layer} with parameters $\Phi$.
Composing them gives a model for the contaminated labels,
\begin{align} \label{eq:composite-model}
  \P{\tY = k \mid X = x; \theta, \Phi}
  = \sum_{l=1}^{K} T_{kl}(\Phi)\, \pi(x,l; \theta),
  \qquad k \in [K],
\end{align}
which can be fitted by minimizing a cross-entropy loss that scores $\pi(\cdot\,;\theta)$ against the clean labels in $\mathcal{D}^{\mathrm{cl}}$ and the composite model~\eqref{eq:composite-model} against the contaminated labels in $\mathcal{D}^{\mathrm{train}}$:
\begin{align} \label{eq:loss}
  \mathcal{L}(\theta, \Phi)
  = \frac{-1}{n_{\mathrm{cl}} + n_{\mathrm{train}}}
    \left[
      \sum_{i \in \mathcal{D}^{\mathrm{cl}}}
      \log \pi(X_i^{\mathrm{cl}}, Y_i^{\mathrm{cl}}; \theta)
      +
      \sum_{i \in \mathcal{D}^{\mathrm{train}}}
       \log \left( \sum_{l=1}^{K} T_{\tY_i l}(\Phi)\, \pi(X_i,l; \theta) \right)
    \right].
\end{align}
Intuitively, the first term anchors the backbone to the conditional distribution of $Y$ given $X$, which is shared by $\mathcal{Q}$ and $\mathcal{P}$, leaving the contamination layer to absorb the discrepancy between clean and contaminated labels. The estimate $\hat{T} = T(\hat\Phi)$ is then inverted and plugged into Algorithm~\ref{alg:adaptive-conformal-marg}.
Appendix~\ref{appendix:nn-details} describes the implementation using a neural network as backbone classifier, including a softmax parametrization ensuring $T(\Phi)$ is always a valid stochastic matrix, its lower-dimensional counterpart for known parametric families, an alternating optimization scheme analogous to the EM algorithm, and a regularization term encouraging $\hat{T}$ to be invertible, which we found useful under severe class imbalance.

\section{Numerical Experiments with Synthetic Data}
\label{sec:numerical_experiments}

\subsection{Setup and Methods under Comparison}
\label{sec:setup_and_methods_under_comparison}
We simulate data with $K=4$ labels and $d=20$ features from a Gaussian mixture distribution using the standard \texttt{make\_classification} function from the Scikit-Learn Python package \citep{pedregosa2011scikit}.
This creates $2K$ clusters of normally distributed points, with random covariance, centered about the vertices of a 10-dimensional hypercube with sides of length 2, and then randomly assigns an equal number of clusters to each of the $K$ classes.
This leads to uniform label frequencies: $\rho_k = 1/K$ for all $k \in [K]$.
The noisy labels $\tilde{Y}$ are generated following the two-level randomized response model defined in Appendix~\ref{appendix:two-level-rmm}, for which we consider different values of the parameters $\epsilon$ and $\nu$.

A random forest classifier implemented by Scikit-Learn is trained on $10,000$ independent observations with contaminated labels generated as described above.
The classifier is then applied to an i.i.d.~calibration data set of size $n$, whose labels are also similarly contaminated, evaluating generalized inverse quantile non-conformity scores \citep{romano2020classification} as detailed in Appendix~\ref{app:adaptive-scores}.
These scores are transformed into prediction sets for 2,000 independent unlabeled test points using the following four alternative approaches.

(i) The {\em Standard} conformal approach that ignores label noise.
(ii) {\em Adaptive}: Algorithm~\ref{alg:adaptive-conformal-marg}, applied with the finite-sample correction factor $\delta^{\mathrm{FS}}(n)$ from Section~\ref{sec:adaptive_pred_finitesample}.
(iii) {\em Adaptive (simplified)}: a less efficient implementation of the latter, with $\delta^{\mathrm{FS}}(n)$ evaluated based on the parameter $\beta$ which optimizes the correction for the randomized response model, regardless of whether that may be optimal for the true contamination model at hand. This helps highlight the distinct importance of solving the optimization problem described in Section~\ref{sec:adaptive_pred_finitesample_generic}.
(iv) {\em Adaptive (asymptotic)}: Algorithm~\ref{alg:adaptive-conformal-marg} applied with the estimated asymptotic correction factor $\hat{\delta}^{\mathrm{asy}}(n)$ from Section~\ref{sec:adaptive_pred_asymptotic}.
All methods target 90\% marginal coverage ($\alpha = 0.1$) and the adaptive approaches use perfect knowledge of the transition matrix $T$.

In Section~\ref{sec:simulations-label-cond}, we also compare with the adaptive method of \citet{sesia2025adaptive}, which provides label-conditional coverage. In Section~\ref{sec:simulations-estimated-contamination} we assess the impact of estimating $T$.

\subsection{The Impact of the Label Contamination Strength}

\begin{figure}[!htb]
\centering
\includegraphics[width=\linewidth]{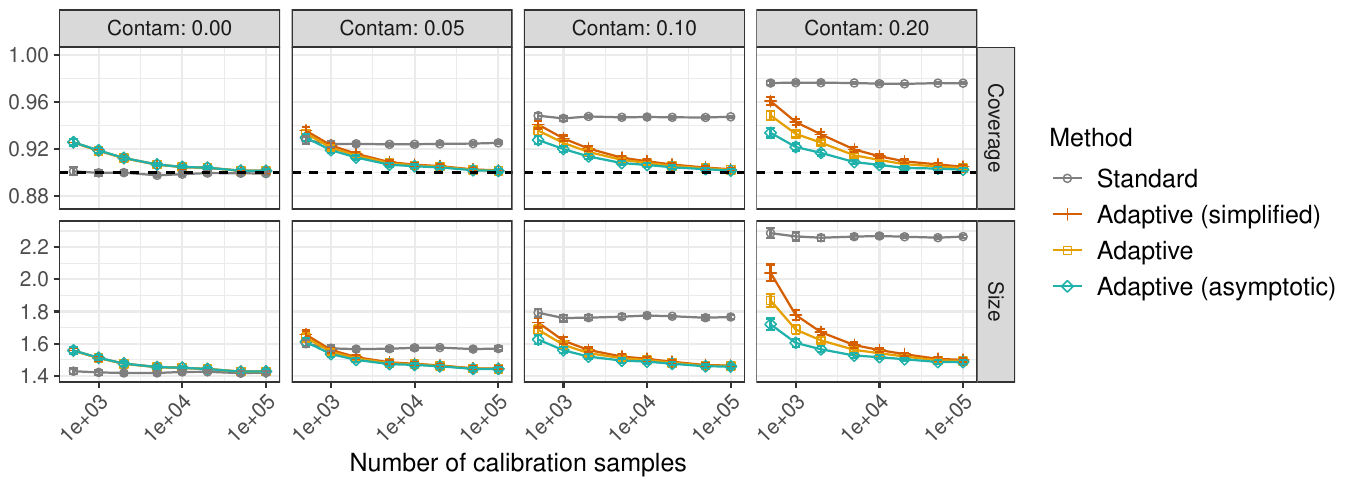}
\caption{Performances of different methods on simulated data with labels contaminated by a two-level randomized response model. The results are stratified based on the strength $\epsilon$ of the label contamination process. The number of classes is $K=4$.}
\label{fig:exp1_synthetic1_ntrain10000_K4_nu0.2_marginal_RRB_optimisticFALSE}
\end{figure}

Figure~\ref{fig:exp1_synthetic1_ntrain10000_K4_nu0.2_marginal_RRB_optimisticFALSE} compares the four methods in terms of coverage and prediction set size, averaged over 20 independent repetitions of each experiment.
The results are shown as a function of $n$ and stratified by the strength $\epsilon$ of the label contamination process. In these experiments, the parameter $\nu$ of the contamination model is fixed equal to 0.2, which corresponds to a moderate deviation from the simple randomized response model.

As expected, the {\em Standard} approach is too conservative when $\epsilon > 0$. Notably, its over-conservativeness persists regardless of increases in the calibration sample size. In contrast, all adaptive methods successfully produce increasingly more informative prediction sets as the calibration sample size grows.
Note that Figure~\ref{fig:exp1_synthetic1_ntrain10000_K4_nu0.2_marginal_RRB_optimisticFALSE} shows no significant performance differences among the three alternative implementations of our adaptive method. This is likely due to the label contamination process being not too far from the randomized response model.
Below, we will explore how deviations from the randomized response model impact the performances of different implementations.
Figures~\ref{fig:exp1_synthetic1_ntrain10000_K4_nu0.2_marginal_uniform_optimisticFALSE} and~\ref{fig:exp1_synthetic1_ntrain10000_K4_nu0.2_marginal_block_optimisticFALSE} in Appendix~\ref{appendix:additional-numerical-results} provide additional results with similar conclusions, from experiments using the randomized response model and the block-randomized response model as contamination models.

\subsection{The Impact of the Label Contamination Model}

\begin{figure}[!htb]
\centering
\includegraphics[width=\linewidth]{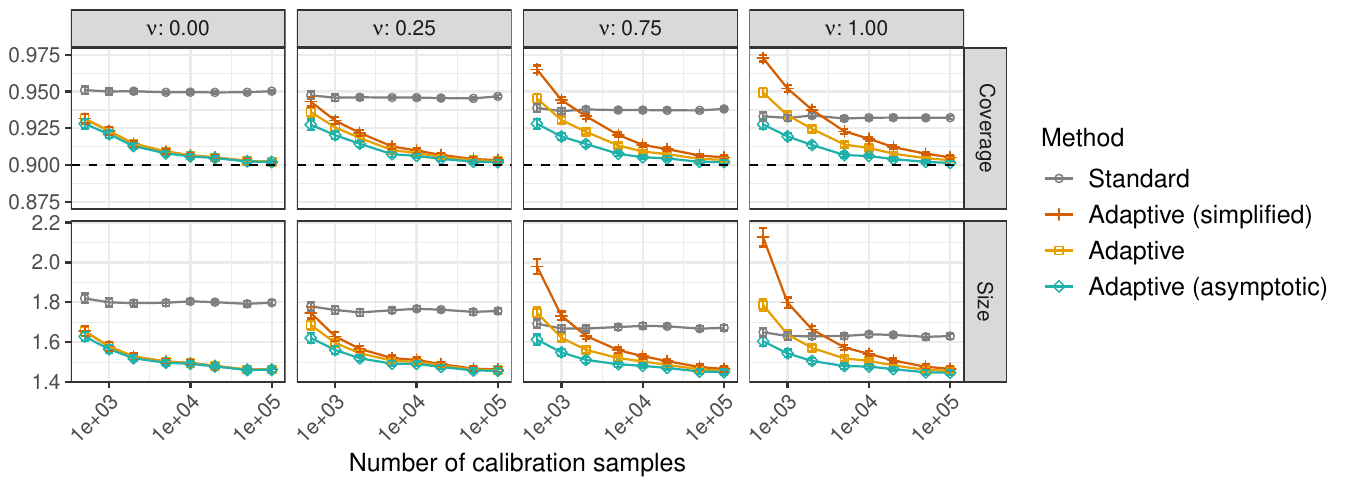}
\caption{Performances of different methods on simulated data with labels contaminated by a two-level randomized response model. The results are stratified based on the contamination model parameter $\nu$. Other details are as in Figure~\ref{fig:exp1_synthetic1_ntrain10000_K4_nu0.2_marginal_RRB_optimisticFALSE}.}
\label{fig:exp2_synthetic1_ntrain10000_K4_eps0.1_marginal_RRB_optimisticFALSE}
\end{figure}

Figure~\ref{fig:exp2_synthetic1_ntrain10000_K4_eps0.1_marginal_RRB_optimisticFALSE} reports on experiments similar to those of Figure~\ref{fig:exp1_synthetic1_ntrain10000_K4_nu0.2_marginal_RRB_optimisticFALSE} but conducted fixing $\epsilon = 0.1$ while varying the parameter $\nu \in [0,1]$, which controls the discrepancy between the two-level randomized response model and the simple randomized response model ($\nu = 0$).

Unsurprisingly, {\em Adaptive (simplified)} becomes noticeably more conservative than the fully optimized implementation of the {\em Adaptive} method as $\nu$ increases, particularly for small $n$.
This occurs because {\em Adaptive (simplified)} relies on a finite-sample correction factor optimized for the randomized response model. The {\em Asymptotic} method increasingly outperforms the other adaptive implementations as $\nu$ grows, primarily because the latter depend on a finite-sample correction factor that, despite our efforts to optimize the constants in the proof of Theorem~\ref{thm:finite-sample-factor}, still relies on theoretical chaining techniques that become less tight when the contamination model deviates significantly from the randomized response model.
Exploring alternative mathematical tools to derive even tighter finite-sample corrections remains an open question.
Figure~\ref{fig:Aexp2_synthetic1_ntrain10000_K4_eps0.1_marginal_RRB_optimisticFALSE} in Appendix~\ref{appendix:additional-numerical-results} gives an alternative visualization of these results, by displaying coverage as function of $\nu$ for different calibration sample sizes.

Appendix~\ref{appendix:impact-number-classes} focuses on the impact of the number of classes: the performance of the {\em Adaptive (simplified)} and {\em Adaptive} implementations generally deteriorates as $K$ grows, particularly when $n$ is small and the label contamination strength is high, while the {\em Asymptotic} implementation outperforms the {\em Standard} method and the other corrections, especially in scenarios with significant deviations from the randomized response model.

\subsection{The Advantage of Optimistic Calibration}\label{sec:simulations-optimistc-calibration}

\begin{figure}[!htb]
\centering
\includegraphics[height=9cm]{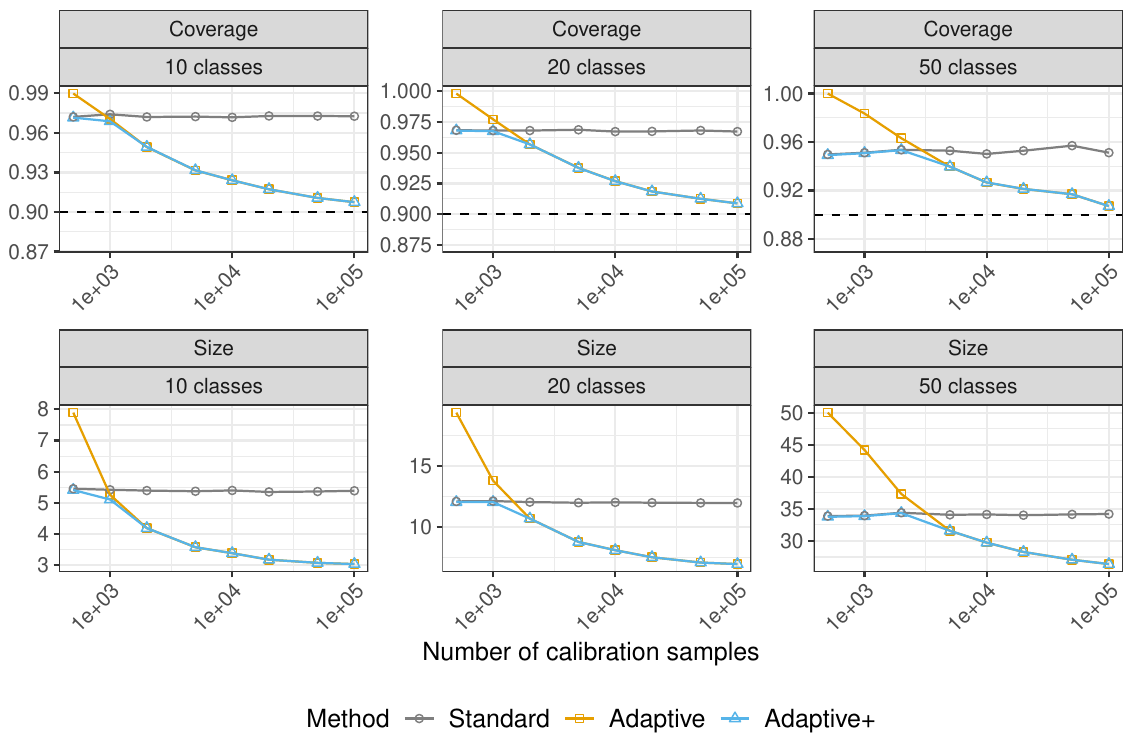}
\caption{Performances of conformal prediction methods on simulated data with contaminated labels. The contamination process is the two-level randomized response model with $\epsilon=0.2$ and $\nu = 0.8$. The reported empirical coverage and average size of the prediction sets are stratified based on the number of possible labels $K$. Other details are as in Figure~\ref{fig:exp1_synthetic1_ntrain10000_K4_nu0.2_marginal_RRB_optimisticFALSE}.}
\label{fig:exp3_synthetic1_ntrain10000_eps0.200000_nu0.8_marginal_RRB_optimisticTRUE}
\end{figure}

Figure~\ref{fig:exp3_synthetic1_ntrain10000_eps0.200000_nu0.8_marginal_RRB_optimisticTRUE} compares the performance of the {\em Adaptive} method from previous experiments with its optimistic counterpart, {\em Adaptive+}, introduced in Section~\ref{sec:adaptive-marginal} and detailed in Appendix~\ref{sec:optimistic-calibration}.
The results indicate that {\em Adaptive+} performs well in practice: it produces more informative prediction sets compared to the {\em Standard} and {\em Adaptive} methods while maintaining valid coverage.
This advantage is particularly evident in scenarios where the {\em Adaptive} method is outperformed by the {\em Standard} method, such as when $n$ is small or $K$ is large, particularly in settings where the contamination model deviates from the randomized response model and the label contamination strength is high.

\subsection{The Advantage of Targeting Marginal Coverage} \label{sec:simulations-label-cond}

In this section, we include in the comparison the ``optimistic'' implementation of the adaptive method proposed by \citet{sesia2025adaptive}, called {\em Adaptive+ (label-cond)}, which constructs conformal prediction sets with label-conditional rather than marginal coverage while accounting for label noise. To facilitate a direct comparison, all methods are still evaluated using the same two metrics: empirical marginal coverage and average prediction set size.

\begin{figure}[!htb]
\centering
\includegraphics[height=9cm]{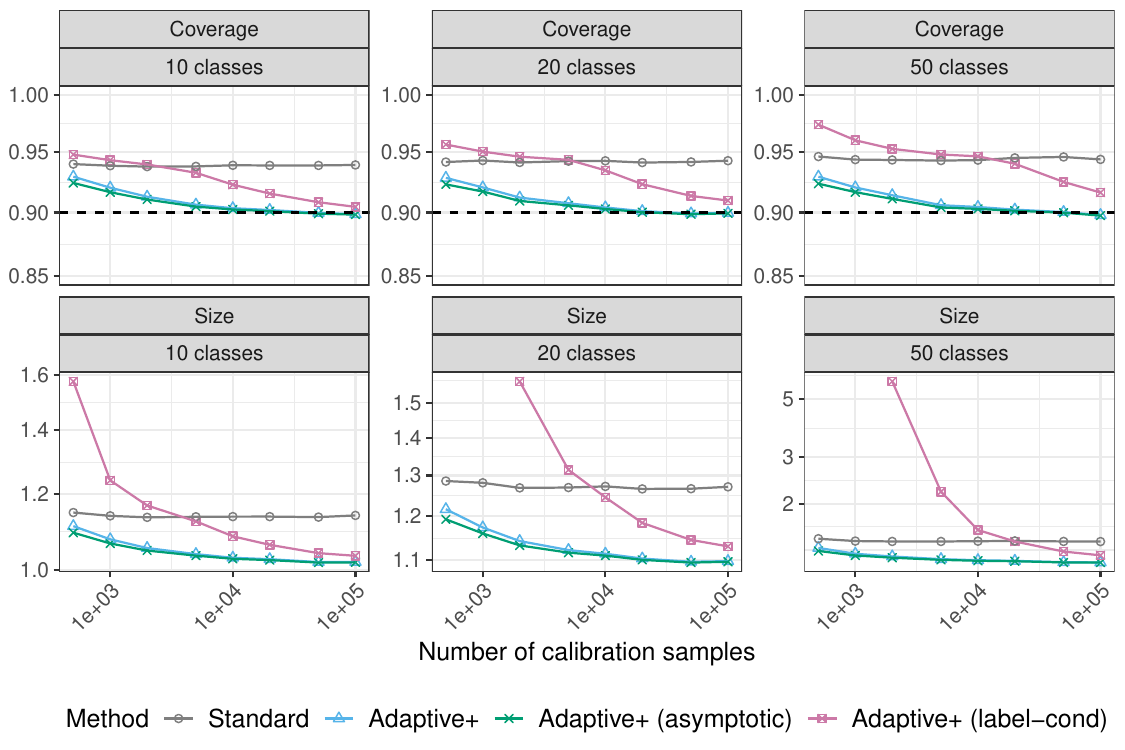}
\caption{Performances of different methods on simulated data with contaminated labels. The label contamination process is the two-level randomized response model with $\epsilon=0.05$ and $\nu = 0.2$. The vertical axis for Size is truncated to highlight differences between the \emph{Standard} and marginal \emph{Adaptive+} methods. Other details are as in Figure~\ref{fig:exp3_synthetic1_ntrain10000_eps0.200000_nu0.8_marginal_RRB_optimisticTRUE}.}
\label{fig:exp4_synthetic1_ntrain10000_eps0.050000_nu0.2_marginal_RRB_optimisticTRUE}
\end{figure}

The results of this comparison, presented in Figure~\ref{fig:exp4_synthetic1_ntrain10000_eps0.050000_nu0.2_marginal_RRB_optimisticTRUE}, demonstrate that the advantage of our method, which focuses on achieving marginal rather than label-conditional coverage, becomes more pronounced in problems with many classes and low calibration sample sizes, where the label-conditional adaptive method fails to produce informative prediction sets.

Additional results with conclusions qualitatively similar to Figure~\ref{fig:exp4_synthetic1_ntrain10000_eps0.050000_nu0.2_marginal_RRB_optimisticTRUE} are presented in Appendix~\ref{appendix:comparison-with-label-cond}.
In particular, Figure~\ref{fig:exp201_synthetic4_ntrain10000_K4_eps0.1_nu0.2_marginal_RRB_optimisticTRUE} shows that conformal prediction sets with label-conditional coverage can become very large as the class imbalance increases, while our method remains informative even under extreme imbalance.

\subsection{The Impact of Estimating the Contamination Model}
\label{sec:simulations-estimated-contamination}

We now evaluate our method when $T$ is estimated from data as explained in Section~\ref{sec:contamination_estimation}.
The data generation and classifier training follow Section~\ref{sec:setup_and_methods_under_comparison}, with $K=4$ and contamination given by a randomized response model with $\epsilon=0.2$.
The clean set $\mathcal{D}^{\mathrm{cl}}$ is obtained by drawing $10\,n_{\mathrm{cl}}$ independent observations and retaining the $10\%$ to which the classifier assigns the highest predicted probability, thus mimicking a realistic scenario in which clean labels are easier to obtain for easier observations.

Alongside {\em Standard} and {\em Adaptive+} (which uses the true $T$), we consider two estimation-based approaches: {\em Adaptive+ (c/n)}, applying the paired estimator of Section~\ref{sec:contamination_estimation-paired} to $\mathcal{D}^{\mathrm{cl}}$; and {\em Adaptive+ (NN)}, applying the unpaired estimator of Section~\ref{sec:contamination_estimation-unpaired} to $\mathcal{D}^{\mathrm{cl}}$ together with $n_{\mathrm{train}}$ noisy observations, using an MLP backbone with two hidden layers of 16 and 8 units and a contamination layer parametrized by the single parameter $\epsilon \in [0,1)$.
We also report {\em Standard (clean, simple)}, obtained by calibrating directly on the clean labels in $\mathcal{D}^{\mathrm{cl}}$.

\begin{figure}[!htb]
\centering
\includegraphics[width=\linewidth]{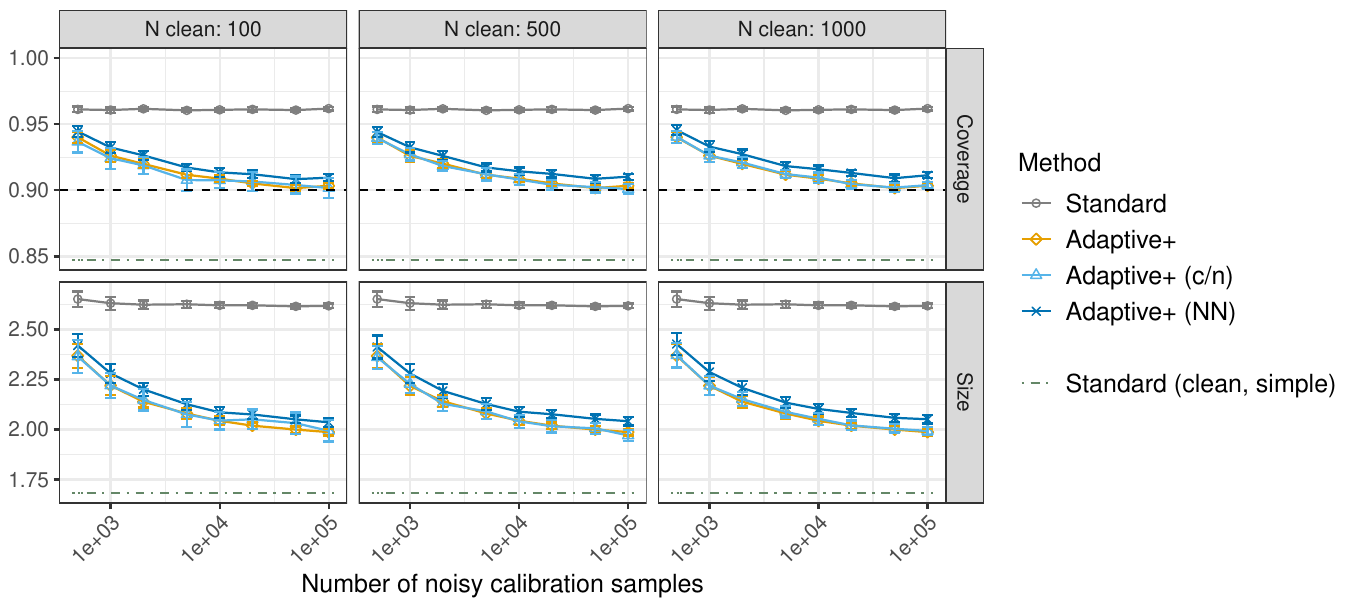}
\caption{Performance of different conformal methods on simulated data with contaminated labels, stratified by the size $n_{\mathrm{cl}}$ of the clean sample used to estimate the contamination model. {\em Adaptive+} uses the true $T$, while {\em Adaptive+ (c/n)} and {\em Adaptive+ (NN)} estimate it from paired and unpaired clean labels, respectively. Here, $n_{\mathrm{train}}=5000$ and $K=4$.}
\label{fig:exp401_synthetic6_K4_nt5000_marginal_uniform_expeasyFALSE}
\end{figure}

Figure~\ref{fig:exp401_synthetic6_K4_nt5000_marginal_uniform_expeasyFALSE} stratifies the results by $n_{\mathrm{cl}}$.
All {\em Adaptive+} variants approach nominal coverage as $n$ grows. {\em Adaptive+ (c/n)} is nearly indistinguishable from the method using the true $T$ for every $n_{\mathrm{cl}}$, indicating that a small paired sample suffices; {\em Adaptive+ (NN)} is slightly more conservative but remains close to both.
By contrast, {\em Standard (clean, simple)} fails to provide valid coverage, as anticipated in Section~\ref{sec:contamination_estimation}.

Appendix~\ref{appendix:additional_results_impact_contamination_estimation} reports on experiments where we vary the number of noisy training samples and the contamination process, leading to similar conclusions (Figures~\ref{fig:exp403_synthetic6_K4_ncl500_marginal_uniform_expeasyFALSE}---~\ref{fig:exp406_synthetic6_K4_nt10000_ncl500_marginal_expeasyFALSE}).
As $n_{\mathrm{train}}$ grows, the performance of {\em Adaptive+ (NN)} approaches that of {\em Adaptive+} with known contamination as long as the noisy sample is large enough (e.g., 5{,}000 observations), even with as few as $n_{\mathrm{cl}}=100$ clean observations and a relatively complicated contamination process.
Appendix~\ref{appendix:effectiveness_of_contamination_estimation} further examines the accuracy of the estimated $T$ itself.

\section{Application to Real Data}
\label{sec:case_study}

We apply our method to the BigEarthNet \citep[BEN]{sumbul2019bigearthnet} data set, deferring a similar analysis of CIFAR-10H \citep{peterson2019human} to Appendix~\ref{appendix:cifar}.
Both data sets contain both clean and contaminated labels, allowing us to assess the robustness of our method when Assumption~\ref{assumption:linear-contam} may not hold exactly and $T$ must be estimated.

Both analyses proceed as follows.
After randomly splitting the data into training, calibration, and test subsets, we identify a set $\mathcal{D}^{\mathrm{cl}}$ of $n_{\mathrm{cl}}$ clean observations among the training data, selected as those to which the classifier assigns the highest predicted probability, and treat the remaining $n_{\mathrm{train}}$ training observations as noisy.
This mimics a realistic scenario in which ground-truth annotations are easier to obtain for easier observations, so that $\mathcal{D}^{\mathrm{cl}}$ is not representative of the test population.
As the contaminated labels in $\mathcal{D}^{\mathrm{cl}}$ are ignored, $T$ is estimated with the unpaired approach of Section~\ref{sec:contamination_estimation-unpaired}, using a contamination layer parametrizing a general transition matrix and a backbone with the same architecture and training parameters as in Section~\ref{sec:simulations-estimated-contamination}, applied to features extracted from the penultimate layer of the pre-trained classifier.
The resulting $\hat{T}$ is then used as a plug-in, and our adaptive method is applied to the calibration data without accessing any clean labels.

BigEarthNet consists of image patches derived from Sentinel-2 satellite imagery (Copernicus Program, European Space Agency), where each 120x120 pixel patch represents a 1.2x1.2 km geographic area. Each patch is associated with multiple land cover classes from the 19-label CORINE Land Cover (CLC) database.
The original annotations were later refined in the Refined BigEarthNet (REBEN) version \citep{clasen2024reben}, which uses the updated CLC2018 database to correct some of the inaccurate labels present in the original data set. As a result, each patch is associated with two sets of labels: one from the original BigEarthNet (BEN 1.0), which can be seen as noisier, and one from REBEN, which is more reliable. We treat the BEN 1.0 labels as a corrupted version of the REBEN labels and construct adaptive prediction sets for the latter.

To simplify the analysis, the 19 original labels are grouped into five broader and mutually exclusive classes: Coast, Waters and Wetlands; Arable Land; Agriculture; Vegetation; and Urban Fabric.
Each patch is assigned a single label, and if not all its original labels from the CLC2018 database fall within the same broader category, the patch is labeled as Mixed. This leads to a classification problem with six classes, whose relative frequencies are highly imbalanced, ranging from approximately 0.69 to 0.001.
In this setting, targeting label-conditional coverage would be impractical, as very rare classes would require a very large sample size, while marginal coverage remains feasible.

To compute the non-conformity scores, we train a ResNet-34 on 25,000 patches with noisy labels, adapting the code of \citet{jerpint-bigearthnet}, whose model was designed for multi-label classification on BigEarthNet but is easily modified for our single-label task.
The model computes non-conformity scores on a separate calibration set, drawn from a pool of approximately 270,000 patches, with the calibration set size varied as a control parameter.
Following the strategy described above, the set of $n_{\mathrm{cl}}=500$ clean observations is selected among $5{,}000$ patches, with the remaining $n_{\mathrm{train}}=4{,}500$ treated as noisy training data.
Under the severe class imbalance of this data set, training the contamination network occasionally converged to singular estimates of $T$; we therefore use the regularization described in Appendix~\ref{appendix:nn-invertibility}, with $\gamma=0.1$.

We compare the {\em Standard} conformal approach with the optimistic implementation of our {\em Adaptive} method described in Appendix~\ref{sec:optimistic-calibration}, applied with both the finite-sample and the asymptotic approximations of $\delta^*(n)$.
We also include the {\em Adaptive+ (label-cond)} method of \citet{sesia2025adaptive}, which targets label-conditional coverage, and {\em Standard (clean, simple)}, obtained by calibrating directly on the clean labels in $\mathcal{D}^{\mathrm{cl}}$.

\begin{figure}[!htb]
\centering
\includegraphics[width=\linewidth]{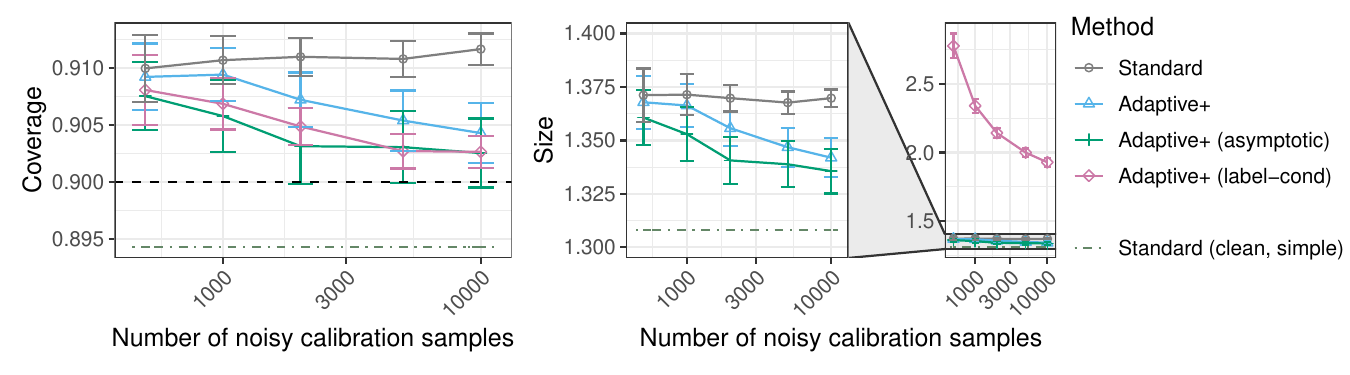}
\caption{Performance of different methods on the BigEarthNet data set with contaminated labels, with $K=6$. The dashed line indicates the nominal 90\% marginal coverage level. In the middle panel, the average prediction set size is zoomed in to better illustrate differences between the \emph{Standard} and marginal \emph{Adaptive+} methods, which would otherwise be obscured by the full scale needed to accommodate the \emph{Adaptive+ (label-cond)} method.}
\label{fig:bigearthnet_oracle_K6_marginal_optimisticTRUE_none_lc}
\end{figure}

Figure~\ref{fig:bigearthnet_oracle_K6_marginal_optimisticTRUE_none_lc} compares the average coverage and size of the prediction sets on a random test set of $2{,}000$ patches, with calibration sample sizes ranging from 500 to $10{,}000$. Each experiment is repeated 100 times with independent random splits of the data into training, calibration, and test sets, re-estimating the contamination matrix every time.
While all adaptive methods correct the over-conservativeness of the standard method and reach valid coverage as the calibration sample size grows, only the marginal adaptive methods produce more informative prediction sets than {\em Standard}, increasingly so as $n$ grows. By contrast, the label-conditional prediction sets are much larger than all others, especially for small $n$.

\section{Discussion} \label{sec:conclusion}

The proposed method leverages data with noisy labels to calibrate informative prediction sets with marginal coverage, even in challenging settings with many labels and potentially high class imbalance.
While our theoretical results assume a known contamination model satisfying Assumption~\ref{assumption:linear-contam}, the method performs well in the applications considered even when that assumption may not hold exactly and $T$ is replaced by an estimate.

Our estimator of $T$ requires a small set of clean observations, possibly drawn from a distribution differing from the target one by a covariate shift.
This approach could be extended further; since $T$ does not depend on $X$ under Assumption~\ref{assumption:linear-contam}, any estimate obtained from a separate but similarly contaminated data set could in principle be transferred to our setting even if the two distributions differ beyond covariate shift, as long as $\mathcal{P}_{\tilde{Y} \mid Y}$ is shared.
In remote sensing, for example, many data sets lack the refinement process applied to BEN \citep{komann2022seasonet,schmitt2019sen12ms} but share the same Sentinel-2 imaging technology and CLC2018-based labeling, so it may be reasonable to transfer an estimate of $T$ obtained from the clean labels in BEN---enabling the use of abundant low-quality data.

Further opportunities for future research include deriving similar guarantees under weaker assumptions, for instance accounting for uncertainty in $T$ \citep{sesia2025adaptive}, and seeking even tighter finite-sample bounds for the correction factor.
In particular, our correction factor does not vanish as the contamination level tends to zero, unlike the (heuristic) one of \citet{clarkson2024split}, because we bound the empirical process in~\eqref{eq:hat-psi} uniformly over all candidate thresholds, but one could attempt to localize the calibrated threshold within a window whose width shrinks with the contamination level. A localized version of our analysis along these lines may yield a rigorous correction that vanishes with the contamination, but this would likely require additional regularity conditions in the spirit of Assumption~\ref{assumption:regularity-dist}, and we leave its development to future work.


\section*{Software Availability}

An open-source software implementation of the methods described in this paper, along  with the code needed to reproduce all numerical results, are available online at \url{https://github.com/tbortolotti/marginal-noise-adaptive-conformal}.

\section*{Acknowledgments}
The authors thank three anonymous referees and an associate editor for their constructive feedback on an earlier version of this manuscript.
T.~B., A.~M., and S.~V.~were partly supported by ACCORDO Attuativo ASI-POLIMI ``Attivit{\`a} di Ricerca e Innovazione" n.~2018-5-HH.0, collaboration agreement between the Italian Space Agency and Politecnico di Milano, and by the initiative ``Dipartimento di Eccellenza 2023–2027", MUR, Italy, Dipartimento di Matematica, Politecnico di Milano. They also acknowledge the financial support of IREA-CNR (Istituto per il Rilevamento Elettromagnetico dell'Ambiente del Consiglio Nazionale delle Ricerche) for funding a PhD grant in cooperation with Politecnico di Milano.
M.~S.~was partly supported by NSF grant DMS 2210637, by a Capital One CREDIF Research Award, and by a Google Research Scholar award.
\bibliographystyle{Chicago}

\bibliography{bibliography.bib}

\newpage

\appendix
\setcounter{theorem}{0}
\setcounter{lemma}{0}
\setcounter{proposition}{0}
\setcounter{equation}{0}
\setcounter{figure}{0}
\setcounter{algocf}{0}

\renewcommand{\thetheorem}{A\arabic{theorem}}
\renewcommand{\thelemma}{A\arabic{lemma}}
\renewcommand{\theproposition}{A\arabic{proposition}}
\renewcommand{\thecorollary}{A\arabic{corollary}}
\renewcommand{\theequation}{A\arabic{equation}}
\renewcommand{\thefigure}{A\arabic{figure}}
\renewcommand{\thealgocf}{A\arabic{algocf}}

\newcommand{\cA}{\mathcal{A}}
\newcommand{\cC}{\mathcal{C}}
\newcommand{\cD}{\mathcal{D}}
\newcommand{\cF}{\mathcal{F}}
\newcommand{\cI}{\mathcal{I}}
\newcommand{\hDelta}{\hat{\Delta}}

\section{Review of Existing Methods}
\label{appendix:standard-marg}
\subsection{Standard Conformal Classification with Marginal Coverage}

\begin{algorithm}[H]
\DontPrintSemicolon

\KwIn{Data set $\{(X_i, \tilde{Y}_i)\}_{i=1}^{n}$ with observable labels $\tilde{Y}_i \in [K]$.}
\myinput{Unlabeled test point with features $X_{n+1}$.}
\myinput{Pre-trained $K$-class classification model $\hat{\pi}$. Prediction function $\mathcal{C}$.}
\myinput{Desired miscoverage probability $\alpha \in (0,1)$.}

Compute the scores $\hat{s}(X_i, \tilde{Y}_i)$ using \eqref{eq:conf-scores}, for all $i \in \mathcal{D}$.\;
Define $\htau^{\mathrm{std}}$ as the $\lceil (1+n)\cdot(1-\alpha) \rceil$ smallest value in $ \{\hat{s}(X_i,\tilde{Y}_i)\}_{i \in \mathcal{D}} \cup \{1\}$.\;
\nonl
\textbf{Output: } Conformal prediction set $\hat{C}^{\mathrm{std}}(X_{n+1}) = \mathcal{C}(X_{n+1}, \htau^{\mathrm{std}};  \hat{\pi})$ for $\tilde{Y}_{n+1}$.\;

\caption{Standard conformal classification with marginal coverage.} \label{alg:standard-marg}
\end{algorithm}

\begin{proposition}[e.g., from \cite{lei2013distribution} or \cite{romano2020classification}] \label{prop:standard-coverage-marginal}
If the data pairs $(X_i,\tilde{Y}_i)$, for all $i \in [n+1]$, are exchangeable random samples from some joint distribution, the prediction set $\hat{C}^{\mathrm{std}}(X_{n+1})$ output by Algorithm~\ref{alg:standard-marg} has marginal coverage for the observable labels $\tilde{Y}$; i.e.,
$\P{ \tilde{Y}_{n+1} \in \hat{C}^{\mathrm{std}}(X_{n+1}) } \geq 1-\alpha$.
Further, if all scores $\hat{s}(X_i,\tilde{Y}_i)$ computed by Algorithm~\ref{alg:standard-marg} are almost-surely distinct,
$\P{ \tilde{Y}_{n+1} \in \hat{C}^{\mathrm{std}}(X_{n+1})} \leq 1-\alpha + 1/(n+1)$,
where $n = |\mathcal{D}|$ is the number of data points in the calibration set.
\end{proposition}

\subsection{Generalized Inverse Quantile Non-conformity Scores} \label{app:adaptive-scores}

We briefly review here the definition of the generalized inverse quantile non-conformity scores proposed by~\cite{romano2020classification}, which are used in the empirical demonstrations presented in this paper.
These non-conformity scores are designed to produce more flexible prediction sets that can account for possible heteroscedasticity in the distribution of $Y \mid X$.

For any $x \in \mathbb{R}^{d}$ and $t \in [0,1]$, define
\begin{align} \label{eq:oracle-threshold}
  \hat{Q}(x, \hat{\pi}, t) & = \min \{ k \in \{1,\ldots,K\} : \hat{\pi}_{(1)}(x) + \hat{\pi}_{(2)}(x) + \ldots + \hat{\pi}_{(k)}(x) \geq t \},
\end{align}
where $\hat{\pi}_{(1)}(x) \geq \ldots \geq \hat{\pi}_{(K)}(x)$ are the descending order statistics of $\hat{\pi}(x,1), \ldots, \hat{\pi}(x,K)$.
Intuitively, $\hat{Q}(x, \hat{\pi}, \cdot)$ may be seen as a generalized quantile function.
Similarly, let $\hat{r}(x, \hat{\pi}, k)$ denote the rank of $\hat{\pi}(x,k)$ among $\hat{\pi}(x,1), \ldots, \hat{\pi}(x,K)$.
With this notation, one can also define a corresponding generalized cumulative distribution function:
\begin{align*}
  \hat{\Pi}(x,\hat{\pi},k) = \hat{\pi}_{(1)}(x) + \hat{\pi}_{(2)}(x) + \ldots + \hat{\pi}_{(\hat{r}(x, \hat{\pi}, k))}(x).
\end{align*}
Then, the function $\mathcal{C}$ proposed by~\cite{romano2020classification} can be written as:
\begin{align}  \label{eq:pred-sets-aps}
  \mathcal{C}(x, \tau; \hat{\pi})
  & = \{ k \in [K] : \hat{r}(x, \hat{\pi}, k) \leq \hat{Q}(x, \hat{\pi}, \tau)\},
\end{align}
and the corresponding non-conformity scores defined in~\eqref{eq:conf-scores} can be evaluated efficiently by noting that $\hat{s}(x,k) = \hat{\Pi}(x,k)$; see \cite{romano2020classification} for further details.

The prediction function defined in~\eqref{eq:pred-sets-aps} may be understood by noting that $\mathcal{C}(x, \tau; \hat{\pi})$ lists the most likely classes according to $\hat{\pi}(x)$ up until the first label $l$ for which $\hat{\Pi}(x,\hat{\pi},l) \geq \tau$.
Therefore, in the ideal case where $\hat{\pi}(x,k) = \P{Y = k \mid X=x}$, one can verify that $\mathcal{C}(x, \tau; \hat{\pi})$ is the smallest possible (deterministic) prediction set for $Y$ with perfect object-conditional coverage at level $\tau$, i.e., satisfying $\P{Y \in \mathcal{C}(x, \tau; \hat{\pi}) \mid X=x} \geq \tau$.

\cite{romano2020classification} also developed a more powerful randomized version of \eqref{eq:pred-sets-aps} that enjoys similar theoretical properties while being able to produce even more informative prediction sets, and that is the practical approach followed in the empirical demonstrations presented in this paper.
We review it explicitly here because the additional randomness is what makes the corresponding non-conformity scores continuously distributed.
Letting $u \in (0,1)$ denote the realization of a uniform random variable $U \sim \text{Unif}(0,1)$ independent of everything else, the randomized prediction sets of~\cite{romano2020classification} discard the least likely included label whenever $u \leq V(x, \hat{\pi}, \tau)$, where
\begin{align} \label{eq:aps-overshoot}
  V(x, \hat{\pi}, \tau) = \frac{1}{\hat{\pi}_{(\hat{Q}(x, \hat{\pi}, \tau))}(x)} \left[ \sum_{k=1}^{\hat{Q}(x, \hat{\pi}, \tau)} \hat{\pi}_{(k)}(x) - \tau \right] \in [0,1);
\end{align}
that is,
\[
  \mathcal{C}(x, u, \tau; \hat{\pi}) = \{ k \in [K] : \hat{r}(x, \hat{\pi}, k) \leq \hat{Q}(x, \hat{\pi}, \tau) - \I{u \leq V(x, \hat{\pi}, \tau)} \}.
\]
As shown by~\citet{einbinder2022training}, the non-conformity scores corresponding to these randomized prediction sets can also be written in closed form:
\begin{align} \label{eq:aps-score-rand}
  \hat{s}(x,k,u) = \hat{\Pi}(x,\hat{\pi},k) - u \cdot \hat{\pi}(x,k),
\end{align}
which reduces to the deterministic score $\hat{s}(x,k) = \hat{\Pi}(x,\hat{\pi},k)$ when $u = 0$.
Assume now that $\hat{\pi}(x,k) > 0$ for all $x$ and $k$, as is automatically the case if $\hat{\pi}$ is the output of a soft-max layer.
Conditional on $(X,Y) = (x,k)$, the score $\hat{s}(x,k,U)$ is a strictly decreasing affine function of $U$, and hence it is uniformly distributed on an interval of positive length $\hat{\pi}(x,k)$.
It follows that the distribution of $\hat{s}(X,Y,U)$ is continuous.

\clearpage

\section{Additional Methodological Details}
\label{appendix:additional-content-for-adaptive-methods}

\subsection{The Conservativeness of Standard Conformal Prediction}
\label{appendix:conservativeness-standard}

For any $k,l \in [K]$ and $t \in [0,1]$, recall the definitions of $F_l^k(t)$ and $\tilde{F}_l^k(t)$ from~\eqref{eq:cdf-scores}:
  \begin{align*}
   F_l^k(t) := \P{\hat{s}(X,k) \leq t \mid Y=l}, \qquad
   & \tilde{F}_l^k(t) := \P{\hat{s}(X,k) \leq t \mid \tY = l}.
 \end{align*}

\begin{corollary} \label{cor:coverage-marg}
Consider the same setting of Theorem~\ref{thm:coverage-marginal} and assume Assumption~\ref{assumption:linear-contam} holds.
Suppose also that the cumulative distribution functions defined in \eqref{eq:cdf-scores} satisfy
    \begin{equation} \label{eq:assump_scores-cond-marg}
        \max_{l \neq k} F_k^{l}(t)  \leq F_k^{k}(t),
      \end{equation}
      for all $t\in\mathbb{R}$ and $k\in[K]$.
Then, $\Delta(\htau^{\mathrm{std}}) \geq 0$ almost-surely, and hence the predictions $\hat{C}^{\mathrm{std}}(X_{n+1})$ of Algorithm~\ref{alg:standard-marg} have valid marginal coverage.
\end{corollary}

Intuitively, \eqref{eq:assump_scores-cond-marg} states that the scores $\hat{s}(X,k)$ assigned by the machine learning model tend to be smaller than any other score $\hat{s}(X,l)$ for $l \neq k$ among data points with true label $Y=k$. In other words, this could be interpreted as saying that the correct label is the most likely point prediction of the machine learning model, for each possible class $k \in [K]$.

\subsection{Relation to Weighted Conformal Prediction}
\label{appendix:weighted-cp-comparison}

It is natural to ask whether our method can be recast as a special case of weighted conformal prediction \citep{tibshirani2019conformal}. We show here that it cannot: under our assumptions, the weights required by weighted conformal prediction are not computable, as they depend on the unknown conditional distribution of $Y \mid X$. Making them computable would require assuming those probabilities known, which is unrealistic and would trivialize the entire classification problem.

While originally developed for covariate shift, weighted conformal prediction applies generally to data $Z_1,\ldots,Z_{n+1}$ with a known (possibly unnormalized) joint density $f$ \citep[see, e.g.,][]{sesia2026elements}, assigning to the $i$-th observation a weight given by the function
\begin{align}\label{eq:weights_sesia_favaro}
    w_i^f(z) := \frac{\sum_{\sigma \in \mathcal{S}_{n+1}: \sigma(n+1)=i} f(z_{\sigma(1)},\dots,z_{\sigma(n+1)})}{\sum_{\sigma \in \mathcal{S}_{n+1}} f(z_{\sigma(1)},\dots,z_{\sigma(n+1)})},
\end{align}
where $z=(z_1,\ldots,z_{n+1})$ denotes a realization of the data and $\mathcal{S}_{n+1}$ is the set of all permutations of $[n+1]$.
In our setting the calibration points carry contaminated labels while the test point carries a clean label, so $Z_i = (X_i,\tY_i)$ for $i \in [n]$ and $Z_{n+1}=(X_{n+1},Y_{n+1})$. Therefore, the joint density factorizes as
\begin{align} \label{eq:weighted-factorization}
    f(z_1,\dots,z_{n+1})
    \propto \underbrace{\mathcal{P}_X(x_{n+1})\, \pi(x_{n+1}, y'_{n+1})}_{=: \, h(z_{n+1})}
    \prod_{j=1}^{n} \underbrace{\mathcal{P}_X(x_j) \sum_{l=1}^K T_{y'_j l}\, \pi(x_j, l)}_{=: \, g(z_j)},
\end{align}
writing $\pi(x,l) := \Ps{Y = l \mid X=x}$ and $z_i=(x_i,y'_i)$; the two distinct factors reflect the clean test point and the contaminated calibration points.
Since all $n!$ permutations with $\sigma(n+1)=i$ contribute equally, the numerator of~\eqref{eq:weights_sesia_favaro} equals $n! \, [h(z_i)/g(z_i)] \prod_{j} g(z_j)$, and the factor $n! \prod_j g(z_j)$ cancels with the denominator, yielding
\begin{align} \label{eq:weights-label-noise}
    w_i^f(z) = \frac{\bar{\nu}_i(z)}{\sum_{j=1}^{n+1} \bar{\nu}_{j}(z)},
    \qquad
    \bar{\nu}_i(z) := \frac{h(z_i)}{g(z_i)} = \frac{\pi(x_i, y'_i)}{\sum_{l=1}^K T_{y'_i l} \, \pi(x_i, l)}.
\end{align}
In the noiseless case $T=I$ we recover $\bar{\nu}_i(z) = 1$ and hence the uniform weights $1/(n+1)$ of standard conformal prediction, as expected.
For $T \neq I$, however, the weights~\eqref{eq:weights-label-noise} depend on $\pi$, which cannot be assumed known without giving away the classification problem itself. Weighted conformal prediction is therefore unable to account for label noise even when the contamination mechanism $T$ is known, unlike our method.
This sets the label noise problem apart from other distribution shift problems, including covariate shift \citep{tibshirani2019conformal} and label shift \citep{podkopaev2021distribution}, where the weights needed for weighted conformal prediction are {\em pivotal} quantities \citep{sesia2026elements} that do not depend on the most sensitive unknown parameters of the data distribution.

\subsection{Optimistic Adaptive Calibration} \label{sec:optimistic-calibration}

This appendix describes an alternative method which can outperform both the standard conformal method and the adaptive method introduced in this paper. We show that, under a reasonable stochastic dominance assumption, it is safe to ``greedily'' choose between the standard and the adaptive method depending on which leads to a lower calibrated threshold; that is, a smaller prediction set.

Concretely, we propose applying Algorithm~\ref{alg:adaptive-conformal-marg} with the set $\hat{\mathcal{I}}$ from~\eqref{eq:I-set-marg} replaced by
\begin{align} \label{eq:I-set-marg-optimistic}
  \hat{\mathcal{I}} := \left\{i \in [n] : \frac{i}{n} \geq 1 - \alpha - \max\left\{\hat{\Delta}(S_{(i)}) - \delta(n), -\frac{1-\alpha}{n} \right\}  \right\}.
\end{align}
This preserves marginal coverage up to a relatively negligible gap of order $\mathcal{O}(n^{-3/2})$,
which we find to be practically unobservable in our empirical validations.

\begin{proposition} \label{prop:algorithm-correction-optimistic}
Let $\hat{C}(X_{n+1})$ be the prediction set output by Algorithm~\ref{alg:adaptive-conformal-marg}
applied with the set $\hat{\mathcal{I}}$ in~\eqref{eq:I-set-marg-optimistic} instead
of~\eqref{eq:I-set-marg}, and with a correction factor $\delta^{*}(n) \leq \delta(n) \leq \alpha$.
Define
\begin{gather*}
    \gamma_n := \frac{1}{2\sqrt{2}\, n^{3/2} \sqrt{\log n}},
    \qquad
    \epsilon_n := \sqrt{\frac{\log(2/\gamma_n)}{2n}}, \\
    \mathcal{T}_n := \left\{ t \in [0,1] : 1 - \alpha - \epsilon_n \leq \tilde{F}(t) \leq 1 - \alpha + \frac{2}{n} + \epsilon_n \right\}.
\end{gather*}
Under the setup of Theorem~\ref{thm:algorithm-lower-bound}, assume also that
\begin{align}
\label{eq:optimistic-assumption}
    \inf_{t \in \mathcal{T}_n} \Delta(t) \; \geq \; \delta(n) - \frac{1-\alpha}{n} + \sqrt{\frac{\log n}{2n}} .
\end{align}
Then,
\begin{align*}
    \P{Y_{n+1} \in \hat{C}(X_{n+1})} \; \geq \; 1 - \alpha - (3 + \delta(n))\, \gamma_n =  1 - \alpha.
\end{align*}
\end{proposition}

\begin{corollary}
\label{cor:optimistic-exact}
Under the assumptions of Proposition~\ref{prop:algorithm-correction-optimistic}, if the
correction factor satisfies the slightly stronger condition
$\delta(n) \geq \delta^{*}(n) + (3 + \delta(n))\gamma_n$, then
$\P{Y_{n+1} \in \hat{C}(X_{n+1})} \geq 1 - \alpha$.
\end{corollary}

Note that $\mathcal{T}_n$ is an interval, since $\tilde{F}$ is non-decreasing, and that it
shrinks around the $(1-\alpha)$-quantile of $\tilde{F}$ at rate $\epsilon_n = \mathcal{O}(\sqrt{\log(n)/n})$;
it is precisely the region in which the standard conformal threshold lies with high probability.
The assumption in~\eqref{eq:optimistic-assumption} is related to the stochastic dominance condition~\eqref{eq:assump_scores-cond-marg} in Corollary~\ref{cor:coverage-marg} and typically stronger, but remains reasonable. If the calibration set size is large enough to make $\epsilon_n$ and $\delta(n) + \sqrt{\log(n) / (2n)}$ small, the assumption is not much stronger than~\eqref{eq:assump_scores-cond-marg}, which implies $\inf_{t\in [0,1]}\Delta(t) \geq 0$.

\subsection{Solving the Optimization Problem} \label{app:optimization}

The optimal choice of $\beta$ in~\eqref{eq:delta_n_finitesample} depends on the structure of the contamination model through the inverse transition matrix $W$.
Fortunately, the finite-sample correction term $\delta^{\mathrm{FS}}(n)$ in~\eqref{eq:delta_n_finitesample} can be practically computed by solving a convex optimization problem.

Let us define
\begin{align} \label{eq:delta_n_finitesample_1}
    \delta^{\mathrm{FS}}_1(n) & :=
        \inf_{\beta \in \mathbb{R}^{K+1}} \left\{ c(n) \left(\lvert \beta_0 \rvert +\frac{\sum_{k=1}^{K} \lvert \beta_k \rvert }{K} \right) + \frac{2}{\sqrt{n}} \max_{l \in [K]} \sum_k |\Omega_{kl}| \sqrt{2 \log(2K n + 2)} \right\},
\end{align}
and
\begin{align} \label{eq:delta_n_finitesample_2}
\begin{split}
\delta^{\mathrm{FS}}_2(n)  & := 
    \inf_{\beta \in \mathbb{R}^{K+1}} \bigg\{ c(n) \left(\lvert \beta_0 \rvert +\frac{\sum_{k=1}^{K} \lvert \beta_k \rvert }{K} \right) \\
  & \qquad\qquad\qquad +  \frac{48}{\sqrt{n}} \max_{l \in [K]} \sum_k |\Omega_{kl}| \left( \sqrt{\log (K + 2)} + \frac{1}{\sqrt{\log (K + 2)}} \right)  \bigg\}.
\end{split}
\end{align}
Clearly, $\delta^{\mathrm{FS}}(n) \leq \min \left\{ \delta^{\mathrm{FS}}_1(n), \delta^{\mathrm{FS}}_2(n) \right\}$.
Moreover, the optimization problems in~\eqref{eq:delta_n_finitesample_1} and~\eqref{eq:delta_n_finitesample_2} are both convex.
This suggests tackling the optimization problem in~\eqref{eq:delta_n_finitesample} by solving separately the two convex problems~\eqref{eq:delta_n_finitesample_1} and~\eqref{eq:delta_n_finitesample_2} and then taking the value of $\beta$ corresponding to smallest of the two optimal values.

Intuitively, the more the contamination process resembles the randomized response model, the more similar the behavior of $\delta^{\mathrm{FS}}(n)$ is to that of $c(n)$. 
If the deviation of the contamination model from the randomized response model is non-negligible -- that is, when the matrix $\Omega$ amplifies the second term in~\eqref{eq:delta_n_finitesample} -- the worst case behavior of the correction factor is $\min \{ \mathcal{O}( \sqrt{(K \log K)/n} ), \mathcal{O}( \sqrt{\log (K n) / n} ) \}$, which still scales well with $n$, especially in those cases where $K$ is considerably lower than $n$.
 We argue in Appendix~\ref{appendix:Special-cases} that, for some special contamination models, the correction factors display improved scaling properties than that of the general case presented here; for the two-level randomized response model, for example, the finite-sample correction factor is proved to scale as $\mathcal{O} ( \sqrt{(\log K)/n })$.

\clearpage

\subsection{Simplified Methods for Special Contamination Models}
\label{appendix:Special-cases}

\subsubsection{Block-Randomized Response Model}\label{appendix:block-rrm}

Recall that $T_{kl} = \mathbb{P}[\tY = k \mid Y = l]$. The Block-Randomized response model with $b$ blocks of dimension $m := K/b$ is identified by the matrix $T$ in the form
\begin{equation}\label{eq:mblock-randomized}
    T = (1 - \epsilon) I_K + \frac{\epsilon}{m} \cdot B_{b},
\end{equation}
where $B_{b}$ is a block-diagonal matrix with $b$ constant blocks equal to $J_m$, i.e., the $(m \times m)$ matrix of ones.

It is possible to verify that matrix $W$ corresponding to model~\eqref{eq:mblock-randomized} reads
\begin{equation*}
    W = \frac{1}{1-\epsilon} I_K - \frac{\epsilon}{m(1-\epsilon)} B_{b}.
\end{equation*}
Consider for instance the case $K=4$ and $b=2$, so that $m=2$. Then the matrix $\Omega$ is

{\small
\begin{align*}
  \Omega =
  \scalebox{0.85}{$
  \begin{bmatrix}
    \frac{1}{1 -\epsilon} - \frac{\epsilon}{2(1-\epsilon)} - \beta_0 -\frac{\beta_1}{4}  & -\frac{\epsilon}{2(1-\epsilon)} -\frac{\beta_1}{4}  & -\frac{\beta_1}{4}  & -\frac{\beta_1}{4} \\ 
    - \frac{\epsilon}{2(1-\epsilon)} -\frac{\beta_2}{4}  & \frac{1}{1 -\epsilon} - \frac{\epsilon}{2(1-\epsilon)} - \beta_0 -\frac{\beta_2}{4} & -\frac{\beta_2}{4}  & -\frac{\beta_2}{4} \\ 
    -\frac{\beta_3}{4}  & -\frac{\beta_3}{4}  & \frac{1}{1 -\epsilon} - \frac{\epsilon}{2(1-\epsilon)} - \beta_0 -\frac{\beta_3}{4}  & -\frac{\epsilon}{2(1-\epsilon)} -\frac{\beta_3}{4} \\ 
    -\frac{\beta_4}{4}  & -\frac{\beta_4}{4}  & -\frac{\epsilon}{2(1-\epsilon)} -\frac{\beta_4}{4}  & \frac{1}{1 -\epsilon} - \frac{\epsilon}{2(1-\epsilon)} - \beta_0 -\frac{\beta_4}{4}\\ 
  \end{bmatrix}.
  $}
\end{align*}
}


We are interested in studying the finite-sample correction in~\eqref{eq:delta_n_finitesample} for this special case. Assume that the number $b$ of blocks is fixed, and that $m = K/b$ is an integer. Set $\beta'_0 = \frac{1}{1-\epsilon}$ and $\beta'_k = -\frac{K \epsilon}{m(1-\epsilon)}$ for all $k$, and let $\beta' = (\beta'_0, \beta'_1,\dots, \beta'_K)$. Note that $|\beta'_0| + \frac{1}{K} \sum_k |\beta'_k| = (1+b\epsilon)/(1-\epsilon)$.

Consider the $\|\Omega\|_1$ term in~\eqref{eq:B-definition}, evaluated at $\beta'$:
\begin{align*}
  \|\Omega\|_1
  & = \max_{l \in [K]} \sum_k |\Omega_{kl}| 
    = \frac{K-m}{K} \frac{K \epsilon}{m (1-\epsilon)} 
    = \frac{K-m}{m} \cdot \frac{\epsilon}{1-\epsilon}
    = \left(b-1 \right) \cdot \frac{\epsilon}{1-\epsilon}.
\end{align*}
Therefore, it follows from the definition of the correction term in~\eqref{eq:delta_n_finitesample} that
\begin{align*}
    \delta^{\mathrm{FS}}(n, K)
    &\leq \frac{1 + b \epsilon }{1-\epsilon} \cdot c(n)  + \frac{1}{\sqrt{n}} B(K,n,\beta'),
\end{align*}
where
\begin{equation}
    B(K,n, \beta') \leq \frac{48 (b-1)\epsilon}{1-\epsilon} \left(\sqrt{\log (K + 2)} + \frac{1}{\sqrt{\log (K + 2)}} \right).
    \label{eq:B-definition-RRM}
\end{equation}
This shows that, in the simplified case of the block-randomized response model, the finite-sample correction scales as $\mathcal{O}(\sqrt{(\log K) / n})$, for any fixed $\epsilon$ and $b$.

\begin{figure}
    \centering
    \includegraphics[width=0.85\linewidth]{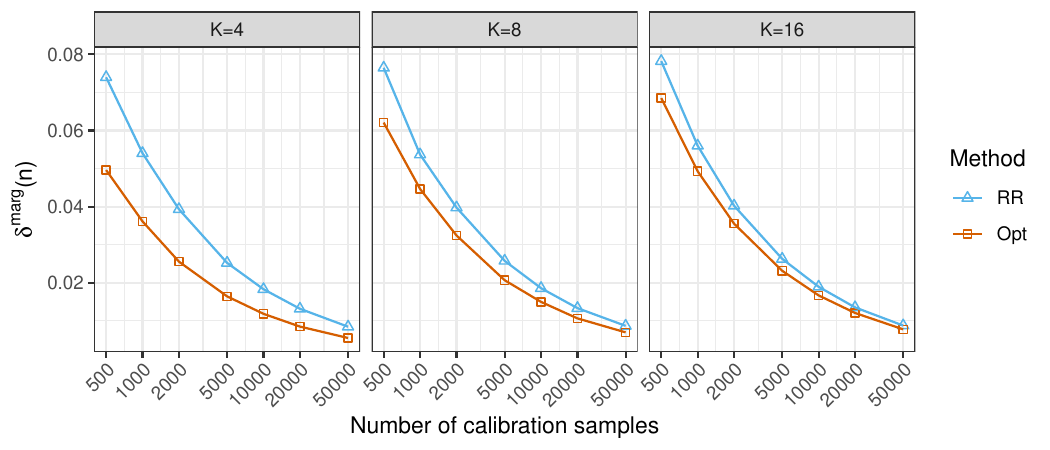}
    \caption{Finite-sample correction for the block randomized response model, estimated using the $\beta$ which optimizes $\delta^{\mathrm{FS}}$ for the randomized response model (RR) and the $\beta$ resulting from the minimization problem in~\eqref{eq:delta_n_finitesample} (Opt). The estimates are displayed as function of the sample size, for different values of $K$. In these experiments, the number of blocks is $b=2$ and $\epsilon = 0.1$.}
    \label{fig:delta_B_Klab}
\end{figure}

\begin{figure}
    \centering
    \includegraphics[width=0.85\linewidth]{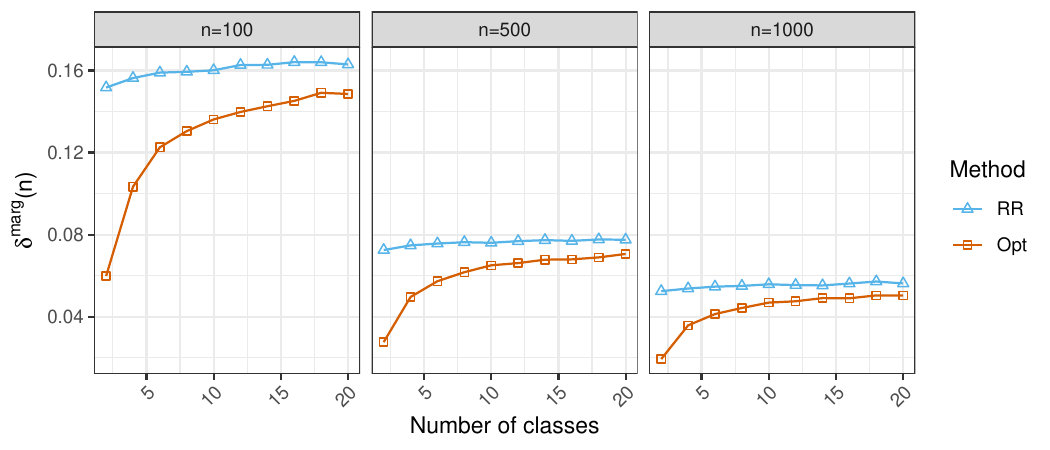}
    \caption{Finite-sample correction for the block randomized response model, estimated using the $\beta$ which optimizes $\delta^{\mathrm{FS}}$ for the randomized response model (RR) and the $\beta$ resulting from the minimization problem in~\eqref{eq:delta_n_finitesample} (Opt). The estimates are displayed as function of the number of classes $K$, for different values of the sample size. In these experiments, the number of blocks is $b=2$ and $\epsilon = 0.1$.}
    \label{fig:delta_B_nlab}
\end{figure}

Figures~\ref{fig:delta_B_Klab} and~\ref{fig:delta_B_nlab} display the comparison between the two estimates of the finite-sample correction for the block randomized response model with two blocks, one obtained by using a basic definition of $\beta$ (RR) and the other obtained by solving the optimization problem in~\eqref{eq:delta_n_finitesample} (Opt). The basic definition is set as the vector of $\beta$'s which solve the minimization problem for the randomized response model described in Section~\ref{sec:adaptive_pred_finitesample_simplified}, namely $\beta_0 = \frac{1}{1-\epsilon}$ and $\beta_k = -\frac{\epsilon}{1-\epsilon}$, $k=1,\dots,K$. The experiment is conducted by setting the number of blocks $b=2$ and $\epsilon = 0.1$. In figure~\ref{fig:delta_B_Klab}, the finite-sample correction is displayed as a function of the sample size, for $K=4,8,16$. As expected, both estimates of $\delta^{\mathrm{FS}}$ decay as $n$ increases and increase only slightly as the number of classes grows. Figure~\ref{fig:delta_B_nlab} displays $\delta^{\mathrm{FS}}$ as a function of $K$, for $n=100,500,1000$. 
The difference between the optimal finite-sample correction and that obtained with the naive beta accentuates for low values of $K$ and of the sample size, and narrows as these two parameters increase.


\subsubsection{Two-Level Randomized Response Model}\label{appendix:two-level-rmm}
The model considered here describes a natural label contamination process involving two clearly defined groups of labels. Let $\epsilon$ and $\nu$ be the two parameters which identify the model, with $\epsilon \in [0,1)$ and $\nu \in [0,1]$. For simplicity, assume that the total number of possible labels $K$ is even. The two-level randomized response model is identified by the transition matrix $T$, which is defined as a $(2 \times 2)$ block matrix in the form
\begin{equation*}
    T =
    \begin{pmatrix}
        (1-\epsilon) I + \frac{\epsilon}{K}(1 + \nu) J  & \frac{\epsilon}{K} (1-\nu) J \\
        \frac{\epsilon}{K} (1-\nu) J & (1-\epsilon) I + \frac{\epsilon}{K}(1 + \nu) J 
    \end{pmatrix}.
\end{equation*}
Above, $I$ denotes the $K/2$-dimensional identity matrix, and $J$ denotes the $K/2$-dimensional constant matrix of ones. For ease of notation, let
\begin{align*}
    f = \epsilon (1+\nu), \quad g=\epsilon (1-\nu), \quad \text{ and } \quad e = \frac{\epsilon(1 + \nu) - \epsilon^2 (1 - \nu)}{1 - \epsilon + f/2}.
\end{align*}
It is possible to verify that the inverse of $T$ is the $2 \times 2$ block matrix $W$ in the form
\begin{equation*}
    W =
    \begin{pmatrix}
        \frac{1}{1-\epsilon}I - \frac{p}{K} J & - \frac{h}{K} J \\
        - \frac{h}{K} J  & \frac{1}{1-\epsilon}I - \frac{p}{K} J
    \end{pmatrix},
\end{equation*}
where
\begin{align*}
    p &= \frac{1}{1-\epsilon} \cdot \frac{e}{1-\epsilon + e/2},\\
    h &= \frac{g}{(1-\epsilon)^2} \cdot \left(1 - \frac{f}{2(1-\epsilon + f/2)}\right) \left(1 - \frac{e}{2(1-\epsilon + e/2)}\right).
\end{align*}

If we consider for instance the case $K=4$, then the matrix $\Omega$ is

\begin{align*}
  \Omega =
  \scalebox{0.85}{$
  \begin{bmatrix}
    \frac{1}{1 -\epsilon} - \beta_0 - \frac{1}{4}(p+\beta_1) & -\frac{1}{4}(p+\beta_1) & -\frac{1}{4}(h+\beta_1) & -\frac{1}{4}(h+\beta_1) \\ 
    -\frac{1}{4}(p+\beta_2) & \frac{1}{1 -\epsilon} - \beta_0 - \frac{1}{4}(p+\beta_2) & -\frac{1}{4}(h+\beta_2) & -\frac{1}{4}(h+\beta_2) \\ 
    -\frac{1}{4}(h + \beta_3) & -\frac{1}{4}(h + \beta_3) & \frac{1}{1 -\epsilon} - \beta_0 - \frac{1}{4}(p+\beta_3) & -\frac{1}{4}(p+\beta_3) \\ 
    -\frac{1}{4}(h + \beta_4) & -\frac{1}{4}(h + \beta_4) & -\frac{1}{4}(p+\beta_4) & \frac{1}{1 -\epsilon} - \beta_0 -\frac{1}{4}(p + \beta_4) \\ 
  \end{bmatrix}.
  $}
\end{align*}

Set $\beta'_0 = \frac{1}{1-\epsilon}$ and $\beta'_k = -p$ for all $k$, and let $\beta' = (\beta'_0,\beta'_1,\dots,\beta'_K)$. Note that
\begin{equation*}
     |\beta'_0| + \frac{1}{K} \sum_k |\beta'_k| = \frac{1}{1-\epsilon} + p.
\end{equation*}
Then, the $\|\Omega\|_1$ term in~\eqref{eq:B-definition} evaluated at $\beta'$ reads
\begin{align*}
  \|\Omega\|_1 = \max_{l \in [K]} \sum_{k=1}^{K} |\Omega_{kl}|
  = \frac{K}{2} \frac{1}{K} |h-p|
    = \frac{1}{2} |h-p|,
\end{align*}
Note that the quantity $|h-p|$ is independent of $K$ and $n$, and that for any $\epsilon \in [0,1)$ and for any $\nu \in [0,1]$, $|h-p| < \infty$. In fact, one may verify that $h\geq 0$ and that it is a decreasing function of $\nu$, while $p \geq \frac{\epsilon}{1-\epsilon} \geq h$ increases with $\nu$. Then, $|h-p|\leq p(\nu=1) - h(\nu=1) = \frac{2 \epsilon}{1-\epsilon}$. This implies that for every $\nu \in [0,1]$ and for every $\epsilon$ that is model-wise meaningful (e.g., $\epsilon$ in $[0,0.5)$), $|h-p|$ is bounded and small. For example, for $\epsilon = 0.2$, $|h-p| \leq 0.5$ for all $\nu$. The same argument, using $h(\nu=1)=0$, shows $p \leq p(\nu=1) = \frac{2\epsilon}{1-\epsilon}$, and hence
\begin{equation*}
    |\beta'_0| + \frac{1}{K} \sum_k |\beta'_k| = \frac{1}{1-\epsilon} + p \leq \frac{1+2\epsilon}{1-\epsilon}.
\end{equation*}

\begin{figure}
    \centering
    \includegraphics[width=0.9\linewidth]{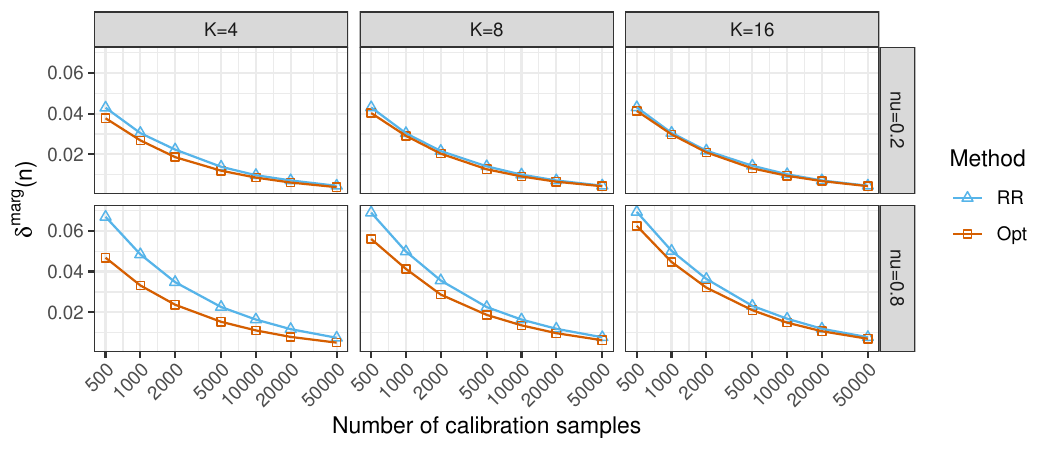}
    \caption{Finite-sample correction for the two-level randomized response model, estimated using the $\beta$ which optimizes $\delta^{\mathrm{FS}}$ for the randomized response model (RR) and the $\beta$ resulting from the minimization problem in~\eqref{eq:delta_n_finitesample} (Opt). The estimates are displayed as function of the sample size, for different values of $K$ and for $\nu \in \{0.2,0.8\}$. Parameter $\epsilon$ is set to $\epsilon = 0.1$.}
    \label{fig:delta_2levelRR_Klab}
\end{figure}

\begin{figure}
    \centering
    \includegraphics[width=0.9\linewidth]{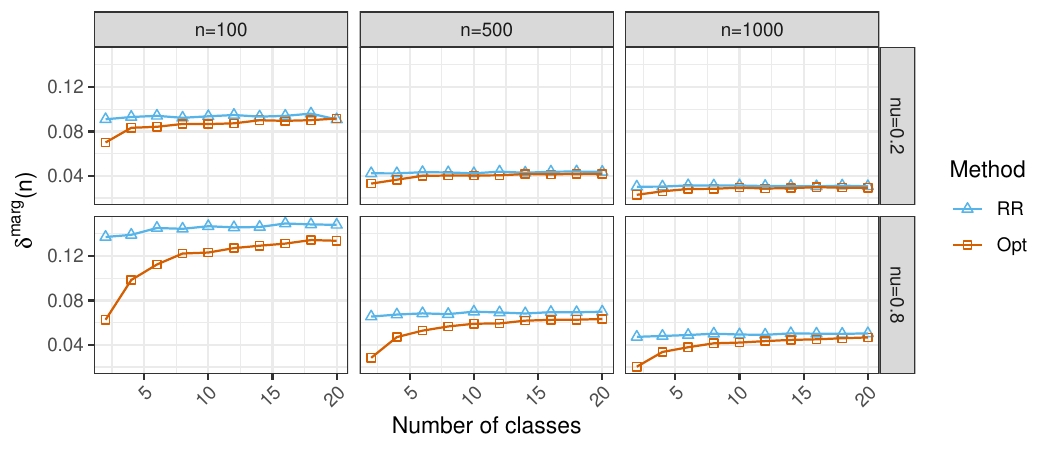}
    \caption{Finite-sample correction for the two-level randomized response model, estimated using the $\beta$ which optimizes $\delta^{\mathrm{FS}}$ for the randomized response model (RR) and the $\beta$ resulting from the minimization problem in~\eqref{eq:delta_n_finitesample} (Opt). The estimates are displayed as function of the number of classes $K$, for different values of the sample size and for $\nu=0.2,0.8$. Parameter $\epsilon$ is set to $\epsilon = 0.1$.}
    \label{fig:delta_2levelRR_nlab}
\end{figure}

\begin{figure}
    \centering
    \includegraphics[width=0.9\linewidth]{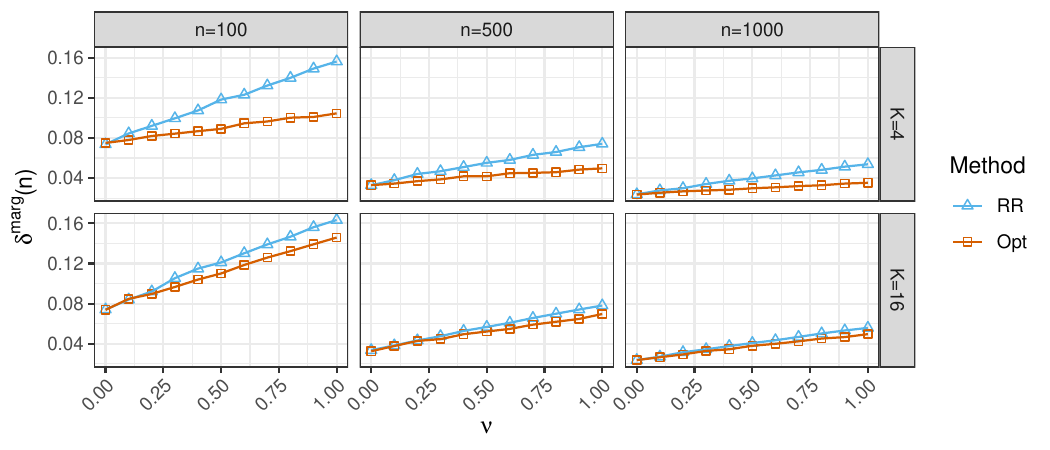}
    \caption{Finite-sample correction for the two-level randomized response model, estimated using the $\beta$ which optimizes $\delta^{\mathrm{FS}}$ for the randomized response model (RR) and the $\beta$ resulting from the minimization problem in~\eqref{eq:delta_n_finitesample} (Opt). The estimates are displayed as function of $\nu$, for different values of the sample size and for $K=16$. Parameter $\epsilon$ is set to $\epsilon = 0.1$.}
    \label{fig:delta_2levelRR_nu_var}
\end{figure}

\begin{figure}
    \centering
    \includegraphics[width=0.9\linewidth]{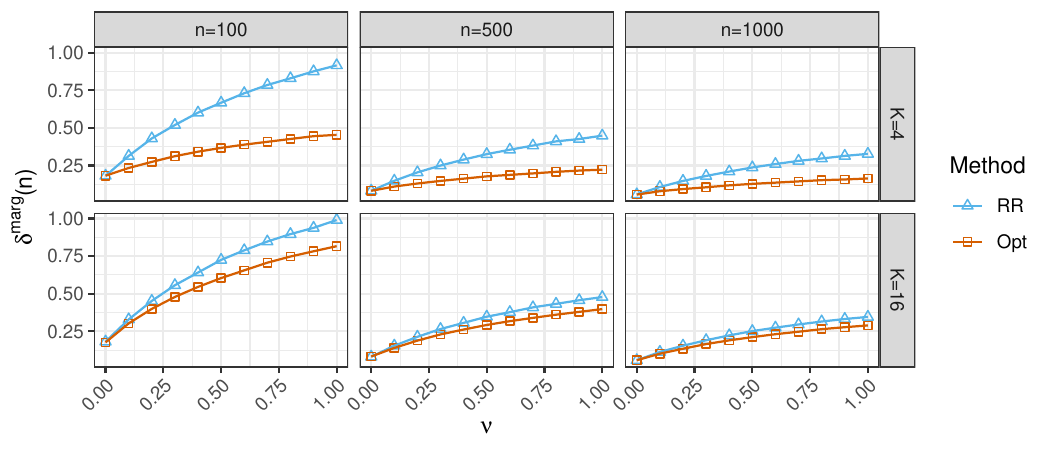}
    \caption{Finite-sample correction for the two-level randomized response model, estimated using the $\beta$ which optimizes $\delta^{\mathrm{FS}}$ for the randomized response model (RR) and the $\beta$ resulting from the minimization problem in~\eqref{eq:delta_n_finitesample} (Opt). The estimates are displayed as function of $\nu$, for different values of the sample size and for $K=16$. Parameter $\epsilon$ is set to $\epsilon = 0.5$.}
    \label{fig:delta_2levelRR_nu_var_0.5}
\end{figure}

Then, the finite-sample correction in the specific case of the two-level randomized response model scales as $\mathcal{O}(\sqrt{(\log K) / n})$ and specifically reads
\begin{align*}
    \delta^{\mathrm{FS}}(n, K) 
    &\leq c(n) \left(|\beta'_0|+\frac{\sum_{k=1}^{K} |\beta'_k|}{K} \right) + \frac{1}{\sqrt{n}} B(K,n,\beta') \\
    &\leq \frac{1+2\epsilon}{1-\epsilon} \cdot c(n) + \frac{48}{\sqrt{n}} \cdot \frac{\epsilon}{1-\epsilon} \cdot \left( \sqrt{\log (K + 2)} + \frac{1}{\sqrt{\log (K + 2)}} \right).
\end{align*}
Figures~\ref{fig:delta_2levelRR_Klab}--\ref{fig:delta_2levelRR_nlab} display the comparison between the estimates of the finite-sample correction $\delta^{\mathrm{FS}}$ for the two-level randomized response model, obtained by using a basic definition of $\beta$ (RR) and by solving the optimization problem in~\eqref{eq:delta_n_finitesample} (Opt). As in Appendix~\ref{appendix:block-rrm}, the basic definition is set as the vector of $\beta$'s which solve the minimization problem for the randomized response model, namely $\beta_0 = \frac{1}{1-\epsilon}$ and $\beta_k = -\frac{\epsilon}{1-\epsilon}$, $k=1,\dots,K$. The comparison is done in two scenarios, identified by two values of $\nu$. When $\nu=0.2$, the two-level randomized response model is close to the randomized response model and the finite-sample correction obtained with the naive $\beta$ estimate is close to the optimal finite-sample correction, for all $K \geq 4$ and for all $n$. When $\nu=0.8$, the two-level randomized response model deviates from the randomized response model and the optimal finite-sample correction improves the one obtained with the naive $\beta$, although the difference between the two estimates narrows both as $K$ increases and as $n$ increases. Figures~\ref{fig:delta_2levelRR_nu_var}--\ref{fig:delta_2levelRR_nu_var_0.5} display the behavior of the finite-sample corrections obtained with the two methods, as the contamination model deviates from the randomized response model. The closer $\nu$ is to 1, the larger is the deviation from the randomized response model. As expected, the difference between the two finite-sample corrections becomes more significant as $\nu$ increases, and is further magnified with an increase in $\epsilon$, namely the parameter that quantifies the level of random noise (see Figure~\ref{fig:delta_2levelRR_nu_var_0.5}). These results show that calculating the optimized $\delta^{\mathrm{FS}}$ proves more effective the more the model deviates from the randomized response model, especially when the number of classes is low and when the amount of random level noise is high. Again, an increase in sample size reduces the gap between the two finite-sample corrections.

Overall, these experiments suggest a convenient practical rule. When the number of classes and the sample size are large, one can simplify the estimation routine by avoiding solving the optimization problem, and evaluating the finite-sample correction at the $\beta$ vector corresponding to the randomized response model. However, the optimal finite-sample correction should be employed when the contamination model deviates from the randomized response model and the amount of label contamination is high.

\clearpage

\subsection{Comparison with Clarkson et al. (2024)}
\label{appendix:clarkson-comparison}

This appendix compares the method proposed in this paper with that of \citet{clarkson2024split}, which also builds on \citet{sesia2025adaptive} to target marginal coverage under label noise. As summarized in Section~\ref{sec:contributions}, the two approaches differ in three respects: (i) the parametrization of the contamination process and the assumptions on the label frequencies (Appendix~\ref{appendix:clarkson-comparison-parametrization}); (ii) the rigor of the finite-sample coverage guarantee (Appendix~\ref{appendix:clarkson-comparison-guarantee}); and (iii) the scaling of the correction factor with the number of classes (Appendix~\ref{appendix:clarkson-comparison-analysis}). 
Appendix~\ref{appendix:clarkson-comparison-models} compares the methods analytically on the three contamination models considered in this paper.

\subsubsection{Parametrization of the Contamination Process} \label{appendix:clarkson-comparison-parametrization}
\citet{clarkson2024split} parametrize the contamination through the matrix $R_{ij} := \mathbb{P}[Y = i \mid \tilde{Y} = j]$, and assume that $R$ and the label marginals $\rho:=(\rho_1,\dots,\rho_K)$ and $\tilde\rho:=(\tilde{\rho}_1,\dots, \tilde{\rho}_K)$ are all known.
In contrast, we parametrize the label contamination process through the transition matrix $T$, and we assume knowledge of $T$ alone.
The two points of view are connected by Bayes' theorem, which implies
\begin{align}
    R = \Lambda T^\top \tilde{\Lambda}^{-1},
    \qquad
    R^{-1}_{ji} = \frac{\tilde{\rho}_j}{\rho_i} \, W_{ij},
    \label{eq:R-T-bayes}
\end{align}
with $\Lambda = \mathrm{diag}(\rho_1,\dots,\rho_K)$, $\tilde{\Lambda} = \mathrm{diag}(\tilde\rho_1,\dots,\tilde\rho_K)$ and $W = T^{-1}$.
Therefore, the two parametrizations carry the same information once the marginals are given, which makes our assumptions weaker.

Moreover, our modeling choice is arguably more practical and intuitive, because Assumption~\ref{assumption:linear-contam} factorizes the joint law of $(X, Y, \tilde{Y})$ into the target distribution $\mathcal{P}_{X,Y}$ and the contamination process $\mathcal{P}_{\tilde{Y} \mid Y}$, whose parameter $T$ varies independently of the former.
Thus, $T$ describes the corruption mechanism alone and remains the same across populations that differ in the distribution of $(X,Y)$.
This is what makes $T$ estimable from a small auxiliary sample that need not be representative of the target population (Section~\ref{sec:contamination_estimation}), and it cleanly separates the target of inference from the nuisance component being modeled.
By contrast, \eqref{eq:R-T-bayes} shows that $R$ entangles the contamination mechanism with the label marginals, making it population-specific and not estimable without knowledge of the clean label distribution $\rho$.

The approach of \citet{clarkson2024split} also requires estimating $\tilde{\rho}$ in practice, but does not account for the associated uncertainty.
By contrast, the sampling error of the empirical label frequencies $\hat{\rho}_l$ is absorbed into the empirical process $\hat\psi$ of~\eqref{eq:hat-psi}, whose expected supremum is bounded by our $\delta^{\mathrm{FS}}(n)$ in~\eqref{eq:delta_n_finitesample}; this source of uncertainty is therefore accounted for by our theory.

\subsubsection{Coverage Guarantee} \label{appendix:clarkson-comparison-guarantee}

Substituting~\eqref{eq:R-T-bayes} into the estimator of the coverage inflation factor proposed by \citet{clarkson2024split} shows that it coincides with our $\hat\Delta$ upon replacing the empirical label frequencies $\hat\rho$ with the assumed-known $\tilde\rho$.
Their method \citep[Eq.~(30)]{clarkson2024split} then selects the index
\begin{align} \label{eq:clarkson-i}
  \hat{i}^{\mathrm{Clarkson}} = \min \left\{ i \in [n]: \frac{i}{n+1} \geq 1 - \alpha - \hat\Delta(S_{(i)}) + \delta^{\mathrm{Clarkson}}(n) \right\},
\end{align}
which is Algorithm~\ref{alg:adaptive-conformal-marg} with $n$ replaced by $n+1$ and $\delta(n) = \delta^{\mathrm{Clarkson}}(n)$, where
\begin{align}
    \delta^{\mathrm{Clarkson}}(n) &= \sum_{i=1}^K \left( |w_i^{(1)}| b(n,i) + \sum_{j:j\neq i} |w_{ij}^{(2)}| b(n,j)\right),
    \label{eq:clarkson}
\end{align}
with $w_i^{(1)} = R_{ii}^{-1} \rho_i - \tilde{\rho}_i$, $w_{ij}^{(2)} = \rho_i R_{ji}^{-1}$ and $b(n,j) = (1-\tilde{\rho}_j)^n + \sqrt{\pi / (n \tilde{\rho}_j)}$, is an upper bound on $\EV{\sup_t |\hat\Delta(t) - \Delta(t)|}$ obtained by applying the DKW inequality separately to each empirical class-conditional CDF $\hat F^k_l$ \citep[Theorem~4]{clarkson2024split}.

The coverage guarantee claimed in \citet{clarkson2024split} for this method, however, does not follow from their argument, which treats the index $\hat{i}^{\mathrm{Clarkson}}$ in~\eqref{eq:clarkson-i} as if it were fixed rather than data-dependent.
This makes their finite-sample correction heuristic as originally written.
In fact, whenever $\hat{i}^{\mathrm{Clarkson}}$ exists (written henceforth as $\hat{i} = \hat{i}^{\mathrm{Clarkson}}$), the coverage of their prediction set satisfies
\begin{align*}
  & \P{Y_{n+1} \in \hat{C}^{\mathrm{Clarkson}}(X_{n+1})} \\
  & \qquad = \EV{F(S_{(\hat{i})})} \\
  & \qquad = \EV{\tilde{F}(S_{(\hat{i})})} + \EV{\hat\Delta(S_{(\hat{i})})} - \EV{\hat\Delta(S_{(\hat{i})}) - \Delta(S_{(\hat{i})})} \\
  & \qquad \geq 1 - \alpha + \delta^{\mathrm{Clarkson}}(n) + \EV{\tilde{F}(S_{(\hat{i})}) - \frac{\hat{i}}{n+1}} - \EV{\hat\Delta(S_{(\hat{i})}) - \Delta(S_{(\hat{i})})} \\
  & \qquad \geq 1 - \alpha + \EV{\tilde{F}(S_{(\hat{i})}) - \frac{\hat{i}}{n+1}},
\end{align*}
where the first inequality follows from the definition of $\hat{i}$, and the second one from their Theorem~4, which establishes that
\begin{align*}
\EV{\left| \hat\Delta(S_{(\hat{i})}) - \Delta(S_{(\hat{i})}) \right|} \leq \delta^{\mathrm{Clarkson}}(n);
\end{align*}
their Theorem~4 is stated for a fixed index $i$, but its proof bounds $\mathbb{E}[\sup_t |\hat\Delta(t) - \Delta(t)|]$ and hence implies the result used above.

The missing step in \citet{clarkson2024split} is to control $\mathbb{E}[\tilde{F}(S_{(\hat{i})}) - \hat{i}/(n+1)]$, which their argument implicitly treats as zero.
This would be correct if $\hat{i}$ were replaced by a
fixed index $i \in [n]$, since then $\mathbb{E}[\tilde{F}(S_{(i)})] = i/(n+1)$,
but the identity fails in general for $\hat{i}$ in~\eqref{eq:clarkson-i}, which is a data-dependent index.

Nonetheless, the term can be bounded from below, uniformly over all indices, following a similar strategy to the one we adopt in the proofs of Theorems~\ref{thm:finite-sample-factor-simplified} and~\ref{thm:finite-sample-factor}. For any index $\hat{i} \in [n]$,
\begin{align*}
  \tilde{F}(S_{(\hat{i})}) - \frac{\hat{i}}{n+1}
  \;\geq\; \tilde{F}(S_{(\hat{i})}) - \frac{\hat{i}}{n}
  \;=\; -\left( \hat{F}(S_{(\hat{i})}) - \tilde{F}(S_{(\hat{i})}) \right)
  \;\geq\; -\sup_{t \in \mathbb{R}} \left( \hat{F}(t) - \tilde{F}(t) \right).
\end{align*}
Taking expectations yields
\begin{align*}
  \mathbb{E}\left[ \tilde{F}(S_{(\hat{i})}) - \frac{\hat{i}}{n+1} \right]
  \;\geq\; - \mathbb{E}\left[ \sup_{t \in \mathbb{R}}
  \left( \hat{F}(t) - \tilde{F}(t) \right) \right]
  \;=\; -c(n),
\end{align*}
with $c(n)$ as in~\eqref{eq:c-definition}.

Therefore, the argument in~\citet{clarkson2024split} establishes only
\begin{align*}
  \P{Y_{n+1} \in \hat{C}^{\mathrm{Clarkson}}(X_{n+1})}
  & \geq 1 - \alpha - c(n),
\end{align*}
and not $\P{Y_{n+1} \in \hat{C}^{\mathrm{Clarkson}}(X_{n+1})} \geq 1-\alpha$.

This explains why $\delta^{\mathrm{Clarkson}}(n)$ vanishes with the contamination while our $\delta^{\mathrm{FS}}(n)$ does not:
the term $c(n)$, which they omit from their finite sample correction term while we include in our $\delta^{\mathrm{FS}}(n)$, is of order $1/\sqrt{n}$ regardless of the contamination level.
While a fair empirical comparison of our finite-sample correction factor with their is not possible, due to the heuristic nature of the latter, it is interesting to note that $\delta^{\mathrm{FS}}(n)$ still scales more favorably with the number of classes than $\delta^{\mathrm{Clarkson}}(n)$, and is smaller in many regimes of practical interest, owing to our tighter analysis of the estimation error of $\hat\Delta$; we make this precise next.

\subsubsection{Comparison of the Two Finite-Sample Corrections} \label{appendix:clarkson-comparison-analysis}

Under common contamination mechanisms (made precise below), the
heuristic finite-sample correction factor of \citet{clarkson2024split} scales as $\sqrt{K_{\mathrm{eff}}/n}$, where
$K_{\mathrm{eff}} := ( \sum_{j=1}^K \sqrt{\tilde\rho_j} )^2 \in [1, K]$ is an effective number of classes equal to $K$ for balanced
marginals, whereas our $\delta^{\mathrm{FS}}(n)$ scales as $\sqrt{\log K/n}$ and is smaller whenever
there are many populated classes and the contamination is not vanishingly small. We now make this precise.

In order to compare $\delta^{\mathrm{Clarkson}}(n)$ from~\eqref{eq:clarkson} with $\delta^{\mathrm{FS}}(n)$, it is to rewrite the former in terms of $W$. Substituting~\eqref{eq:R-T-bayes} into~\eqref{eq:clarkson} gives
\begin{align*}
    w_i^{(1)} &= R_{ii}^{-1} \rho_i - \tilde{\rho}_i = \frac{\tilde{\rho}_i}{\rho_i} W_{ii} \, \rho_i - \tilde{\rho}_i = \tilde{\rho}_i \, (W_{ii} - 1),\\
    w_{ij}^{(2)} &= \rho_i R_{ji}^{-1} = \rho_i \, \frac{\tilde{\rho}_j}{\rho_i} \, W_{ij} = \tilde{\rho}_j \,W_{ij},
\end{align*}
so that, collecting the terms multiplying each $b(n,j)$,
\begin{align*}
    \delta^{\mathrm{Clarkson}}(n) &= \sum_{j=1}^K b(n,j) \left( |w_j^{(1)}| + \sum_{i:i \neq j} |w_{ij}^{(2)}|\right)\\
    &= \sum_{j=1}^K \left( (1-\tilde{\rho}_j)^n + \sqrt{\frac{\pi}{n \tilde{\rho}_j}} \right) \left( \tilde{\rho}_j |W_{jj}-1| + \sum_{i:i\neq j} \tilde{\rho}_j |W_{ij}|\right)\\
    &= \sum_{j=1}^K \left( \tilde{\rho}_j (1-\tilde{\rho}_j)^n + \sqrt{\frac{\pi}{n} \tilde{\rho}_j} \right) \left( |W_{jj}-1| + \sum_{i:i\neq j} |W_{ij}|\right) \\
    &= \sum_{j=1}^K \left( \tilde{\rho}_j (1-\tilde{\rho}_j)^n + \sqrt{\frac{\pi}{n} \tilde{\rho}_j} \right) \lVert  (W - I)_j\rVert_1, \numberthis \label{eq:clarkson_rewritten}
\end{align*}
where $(W-I)_j$ is the $j$-th column of $W-I$. This form reveals two interesting features.

First, $\delta^{\mathrm{Clarkson}}(n)$ is proportional to the column norms of $W-I$, which
measure how far the contamination is from leaving the labels untouched; it therefore vanishes
as the contamination disappears, and the method applies no correction at all when $T=I$.
This is an attractive property that our bound does not share.
Second, since every term in~\eqref{eq:clarkson_rewritten} is nonnegative, dropping the
geometric part gives the lower bound
\begin{align} \label{eq:clarkson-scaling}
\begin{split}
    \delta^{\mathrm{Clarkson}}(n)
    \;\ge\; \sqrt{\frac{\pi}{n}} \sum_{j=1}^K \sqrt{\tilde\rho_j}\, \lVert (W-I)_j \rVert_1
    \;\ge\; \sqrt{\frac{\pi K_{\mathrm{eff}}}{n}}\; \min_{j} \lVert (W-I)_j \rVert_1.
  \end{split}
\end{align}
When the contamination is spread evenly across classes, so that all columns
of $W-I$ have the same $\ell_1$ norm (as for the symmetric models considered below), the minimum in~\eqref{eq:clarkson-scaling} equals
$\lVert W-I\rVert_1 := \max_l \sum_k |(W-I)_{kl}|$, and the correction factor of
\citet{clarkson2024split} is at least $\lVert W-I\rVert_1 \sqrt{\pi K_{\mathrm{eff}}/n}$,
i.e., it grows at least like $\sqrt{K_{\mathrm{eff}}/n}$.

For comparison, evaluating our finite-sample correction factor $\delta^{\mathrm{FS}}(n)$ 
in~\eqref{eq:delta_n_finitesample} at $\beta = (1,0,\dots,0)$, for which $\bar W = I$ and
$\Omega = W - I$, gives
\begin{align}
    \delta^{\mathrm{FS}}(n)
    \;\le\; c(n) \;+\; \frac{2}{\sqrt{n}}\,\lVert W - I\rVert_1 \sqrt{2\log(2Kn+2)}
    \;=\; O\!\left( \frac{1}{\sqrt{n}} + \lVert W - I\rVert_1 \sqrt{\log(Kn)/n} \right),
    \label{eq:delta-n-scaling}
\end{align}
which the optimization over $\beta$ can only improve.
Comparing the lower bound~\eqref{eq:clarkson-scaling} with the upper
bound~\eqref{eq:delta-n-scaling} shows that our finite-sample correction factor is smaller than $\delta^{\mathrm{Clarkson}}(n)$ whenever
\begin{align}
    \sqrt{\pi K_{\mathrm{eff}}}\; \min_j \lVert (W-I)_j\rVert_1
    \;-\; 2\sqrt{2\log(2Kn+2)}\; \lVert W-I\rVert_1
    \;\ge\; \sqrt{n}\, c(n),
    \label{eq:comparison}
\end{align}
for which, since $c(n) \le \sqrt{\pi/(8n)}$, it suffices that the left-hand side be at
least $\sqrt{\pi/8}$. For evenly spread contamination, $\min_j \lVert (W-I)_j\rVert_1 = \lVert W-I\rVert_1$, and~\eqref{eq:comparison} reduces to
$K_{\mathrm{eff}} \gtrsim \log(Kn)$ as long as the contamination is not vanishingly small.

\subsubsection{Comparison for Special Contamination Models} \label{appendix:clarkson-comparison-models}

We now compare $\delta^{\mathrm{Clarkson}}(n)$ with $\delta^{\mathrm{FS}}(n)$ within the three contamination models considered in this paper, assuming for simplicity that the label marginals are uniform. In all three, $T$ is doubly stochastic, so a uniform $\rho$ yields a uniform $\tilde\rho_j = 1/K$ and $K_{\mathrm{eff}} = K$.
In all three cases, $\delta^{\mathrm{Clarkson}}(n)$ grows as $\sqrt{K/n}$, whereas $\delta^{\mathrm{FS}}(n)$ grows at most as $\sqrt{\log K/n}$, or not at all with $K$ under the randomized response model.

\paragraph{Randomized response model.}

Here $W = \frac{1}{1-\epsilon}I - \frac{\epsilon}{K(1-\epsilon)}J$, so that
$\lVert (W-I)_j \rVert_1 = \frac{2\epsilon(K-1)}{K(1-\epsilon)}$ for every $j$, which vanishes
linearly in $\epsilon$ and grows with $K$ toward $2\epsilon/(1-\epsilon)$. Then
\eqref{eq:clarkson_rewritten} takes the closed form
\begin{align*}
    \delta^{\mathrm{Clarkson}}(n) \;=\; \frac{2\epsilon(K-1)}{K(1-\epsilon)}
    \left[\left(1-\frac{1}{K}\right)^n + \sqrt{\frac{\pi K}{n}}\right] = \mathcal{O}\left(\sqrt{\frac{K}{n}}\right),
\end{align*}
whereas our correction is free of $K$ altogether (Theorem~\ref{thm:finite-sample-factor-simplified}):
\begin{align*}
\delta^{\mathrm{FS}}(n,\epsilon,K) = \frac{1+\epsilon}{1-\epsilon} c(n) = \mathcal{O}\left(\frac{1}{\sqrt{n}}\right).
\end{align*}

\paragraph{Two-level randomized response model.}
From Appendix~\ref{appendix:two-level-rmm}, the column of $W-I_K$ corresponding to any class $j$ has diagonal entry
$\frac{\epsilon}{1-\epsilon}-\frac{p}{K}$, the $m-1$ remaining within-block entries $-\frac{p}{K}$,
and the $m$ cross-block entries $-\frac{h}{K}$, with $m=K/2$. Throughout the range of $\epsilon$ and $\nu$ of
interest, $p,h>0$ and $\frac{\epsilon}{1-\epsilon}>\frac{p}{K}$, so
$\lVert (W-I)_j\rVert_1 = \frac{\epsilon}{1-\epsilon}+\frac{p+h}{2}-\frac{2p}{K}$ and 
\begin{align*}
    \delta^{\mathrm{Clarkson}}(n) 
    &= \left[\frac{\epsilon}{1-\epsilon}+\frac{p+h}{2}-\frac{2p}{K}\right]
    \left[\left(1-\frac{1}{K}\right)^n+\sqrt{\frac{\pi K}{n}}\right] = \mathcal{O}\left(\sqrt{\frac{K}{n}}\right).
\end{align*}
In contrast, our correction is (Appendix~\ref{appendix:two-level-rmm})
\begin{align*}
    \delta^{\mathrm{FS}}(n, \epsilon, K)
    &\leq \frac{1+2\epsilon}{1-\epsilon} c(n) + \frac{48}{\sqrt{n}} \frac{\epsilon}{1-\epsilon}
      \left( \sqrt{\log (K + 2)} + \frac{1}{\sqrt{\log (K + 2)}} \right) = \mathcal{O}\left(\sqrt{\frac{\log K}{n}}\right).
\end{align*}

\paragraph{Block randomized response model.}
For $b$ blocks of size $m=K/b$, $W = \frac{1}{1-\epsilon}I_K - \frac{\epsilon}{m(1-\epsilon)}B_b$
with $B_b = \mathrm{blkdiag}(J_m,\dots,J_m)$; see Appendix~\ref{appendix:block-rrm}.
The entries of $W-I_K$ in a column $j$ are $\frac{\epsilon}{1-\epsilon}(1-\frac{1}{m})$ on the
diagonal, $-\frac{\epsilon}{m(1-\epsilon)}$ on the remaining $m-1$ entries of the same block, and
zero outside it, so that
\begin{align*}
    \lVert (W-I)_j\rVert_1 = \frac{2\epsilon(m-1)}{m(1-\epsilon)},
    \quad
    \delta^{\mathrm{Clarkson}}(n) = \frac{2\epsilon(m-1)}{m(1-\epsilon)}
    \left[\left(1-\frac{1}{K}\right)^n+\sqrt{\frac{\pi K}{n}}\right] = \mathcal{O}\left(\sqrt{\frac{K}{n}}\right).
\end{align*}
In contrast, our correction reads (Appendix~\ref{appendix:block-rrm})
\begin{align*}
    \delta^{\mathrm{FS}}(n, K)
    &\leq \frac{1 + b \epsilon }{1-\epsilon} \cdot c(n)  + \frac{1}{\sqrt{n}} \frac{48 (b-1)\epsilon}{1-\epsilon} \left(\sqrt{\log (K + 2)} + \frac{1}{\sqrt{\log (K + 2)}} \right) 
= \mathcal{O}\left(\sqrt{\frac{\log K}{n}}\right).
\end{align*}

\FloatBarrier

\clearpage

\subsection{Richardson Extrapolation} \label{appendix:Richardson-extrapolation}

Richardson extrapolation~\citep{Richardson1911} is a numerical technique which improves the order of accuracy of an estimate by leveraging a sequence of approximations obtained at different discretization steps and extrapolating to the zero step size. In our context, the method turns out to be particularly useful, as it allows us to reduce the computational burden which comes with calculating~\eqref{eq:delta_n_h} for small values of $h$. By order of the Richardson extrapolation, we refer to the number of times that the extrapolation is applied recursively to improve the accuracy of the resulting estimate.

The idea is the following. Let $A(h)$ denote the quantity of interest computed on a grid of
mesh $h$, and suppose its discretization error admits an expansion of the form
$A(h) = A(0) + a_1 h^{p} + a_2 h^{2p} + \dots$ for some known exponent $p>0$ and unknown
coefficients $a_1, a_2, \dots$. Computing $A$ at two step sizes $h$ and $h/2$ and taking the
combination
\begin{align*}
    R_1(h) = \frac{2^{p} A(h/2) - A(h)}{2^{p} - 1} = A(0) + \mathcal{O}(h^{2p})
\end{align*}
cancels the leading error term. Applying the same combination recursively to $R_1(h)$ and
$R_1(h/2)$, with $2^{p}$ replaced by $2^{2p}$, cancels the next term, and so on; the
$k$-th order extrapolate $R_k(h)$ has error $\mathcal{O}(h^{(k+1)p})$ and requires computing
$A$ at $k+1$ step sizes $h, h/2, \dots, h/2^{k}$. 

The choice of $p$ is dictated by the
problem. Here the error
arises from replacing the supremum of a continuous Gaussian process over $[0,1]$ with its
maximum over a grid. Since the increments of a Brownian-bridge-type process over an
interval of length $h$ have standard deviation of order $\sqrt{h}$, the expected supremum
over the grid underestimates the true one by a term of order $\sqrt{h}$
\citep{asmussen1995discretization}, so the natural expansion is in
powers of $\sqrt{h}$ and we take $p = 1/2$.

\subsubsection{Example: The F-Brownian Bridge}
To demonstrate the effectiveness of estimating $\delta(n)$ via a Monte Carlo simulation coupled with Richardson extrapolation, we consider the Brownian bridge as an illustrative example. As the distribution of the supremum is known for the Brownian bridge, its expected value is a known target that enables us to assess how the estimation strategy behaves. These experiments additionally provide us with a convenient rule of thumb for setting the parameters $h$ and $M$ of the Monte Carlo simulation, together with the order of Richardson's extrapolation, which lead to an estimate of $\delta(n)$ that is sufficiently close to its true value.

Let $X_1,\dots,X_n$ be a random sample from a distribution function $F$ on the real line. Let $\hat{F}_n(t) = \frac{1}{n} \sum_{i=1}^n \I{X_i \leq t}$ and $F(t) = \P{X \leq t}$. From Donsker's theorem, we know that $\sqrt{n} [\hat{F}_n(t) - F(t) ]$ converges to the $F$-Brownian bridge process, namely a zero mean Gaussian process with covariance function $c^{B} (t,s) = F(t \wedge s) - F(t)F(s)$. The absolute supremum of the $F$-Brownian bridge process has Kolmogorov distribution, and its expected value is equal to the expected value of the sup of the uniform Brownian bridge; that is,
\begin{equation}
    \EV{\underset{t}{\sup} |B_F(t)|} = \EV{\underset{t}{\sup} |B_{U[0,1]}(t)|} = \sqrt{\frac{\pi}{2}} \log 2.
    \label{eq:bb-target}
\end{equation}

\begin{figure}[!htb]
    \centering
    \includegraphics[width=0.8\linewidth]{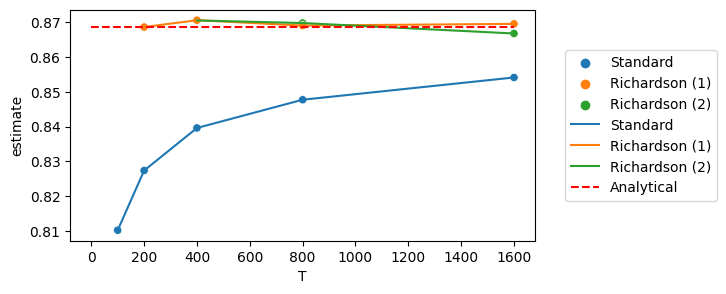}
    \caption{Estimate of the expected value of the supremum of the Brownian Bridge as function of $N = 1/h + 1$. The estimates are obtained via a Monte Carlo simulation with $M=1{,}000{,}000$. The standard method refers to the estimate obtained without employing Richardson extrapolation downstream of the MC simulation. Richardson(1) and Richardson(2) refer to the extrapolation method of order 1 and order 2, respectively. The dashed line marks the analytical value of the expectation of the supremum of the Brownian Bridge.}
    \label{fig:bb-expsup-estimate}
\end{figure}

\begin{figure}[!htb]
    \centering
    \includegraphics[width=0.8\linewidth]{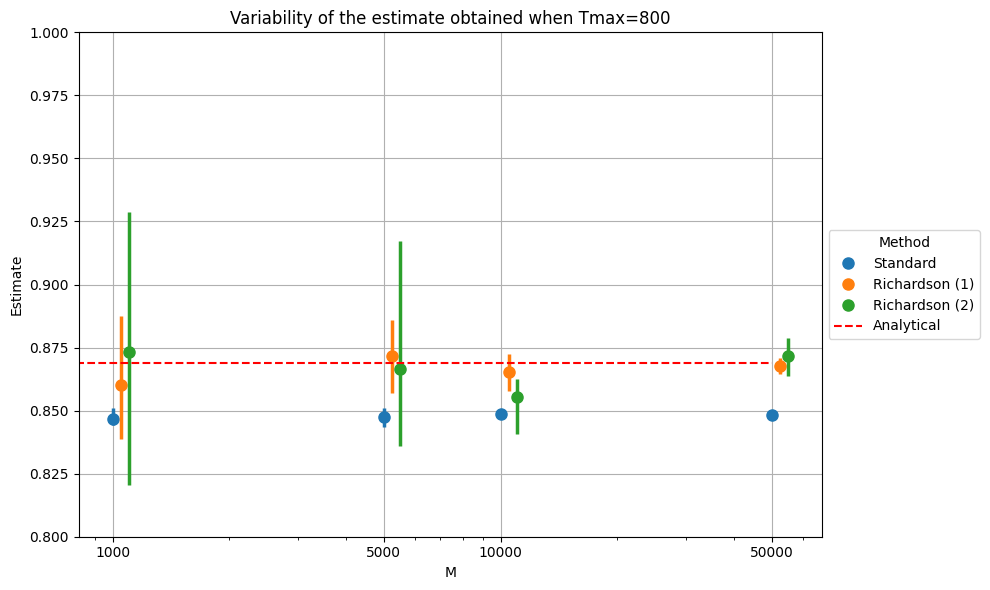}
    \caption{Variability of the estimates of the expected value of the supremum of the Brownian Bridge as function of $M$, obtained with $B=20$ repeated experiments. The estimates are obtained via Monte Carlo simulation and for $h=1/800$. The dashed line marks the analytical value of the expectation of the supremum of the Brownian Bridge.}
    \label{fig:bb-expsup-estimate-variability}
\end{figure}

Figure~\ref{fig:bb-expsup-estimate} displays the estimates of~\eqref{eq:bb-target} obtained via Monte Carlo simulation with $M=1{,}000{,}000$, as function of the number $N$ of points of the grid. The standard method refers to the estimate obtained without employing Richardson extrapolation downstream of the MC simulation. Richardson(1) and Richardson(2) refer to the extrapolation method of order 1 and order 2, respectively. The dashed line marks the target~\eqref{eq:bb-target}. This experiment demonstrates that, for a given value of $h$, the estimate obtained through Richardson extrapolation improves upon the estimate obtained solely from the Monte Carlo simulation performed on the grid corresponding to $h$. Furthermore, reducing $h$ below a certain threshold does not lead to any significant improvement in the estimate obtained through Richardson extrapolation for higher values of $h$. Finally, Richardson(1) and Richardson(2) yield similar estimates in this single realization, especially for small $h$. Figure~\ref{fig:bb-expsup-estimate-variability} shows that this does not extend to their variability across repeated realizations.

Figure~\ref{fig:bb-expsup-estimate-variability} displays the variability of the estimates for each of the considered methods as function of $M$. The variability of the estimates is obtained by running the experiment $B=20$ times, for $h = 1/800$. These results suggest that choosing Richardson(1) over Richardson(2), although not providing any significant improvement in the estimates in mean, generally result in more stable estimates.

\FloatBarrier

\clearpage

\subsection{Estimation of the Contamination Process} \label{appendix:nn-details}

\begin{figure}[!htb]
\centering
    \includegraphics[width=0.8\linewidth]{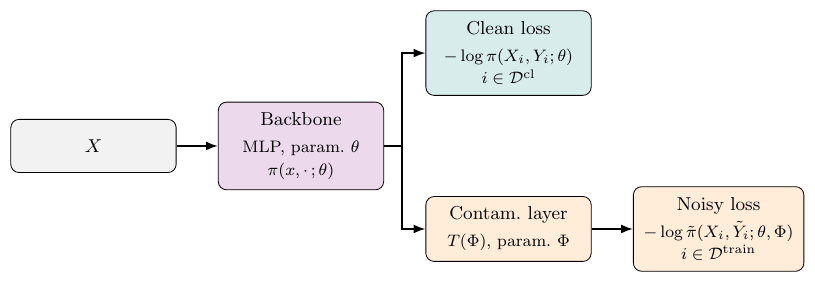}
\caption{Architecture of the neural network used to estimate the contamination matrix $T$. The backbone maps features $X$ to a clean label distribution $\pi(\cdot\,;\theta)$. The clean branch evaluates a cross-entropy loss on $\mathcal{D}^{\mathrm{cl}}$. The noisy branch passes the output through the contamination layer $T(\Phi)$ to produce $\tilde{\pi}(\cdot\,;\theta,\Phi)$, evaluated on $\mathcal{D}^{\mathrm{train}}$. Both $\theta$ and $\Phi$ are optimized jointly.}
\label{fig:nn-architecture}
\end{figure}

This appendix details the implementation of the method outlined in Section~\ref{sec:contamination_estimation-unpaired}, which estimates the transition matrix $T$ by jointly fitting a backbone classifier $\pi(\cdot\,;\theta)$ and a contamination layer $T(\Phi)$ through the loss~\eqref{eq:loss}. The overall architecture is illustrated in Figure~\ref{fig:nn-architecture}.

\subsubsection{Choice of the Backbone}
\label{appendix:nn-backbone}

In the most general case, the backbone can be any differentiable architecture trained from scratch jointly with the contamination layer. In practice, however, a pre-trained classifier $\hat{\pi}$ is often already available, since Algorithm~\ref{alg:adaptive-conformal-marg} requires one to compute the non-conformity scores. A natural and computationally convenient option is then to reuse its learned feature representation if $\hat{\pi}$ is a deep neural network: one extracts the penultimate-layer features of $\hat{\pi}$ for all observations in $\mathcal{D}^{\mathrm{cl}}$ and $\mathcal{D}^{\mathrm{train}}$, and trains a lightweight multi-layer perceptron (MLP) on those features jointly with the contamination layer. This is the approach adopted in the empirical studies of Section~\ref{sec:case_study}, where pre-trained ResNets serve as feature extractors and a small MLP is used as the backbone.

We emphasize that the classifier used to compute the non-conformity scores remains the original pre-trained $\hat{\pi}$; the network described here is trained solely to estimate $T$. Note also that $\hat{\pi}$ approximates the conditional distribution of the \emph{contaminated} label $\tY$ given $X$, whereas the backbone $\pi(\cdot\,;\theta)$ targets that of the \emph{clean} label $Y$; the two models play distinct roles and are not interchangeable.

\subsubsection{Parametrization of the Contamination Layer}
\label{appendix:nn-parametrization}

To ensure that $T(\Phi)$ is a valid stochastic matrix throughout training, without imposing explicit constraints, we parametrize each column $l$ of $T$ via an unconstrained vector $\boldsymbol{\phi}_l \in \mathbb{R}^K$ and apply a column-wise softmax:
\begin{align} \label{eq:T-softmax}
  T_{kl}(\Phi) = \frac{e^{\phi_{kl}}}{\sum_{j=1}^{K} e^{\phi_{jl}}},
  \qquad k, l \in [K],
\end{align}
where $\Phi = [\boldsymbol{\phi}_1 \mid \cdots \mid \boldsymbol{\phi}_K] \in \mathbb{R}^{K \times K}$ collects the unconstrained parameters. Both $T(\Phi)$ and the composite model~\eqref{eq:composite-model} are differentiable functions of $\Phi$, so gradients propagate through the contamination layer without approximation.

When the parametric form of $T$ is known, the general reparametrization~\eqref{eq:T-softmax} can be replaced by a lower-dimensional one, which reduces the number of parameters to be learned and typically speeds up convergence. Under the randomized response model, for instance, the contamination layer reduces to a single learnable scalar $\phi \in \mathbb{R}$, with $\epsilon = \sigma(\phi)$, where $\sigma$ denotes the sigmoid function, and $T(\phi) = (1-\sigma(\phi)) I_K + (\sigma(\phi)/K) \cdot J_K$.

\subsubsection{Invertibility and Regularization}
\label{appendix:nn-invertibility}

Since $T(\Phi)$ and $\pi(\cdot\,;\theta)$ are softmax-parametrized, the gradients $\partial \mathcal{L} / \partial \theta$ and $\partial \mathcal{L} / \partial \Phi$ are always well-defined, regardless of whether the resulting estimate $\hat{T}$ is invertible. Invertibility becomes relevant only after training, since $\hat{W} = \hat{T}^{-1}$ is required by the calibration procedure of Section~\ref{sec:adaptive-marginal}.

In most of our experiments, no further intervention was needed to obtain an invertible estimate. Under severe class imbalance, however, training occasionally converged to singular estimates of $T$. In such cases we replace $\mathcal{L}(\theta,\Phi)$ with the regularized objective
\begin{equation} \label{eq:loss-regularized}
  \mathcal{L}(\theta, \Phi) - \gamma \log \left| \det T(\Phi) \right|,
\end{equation}
which penalizes solutions with small determinant and thus encourages invertible and well-conditioned transition matrices. We set $\gamma = 0.1$ in the BigEarthNet analysis of Section~\ref{sec:case_study}, the only experiment in which this regularization proved necessary.

\subsubsection{Optimization}
\label{appendix:nn-optimization}

The network is optimized via an alternating procedure that decouples the update of the backbone from that of the contamination layer, in a fashion analogous to the Expectation-Maximization algorithm. Each epoch consists of two phases. In the first, the contamination layer is frozen and a fixed number of gradient steps are taken on $\theta$, using all observations in $\mathcal{D}^{\mathrm{cl}} \cup \mathcal{D}^{\mathrm{train}}$. In the second, the backbone is frozen and gradient steps are taken on $\Phi$, using only the contaminated observations in $\mathcal{D}^{\mathrm{train}}$, as the contamination layer does not contribute to the loss on clean observations.

Training is conducted in two stages with decreasing learning rate, to encourage convergence to a stable solution. The contamination layer is initialized so that the initial transition matrix corresponds to a randomized response model with $\epsilon = 10^{-6}$; that is, training starts from an essentially noiseless model and the estimated contamination level increases as needed.

\FloatBarrier

\clearpage

\section{Mathematical Proofs}\label{appendix:mathematical_proofs}

\subsection{Auxiliary Lemmas}

\begin{lemma}\label{lemma:joint-cdf-relation}
For any $t \in [0,1]$, define the $K \times K$ matrices $P(t)$ and $\tilde{P}(t)$ with entries
\begin{align*}
    P_{lk}(t) := \rho_l F_l^{k}(t), \qquad \tilde{P}_{lk}(t) := \tilde{\rho}_l \tilde{F}_l^{k}(t), \qquad l,k \in [K].
\end{align*}
Under Assumption~\ref{assumption:linear-contam}, $\tilde{P}(t) = T P(t)$. Consequently, under Assumption~\ref{assumption:known-contam}, $P(t) = W \tilde{P}(t)$.
\end{lemma}

\begin{proof}[Proof of Lemma~\ref{lemma:joint-cdf-relation}]
Fix $l,k \in [K]$ and $t \in [0,1]$.
We have that:
\begin{align*}
    \tilde{P}_{lk}(t) &= \tilde{\rho}_l \tF_l^k(t) = \P{\tY = l} \P{\hat{s}(X,k)\leq t \mid \tY = l}\\
    &= \P{\hat{s}(X,k)\leq t , \tY = l}\\
    &= \sum_{m=1}^{K}  \P{\hat{s}(X,k)\leq t , \tY = l \mid Y=m} \rho_m \\
    &= \sum_{m=1}^{K}  \P{\hat{s}(X,k)\leq t \mid Y=m} \P{\tY = l \mid Y=m} \rho_m \\
    &= \sum_{m=1}^{K} T_{lm} P_{mk}(t) = [TP(t)]_{lk}.
\end{align*}
The fourth equality follows from Assumption~\ref{assumption:linear-contam}, which states that $X$ and $\tY$ are conditionally independent given $Y$.
It follows that $\tilde{P}(t) = TP(t)$ and that $P(t) = W \tilde{P}(t)$.
\end{proof}

\subsection{Preliminaries}

\begin{proof}[Proof of Theorem~\ref{thm:coverage-marginal}]
For any $k,l \in [K]$ and $t \in [0,1]$, define, as in Section~\ref{sec:adaptive-marginal},
  \begin{align*}
   F_l^k(t) := \P{\hat{s}(X,k) \leq t \mid Y=l}, \qquad
   & \tilde{F}_l^k(t) := \P{\hat{s}(X,k) \leq t \mid \tY = l}.
 \end{align*}
Above, $F_l^k(t)$ is the CDF of $\hat{s}(X,k)$, based on a fixed function $\hat{s}$ applied to a random $X$ from the distribution of $X \mid Y=l$, while $\tF_l^k(t)$ is the corresponding CDF of $\hat{s}(X,k)$, with $X$ conditioned on $\tilde{Y}=l$.
Further, recall, for any $k \in [K]$, $\rho_k := \Ps{Y=k}$ and $\tilde{\rho}_k := \Ps{\tY=k}$, the marginal frequencies of the $k$-th class in the clean and contaminated labels, respectively.

By definition of the non-conformity score function in~\eqref{eq:conf-scores}, $Y_{n+1} \in \hat{C}^{\mathrm{std}}(X_{n+1})$ if and only if $\hat{s}(X_{n+1},Y_{n+1}) \leq \htau^{\mathrm{std}}$.
  Therefore,
  \begin{align*}
    & \P{Y_{n+1} \in \hat{C}^{\mathrm{std}}(X_{n+1})} \\
    & \qquad = \P{\tY_{n+1} \in \hat{C}^{\mathrm{std}}(X_{n+1})} + \left( \P{Y_{n+1} \in \hat{C}^{\mathrm{std}}(X_{n+1})} - \P{\tilde{Y}_{n+1} \in \hat{C}^{\mathrm{std}}(X_{n+1})} \right).
  \end{align*}
The proof is completed by noting that the second term on the right-hand-side above is:
  \begin{align*}
    & \P{Y_{n+1} \in \hat{C}^{\mathrm{std}}(X_{n+1})} - \P{\tilde{Y}_{n+1} \in \hat{C}^{\mathrm{std}}(X_{n+1})} \\
    & \qquad = \sum_{k=1}^{K} \left( \rho_k \cdot \P{Y_{n+1} \in \hat{C}^{\mathrm{std}}(X_{n+1}) \mid Y_{n+1} = k} - \tilde{\rho}_k \cdot  \P{\tY_{n+1} \in \hat{C}^{\mathrm{std}}(X_{n+1}) \mid \tY_{n+1} = k} \right) \\
    & \qquad = \sum_{k=1}^{K} \left( \rho_k \cdot \P{\hat{s}(X_{n+1}, k) \leq \htau^{\mathrm{std}} \mid Y_{n+1} = k} - \tilde{\rho}_k \cdot  \P{ \hat{s}(X_{n+1}, k) \leq \htau^{\mathrm{std}}  \mid \tilde{Y}_{n+1} = k} \right) \\
    & \qquad = \sum_{k=1}^{K} \left( \rho_k \cdot \EV{\P{\hat{s}(X_{n+1}, k) \leq \htau^{\mathrm{std}} \mid Y_{n+1} = k, \mathcal{D}}} \right. \\
    & \qquad \qquad \left. - \tilde{\rho}_k \cdot  \EV{\P{ \hat{s}(X_{n+1}, k) \leq \htau^{\mathrm{std}}  \mid \tilde{Y}_{n+1} = k, \mathcal{D}}} \right) \\
    & \qquad = \EV{ \sum_{k=1}^{K} \left[ \rho_k F_k^{k}(\htau^{\mathrm{std}}) - \tilde{\rho}_k \tF_k^{k}(\htau^{\mathrm{std}})\right] } \\
    & \qquad = \EV{F(\htau^{\mathrm{std}}) - \tF(\htau^{\mathrm{std}})} = \EV{ \Delta(\htau^{\mathrm{std}})}.
  \end{align*}
  The expected value above is taken with respect to the randomness of the data in $\mathcal{D}$, hence including both the training data in $\mathcal{D}^{\mathrm{train}}$ and the calibration data in $\mathcal{D}$.
The upper and lower coverage bounds of Theorem~\ref{thm:coverage-marginal} follow directly because
  \begin{equation*}
      1-\alpha \leq \P{\tY_{n+1} \in \hat{C}^{\mathrm{std}}(X_{n+1})} \leq 1 - \alpha + \frac{1}{n + 1}
  \end{equation*}
  by Proposition~\ref{prop:standard-coverage-marginal}, which applies here since $(X_i, \tY_i)$ are i.i.d.~random samples.
\end{proof}

\begin{proof}[Proof of Proposition~\ref{prop:Delta-marg-linear}]

This result follows directly from Lemma~\ref{lemma:joint-cdf-relation}, which tells us that $P(t) = W \tilde{P}(t)$ where
\begin{align*}
    P_{lk}(t) := \rho_l F_l^{k}(t), \qquad \tilde{P}_{lk}(t) := \tilde{\rho}_l \tilde{F}_l^{k}(t), \qquad l,k \in [K].
\end{align*}
Therefore,
\begin{align*}
  F(t) 
  & = \sum_{k=1}^K \rho_k F_k^k(t) = \sum_{k=1}^K P_{kk}(t)
   = \sum_{k=1}^K \sum_{l=1}^K W_{kl} \tilde{P}_{lk}(t) = \sum_{k=1}^K \sum_{l=1}^K W_{kl} \tilde{\rho}_l \tilde{F}_l^k(t).
\end{align*}
Substituting into $\Delta(t) = F(t) - \tilde{F}(t)$ gives the desired result~\eqref{eq:Delta-marg-linear}.
\end{proof}

\begin{proof}[Proof of Corollary~\ref{cor:coverage-marg}]
  By Theorem~\ref{thm:coverage-marginal}, it suffices to prove $\EV{\Delta(t)} \geq 0$ for all $t \in [0,1]$.
  
  To establish that, first define $M_{kl} := \P{Y=l \mid \tY = k}$.
  Under Assumption~\ref{assumption:linear-contam}, Proposition 1 in \citet{sesia2025adaptive} connects the distribution of $X\mid \tY=k$, namely $\tP_k$, and the distributions of $X \mid Y=l$, namely $P_l$, for all $k,l \in [K]$. That is,
  \begin{align}\label{eq:prop1-sesia}
      \tP_k = \sum_{l=1}^K M_{kl} P_l.
  \end{align}
  
  Note that combining~\eqref{eq:Delta-marg} with~\eqref{eq:prop1-sesia} gives
  \begin{align*}
    \Delta(t) 
    & = F(t) - \tilde{F}(t)\\
    & = \sum_{k=1}^{K} \left[ \rho_k F_k^{k}(t) - \tilde{\rho}_k \tF_k^{k}(t)\right]\\
    & = \sum_{k=1}^{K} \left[ \rho_k  F_k^{k}(t) - \sum_{l=1}^{K} M_{kl}  \tilde{\rho}_k  F_l^{k}(t)  \right].
  \end{align*}
   By combining Bayes' theorem with Assumption~\ref{assumption:linear-contam}, one gets
    $T_{kl} = M_{kl} \cdot \tilde{\rho}_k / \rho_l$.
  Then,
  \begin{align*}
    \Delta(t)
    & = \sum_{l=1}^{K} \rho_l  F_l^{l}(t) -  \sum_{l=1}^{K} \rho_l  \sum_{k=1}^{K}	T_{kl} F_l^{k}(t) \\
    & = \sum_{l=1}^{K} \rho_l  F_l^{l}(t) -  \sum_{l=1}^{K} \rho_l  \left[ T_{ll} F_l^{l}(t) + \sum_{k \neq l} T_{kl} F_l^{k}(t) \right] \\
    & \geq \sum_{l=1}^{K} \rho_l  F_l^{l}(t) -  \sum_{l=1}^{K} \rho_l  \left[ T_{ll} F_l^{l}(t) + \left(\sum_{k \neq l} T_{kl}\right) \max_{k \neq l} F_l^{k}(t) \right] \\
    & = \sum_{l=1}^{K} \rho_l  F_l^{l}(t) -  \sum_{l=1}^{K} \rho_l  \left[ T_{ll} F_l^{l}(t) + \left(\sum_{k=1}^{K} T_{kl} \right) \max_{k \neq l} F_l^{k}(t) - T_{ll} \max_{k \neq l} F_l^{k}(t) \right] \\
    & = \sum_{l=1}^{K} \rho_l  F_l^{l}(t) -  \sum_{l=1}^{K} \rho_l  \left[ T_{ll} F_l^{l}(t) + \max_{k \neq l} F_l^{k}(t) - T_{ll} \max_{k \neq l} F_l^{k}(t) \right] \\
    & = \sum_{l=1}^{K} \rho_l (1-T_{ll}) \left[ F_l^{l}(t) - \max_{k\neq l} F_l^{k}(t) \right] \geq 0,
  \end{align*}
  where the last inequality is given by~\eqref{eq:assump_scores-cond-marg}.
This implies that $\Delta(\htau^{\mathrm{std}}) \geq 0$ almost-surely.
\end{proof}

\subsection{Lower Bound on Coverage of Adaptive Prediction Sets}

\begin{proof}[Proof of Theorem~\ref{thm:algorithm-lower-bound}]

The event $Y_{n+1} \in \hat{C}(X_{n+1})$ occurs if and only if $\hat{s}(X_{n+1},Y_{n+1}) \leq \htau$.
Without loss of generality, we restrict our attention to the event $\hat{\mathcal{I}} \neq \emptyset$; otherwise, coverage is trivially one.
  Therefore, the probability of a miscoverage event conditional on the data in $\mathcal{D}$ can be decomposed as:
  \begin{align*}
    & \P{Y_{n+1} \notin \hat{C}(X_{n+1}) \mid \mathcal{D}} \\
    & \qquad = \P{\tilde{Y}_{n+1} \notin \hat{C}(X_{n+1}) \mid \mathcal{D}} \\
    & \qquad \qquad + \left( \P{Y_{n+1} \notin \hat{C}(X_{n+1}) \mid \mathcal{D}} - \P{\tilde{Y}_{n+1} \notin \hat{C}(X_{n+1}) \mid \mathcal{D}} \right) \\
    & \qquad = 1 - \tilde{F}(S_{(\hat{i})}) - \Delta(S_{(\hat{i})}) \\
     & \qquad = 1 - \hat{F}(S_{(\hat{i})}) - \hat{\Delta}(S_{(\hat{i})}) \\
     & \qquad \qquad + \left[ \hat{F}(S_{(\hat{i})}) - \tilde{F}(S_{(\hat{i})}) \right] + \left[ \hat{\Delta}(S_{(\hat{i})}) - \Delta(S_{(\hat{i})}) \right] \\
     & \qquad = \left\{ 1 - \frac{\hat{i}}{n} - \hat{\Delta}(S_{(\hat{i})}) + \delta(n) \right\} - \delta(n)
       + \sum_{k=1}^{K}\sum_{l=1}^{K} W_{kl} \left[ \hat{\rho}_l \hat{F}_l^{k}(S_{(\hat{i})}) - \tilde{\rho}_l \tilde{F}_l^{k}(S_{(\hat{i})}) \right] \\
     & \qquad \leq \sup_{i \in \hat{\mathcal{I}}} \left\{ 1 - \frac{i}{n} - \hat{\Delta}(S_{(i)}) + \delta(n) \right\} - \delta(n)\\
     & \qquad \qquad + \sup_{t \in [0,1]} \left\{ \sum_{k=1}^{K} \sum_{l=1}^{K} W_{kl} \left[ \hat{\rho}_l \hat{F}_l^{k}(t) - \tilde{\rho}_l \tilde{F}_l^{k}(t) \right] \right\}.
  \end{align*}
By definition of $\hat{\mathcal{I}}$, for all $i \in \hat{\mathcal{I}}$, $ 1 - i/n - \hat{\Delta}(S_{(i)}) + \delta(n) \leq \alpha$,
which implies a.s.
  \begin{align*}
    \sup_{i \in \hat{\mathcal{I}}} \left\{ 1 - \frac{i}{n} - \hat{\Delta}(S_{(i)}) + \delta(n) \right\} \leq \alpha.
  \end{align*}
Therefore,
  \begin{align*}
    \P{Y_{n+1} \notin \hat{C}(X_{n+1})}
      & \leq \alpha + \EV{ \sup_{t \in [0,1]} \left\{ \sum_{k=1}^{K} \sum_{l=1}^{K} W_{kl} \left[ \hat{\rho}_l \hat{F}_l^{k}(t) - \tilde{\rho}_l \tilde{F}_l^{k}(t) \right] \right\} } - \delta(n)\\
     & \qquad = \alpha + \delta^*(n) - \delta(n)\\ 
     & \qquad \leq \alpha,
  \end{align*}
  where the last inequality above follows directly from the assumption that $\delta(n) \geq \delta^*(n)$.
\end{proof}

\begin{proof}[Proof of Proposition~\ref{prop:algorithm-correction-optimistic}]
Throughout the proof, write $\lambda_n := \sqrt{\log(n) / (2n)}$, so that
assumption~\eqref{eq:optimistic-assumption} reads
$\inf_{t \in \mathcal{T}_n} \Delta(t) \geq \delta(n) - (1-\alpha)/n + \lambda_n$, and define
\begin{align*}
    \hat{\Xi}_{F} := \sup_{t \in [0,1]} \left( \hat{F}(t) - \tilde{F}(t) \right).
\end{align*}
Further, let $\hat{\mathcal{I}}_{\mathrm{ad}}$ denote the set in~\eqref{eq:I-set-marg},
let $\hat{i}_{\mathrm{ad}} := \min \hat{\mathcal{I}}_{\mathrm{ad}}$, and let
$i_{+} := \lceil (n+1)(1-\alpha) \rceil$. Note that $\hat{\mathcal{I}}_{\mathrm{ad}} \neq \emptyset$,
because $n \in \hat{\mathcal{I}}_{\mathrm{ad}}$: indeed, $\hat{\Delta}(1) = \Delta(1) = 0$ (see the
computation of $\hat{\psi}(1)$ below) and $\delta(n) \leq \alpha$.
Since $\max\{\hat{\Delta}(S_{(i)}) - \delta(n), -(1-\alpha)/n\} \geq -(1-\alpha)/n$ for all $i \in [n]$,
the set $\hat{\mathcal{I}}$ in~\eqref{eq:I-set-marg-optimistic} equals
$\hat{\mathcal{I}}_{\mathrm{ad}} \cup \{i \in [n] : i \geq i_{+}\}$, and hence
\begin{align}
\label{eq:optimistic-min-rank}
    \hat{i} = \min\{\hat{i}_{\mathrm{ad}}, \, i_{+}\}.
\end{align}

Proceeding as in the proof of Theorem~\ref{thm:algorithm-lower-bound}, and recalling that
$\hat{\psi}(t) = \hat{F}(t) - \tilde{F}(t) + \hat{\Delta}(t) - \Delta(t)$ for all $t \in [0,1]$,
which follows directly from the definitions of $\hat{\psi}$, $\Delta$, and $\hat{\Delta}$, we
obtain:
\begin{align*}
    & \P{Y_{n+1} \notin \hat{C} (X_{n+1}) \mid \mathcal{D}} \\
    & \qquad = \left[ 1 - \frac{\hat{i}}{n} - \max\left\{\hat{\Delta}(S_{(\hat{i} )}) - \delta (n), -\frac{1-\alpha}{n} \right\}\right] - \delta(n)  \\
    & \qquad \qquad + \max\left\{\hat{\psi}(S_{(\hat{i} )}), \; \hat{F}(S_{(\hat{i})}) - \tilde{F}(S_{(\hat{i})}) - \left( \Delta(S_{(\hat{i}) }) - \delta(n) + \frac{1-\alpha}{n}\right) \right\} \\
    & \qquad = 1 - \tilde{F}(S_{(\hat{i})}) - \Delta(S_{(\hat{i})}),
\end{align*}
where the last equality follows by substituting the expression for $\hat{\psi}$ above and
$\hat{F}(S_{(\hat{i})}) = \hat{i}/n$, after which the two maxima cancel.
Since the function $t \mapsto 1 - \tilde{F}(t) - \Delta(t)$ is non-increasing (it is the
conditional miscoverage probability of the prediction set with threshold $t$), it follows
from~\eqref{eq:optimistic-min-rank} and $S_{(\hat{i})} = \min\{S_{(\hat{i}_{\mathrm{ad}})}, S_{(i_{+})}\}$ that
\begin{align}
\label{eq:optimistic-max}
    \P{Y_{n+1} \notin \hat{C} (X_{n+1}) \mid \mathcal{D}}
    = \max\{ M_{\mathrm{ad}}, \, M_{\mathrm{std}} \},
\end{align}
where
\begin{align*}
    M_{\mathrm{ad}} := 1 - \tilde{F}(S_{(\hat{i}_{\mathrm{ad}})}) - \Delta(S_{(\hat{i}_{\mathrm{ad}})}),
    \qquad
    M_{\mathrm{std}} := 1 - \tilde{F}(S_{(i_{+})}) - \Delta(S_{(i_{+})}) .
\end{align*}
We bound each term separately. Regarding $M_{\mathrm{ad}}$, writing
$\tilde{F} + \Delta = \hat{F} + \hat{\Delta} - \hat{\psi}$ and using $\hat{F}(S_{(\hat{i}_{\mathrm{ad}})}) = \hat{i}_{\mathrm{ad}}/n$,
\begin{align*}
    M_{\mathrm{ad}}
    = 1 - \frac{\hat{i}_{\mathrm{ad}}}{n} - \hat{\Delta}(S_{(\hat{i}_{\mathrm{ad}})}) + \hat{\psi}(S_{(\hat{i}_{\mathrm{ad}})}) .
\end{align*}
By definition of $\hat{\mathcal{I}}_{\mathrm{ad}}$ in~\eqref{eq:I-set-marg}, we have
$1 - \hat{i}_{\mathrm{ad}}/n - \hat{\Delta}(S_{(\hat{i}_{\mathrm{ad}})}) \leq \alpha - \delta(n)$, hence
\begin{align}
\label{eq:optimistic-M-ad}
    M_{\mathrm{ad}} \leq \alpha - \delta(n) + \sup_{t \in [0,1]} \hat{\psi}(t) .
\end{align}
Regarding $M_{\mathrm{std}}$, writing $\tilde{F} = \hat{F} - (\hat{F} - \tilde{F})$ and using
$\hat{F}(S_{(i_{+})}) = i_{+}/n \geq (n+1)(1-\alpha)/n$,
\begin{align}
\label{eq:optimistic-M-std}
    M_{\mathrm{std}}
    = 1 - \frac{i_{+}}{n} + \left( \hat{F}(S_{(i_{+})}) - \tilde{F}(S_{(i_{+})}) \right) - \Delta(S_{(i_{+})})
    \leq \alpha - \frac{1-\alpha}{n} + \hat{\Xi}_{F} - \Delta(S_{(i_{+})}) .
\end{align}

Combining these results,
\begin{align*}
  \P{Y_{n+1} \notin \hat{C} (X_{n+1}) \mid \mathcal{D}}
  & \leq \alpha - \delta(n) + \max\left\{ \sup_{t \in [0,1]} \hat{\psi}(t), \hat{\Xi}_{F} - \left[ \Delta(S_{(i_{+})}) - \delta(n) + \frac{1-\alpha}{n} \right] \right\}  \\
    & \leq \alpha - \delta(n) + \sup_{t \in [0,1]} \hat{\psi}(t)
    + \left( \hat{\Xi}_{F} - \left[ \Delta(S_{(i_{+})}) - \delta(n) + \frac{1-\alpha}{n} \right] \right)_{\!+},
\end{align*}
where the second step uses the elementary inequality $\max\{a,b\} \leq a + (b)_+$, which is valid for any $a \geq 0$
and therefore applicable here because $\sup_{t \in [0,1]}\hat{\psi}(t) \geq 0$ almost surely. Indeed,
\begin{align*}
    \sup_{t \in [0,1]}\hat{\psi}(t) &\geq \hat{\psi}(1) \\
    &= \sum_{k} \sum_{l} W_{kl} \left(\hat{\rho}_l - \tilde{\rho}_l\right)\\
    &= \sum_{l} \sum_{k} W_{kl} \hat{\rho}_l - \sum_{l} \sum_{k} W_{kl} \tilde{\rho}_l\\
    &= \sum_{l} \hat{\rho}_l \sum_{k} W_{kl} - \sum_{l} \tilde{\rho}_l \sum_{k} W_{kl}\\
    &= \sum_{l} \hat{\rho}_l - \sum_{l} \tilde{\rho}_l = 1 - 1 = 0,
\end{align*}
where the first equality uses $\hat{F}^k_l(1) = \tilde{F}^k_l(1) = 1$ for all $k,l \in [K]$,
which holds because the non-conformity scores take values in $[0,1]$ (cf.~Definition~\ref{def:pred-function}), 
and where we used the fact that, being the inverse of a stochastic matrix, the columns
of $W$ sum to one; i.e., $\sum_{k} W_{kl} = 1$ for all $l \in [K]$. 

Next, taking the expected value with respect to the data in $\mathcal{D}$ and proceeding similarly
as in the proof of Theorem~\ref{thm:algorithm-lower-bound}, we get
\begin{align}
\label{eq:optimistic-master}
\begin{split}
    \P{Y_{n+1} \notin \hat{C} (X_{n+1})}
    & \leq \alpha - \delta(n) + \delta^{*}(n) + \EV{ \left( \hat{\Xi}_{F} - \left[ \Delta(S_{(i_{+})}) - \delta(n) + \frac{1-\alpha}{n} \right] \right)_{\!+} } \\
    & \leq  \alpha + \EV{ \left( \hat{\Xi}_{F} - \left[ \Delta(S_{(i_{+})}) - \delta(n) + \frac{1-\alpha}{n} \right] \right)_{\!+} },
  \end{split}
\end{align}
where the last inequality above follows directly from the assumption that
$\delta(n) \geq \delta^{*}(n)$.

It remains to bound the expectation in~\eqref{eq:optimistic-master}. Let
$E_n := \{ S_{(i_{+})} \in \mathcal{T}_n \}$. On $E_n$, assumption~\eqref{eq:optimistic-assumption}
gives $\Delta(S_{(i_{+})}) - \delta(n) + (1-\alpha)/n \geq \lambda_n$, so that the positive
part in~\eqref{eq:optimistic-master} is bounded from above by $(\hat{\Xi}_{F} - \lambda_n)_+$.
On $E_n^{c}$, we simply use $\hat{\Xi}_{F} \leq 1$ and $\Delta(t) \geq -1$ for all $t \in [0,1]$
(as $\Delta$ is the difference of two functions with values in $[0,1]$), so that the positive
part is bounded from above by $2 + \delta(n)$. Hence,
\begin{align}
\label{eq:optimistic-split}
    \EV{ \left( \hat{\Xi}_{F} - \left[ \Delta(S_{(i_{+})}) - \delta(n) + \frac{1-\alpha}{n} \right] \right)_{\!+} }
    \leq \EV{ \left( \hat{\Xi}_{F} - \lambda_n \right)_{\!+} } + (2 + \delta(n)) \, \P{E_n^{c}} .
\end{align}
To bound $\P{E_n^c}$, note that $\hat{F}(S_{(i_{+})}) = i_{+}/n \in [1-\alpha, \, 1 - \alpha + 2/n]$,
since $(n+1)(1-\alpha) \leq i_{+} < (n+1)(1-\alpha) + 1$. Therefore, on the event
$\sup_{t \in [0,1]} |\hat{F}(t) - \tilde{F}(t)| \leq \epsilon_n$, we have
$1 - \alpha - \epsilon_n \leq \tilde{F}(S_{(i_{+})}) \leq 1 - \alpha + 2/n + \epsilon_n$, that is,
$S_{(i_{+})} \in \mathcal{T}_n$. As $\hat{F}$ is the empirical distribution
function of the $n$ i.i.d.\ scores $\hat{s}(X_i, \tilde{Y}_i)$, whose common distribution
function is $\tilde{F}$, the two-sided DKW inequality with Massart's sharp constant yields
\begin{align*}
    \P{E_n^{c}} \leq \P{ \sup_{t \in [0,1]} |\hat{F}(t) - \tilde{F}(t)| > \epsilon_n } \leq 2 e^{-2 n \epsilon_n^2} = \gamma_n ,
\end{align*}
by the choice of $\epsilon_n$.

Finally, the one-sided DKW inequality with Massart's sharp constant gives
$\P{\hat{\Xi}_{F} > z} \leq e^{-2nz^2}$ for every $z > 0$. Hence, using
$z / \lambda_n \geq 1$ on the domain of integration,
\begin{align*}
    \EV{ \left( \hat{\Xi}_{F} - \lambda_n \right)_{\!+} }
    = \int_{\lambda_n}^{\infty} \P{\hat{\Xi}_{F} > z} \, dz
    \leq \int_{\lambda_n}^{\infty} e^{-2nz^2} dz
    \leq \int_{\lambda_n}^{\infty} \frac{z}{\lambda_n} e^{-2nz^2} dz
    = \frac{1}{4n\lambda_n} e^{-2n\lambda_n^2} .
\end{align*}
The choice $\lambda_n = \sqrt{\log(n)/(2n)}$ yields $2n\lambda_n^2 = \log n$ and
$4n\lambda_n = 2\sqrt{2}\sqrt{n \log n}$, so that
\begin{align*}
    \EV{ \left( \hat{\Xi}_{F} - \lambda_n \right)_{\!+} }
    \leq \frac{1}{2\sqrt{2}\sqrt{n \log n}} \cdot \frac{1}{n}
    = \frac{1}{2\sqrt{2}\, n^{3/2} \sqrt{\log n}} = \gamma_n .
\end{align*}
Combining this with~\eqref{eq:optimistic-split} and~\eqref{eq:optimistic-master} completes the proof.
\end{proof}

\begin{proof}[Proof of Corollary~\ref{cor:optimistic-exact}]
Immediate from the first inequality in~\eqref{eq:optimistic-master} together with the bound
$\EV{(\hat{\Xi}_{F} - \lambda_n)_+} \leq \gamma_n$ established in the proof of
Proposition~\ref{prop:algorithm-correction-optimistic}.
\end{proof}

\subsection{Finite-Sample Correction Factor}

\begin{proof}[Proof of Theorem~\ref{thm:finite-sample-factor-simplified}]
Consider $\delta^{*}(n)$ defined in~\eqref{eq:hat-psi}.
We want to show that
\begin{align*}
    \frac{1+\epsilon}{1-\epsilon} \, c(n) \geq \delta^*(n),
\end{align*}
if $T = (1-\epsilon) \cdot I + \epsilon/K \cdot J$ for some $\epsilon \in [0,1)$.

To start, it is helpful to parametrize $W$ in terms of coefficients $\beta_0, \beta_1$; that is
\begin{equation*}
    W = T^{-1} = \frac{1}{1-\epsilon} I - \frac{\epsilon}{K(1-\epsilon)} J = \beta_0 I + \frac{\beta_1}{K} J,
\end{equation*}
where we define $\beta_0 := \frac{1}{1-\epsilon}$ and $\beta_1 := - \frac{\epsilon}{1-\epsilon} $.

Note that
\begin{align*}
  \hat{\psi}(t) 
  & = \frac{1}{n} \sum_{i=1}^{n} \sum_{k=1}^{K} \sum_{l=1}^{K} W_{kl} \left[  \I{\hat{s}(X_i,k) \leq t, \tilde{Y}_i = l} - \tilde{\rho}_l \tilde{F}_l^{k}(t) \right]  \\
  & =  \frac{1}{n} \sum_{i=1}^{n} \sum_{k=1}^{K} \sum_{l=1}^{K} \left( \beta_0 \I{k=l} + \beta_1/K \right) \left[  \I{\hat{s}(X_i,k) \leq t, \tilde{Y}_i = l} - \tilde{\rho}_l \tilde{F}_l^{k}(t) \right]  \\
  & = \beta_0 \frac{1}{n} \sum_{i=1}^{n} \sum_{k=1}^{K} \left[  \I{\hat{s}(X_i,k) \leq t, \tilde{Y}_i = k} - \tilde{\rho}_k \tilde{F}_k^{k}(t) \right]  \\
  & \qquad +  \frac{1}{n} \sum_{i=1}^{n} \sum_{k=1}^{K} (\beta_1/K) \sum_{l=1}^{K} \left[  \I{\hat{s}(X_i,k) \leq t, \tilde{Y}_i = l} - \tilde{\rho}_l \tilde{F}_l^{k}(t) \right]  \\
  & = \beta_0 \frac{1}{n} \sum_{i=1}^{n} \left[  \I{\hat{s}(X_i,\tilde{Y}_i) \leq t} - \tilde{F}(t) \right]  \\
  & \qquad + \frac{1}{n} \sum_{i=1}^{n} \sum_{k=1}^{K} (\beta_1/K) \left[  \I{\hat{s}(X_i,k) \leq t} - \tilde{F}^{k}(t) \right]\\
  & = \hat{\psi}_1(t) + \hat{\psi}_2(t),
\end{align*}
where
\begin{align*}
    \hat{\psi}_1(t)
    & := \beta_0 \frac{1}{n} \sum_{i=1}^{n} \left[  \I{\hat{s}(X_i,\tilde{Y}_i) \leq t} - \tilde{F}(t) \right], \\
    \hat{\psi}_2(t)
    & := \frac{1}{n} \sum_{i=1}^{n} \sum_{k=1}^{K} (\beta_1/K) \left[  \I{\hat{s}(X_i,k) \leq t} - \tilde{F}^{k}(t) \right],\\
  \tilde{F}^{k}(t)
  & := \P{\hat{s}(X_i,k) \leq t}, \\
  \tilde{F}(t)
  & := \P{\hat{s}(X_i,\tilde{Y}_i) \leq t}.
\end{align*}

We now aim to bound the expected value of the supremum of $\hat{\psi}$, and we proceed by bounding separately the expected value of the suprema of $\hat{\psi}_1(t)$ and $\hat{\psi}_2(t)$.

We start with $\hat{\psi}_1(t)$.
First, note that
\begin{align*}
    \hat{\psi}_1(t) = \beta_0 \; \left( \hat{F}(t) - \tilde{F}(t) \right).
\end{align*}

Let $U_1,\ldots, U_n$ be i.i.d uniform random variables on $[0,1]$, and denote their order statistics as $U_{(1)},\ldots,U_{(n)}$. Since $\tilde{F}$ is continuous, as implied by Assumption~\ref{assumption:score-cdf-continuous}, the following equality in distribution holds
\begin{equation*}
    \sup_{t \in [0,1]} \left( \hat{F}(t) - \tilde{F}(t)\right) \overset{d}{=} \sup_{i \in [n]} \left\{  \frac{i}{n} - U_{(i)} \right\}.
\end{equation*}

Following from the definition of $c(n)$ in~\eqref{eq:c-definition}, this implies that
\begin{equation*}
    \EV{ \sup_{t \in [0,1]} \left( \hat{F}(t) - \tilde{F}(t)\right) } = \EV{\sup_{i \in [n]} \left\{ \frac{i}{n} - U_{(i)}\right\}} = c(n).
\end{equation*}

Therefore, we get
\begin{equation*}
  \EV{\sup_{t \in [0,1]}  \hat{\psi}_1(t)} = \beta_0 \EV{\sup_{t \in [0,1]} \;  \left( \hat{F}(t) - \tilde{F}(t)\right)} = \beta_0 c(n).
\end{equation*}

We can deal with $\hat{\psi}_2(t)$ similarly.
First, note that
\begin{align*}
    \hat{\psi}_2(t) = \sum_{k=1}^K \frac{\beta_1}{K} \left( \hat{F}^k(t) - \tilde{F}^k(t) \right),
\end{align*}
where $\hat{F}^k(t) = \frac{1}{n} \sum_{i=1}^n \I{\hat{s}(X_i,k) \leq t}$. Recall that $\beta_1 < 0$.
Then,
\begin{align*}
  \EV{ \sup_{t \in [0,1]} \hat{\psi}_2(t) } &= \EV{ \sup_{t \in [0,1]}  \left( \sum_{k=1}^{K} \frac{\beta_1}{K} \left( \hat{F}^{k}(t) - \tilde{F}^{k}(t) \right) \right) }\\
  & = \EV{ \sup_{t \in [0,1]} \left(\sum_{k=1}^{K} \frac{|\beta_1|}{K} \frac{\beta_1}{|\beta_1|} \left(\hat{F}^{k}(t) - \tilde{F}^{k}(t) \right)\right) }\\
  & \leq \EV{\sum_{k=1}^{K} \frac{|\beta_1|}{K} \; \sup_{t \in [0,1]} \frac{\beta_1}{|\beta_1|} \left(\hat{F}^{k}(t) - \tilde{F}^{k}(t) \right) }\\
  & = \frac{|\beta_1|}{K} \sum_{k=1}^{K} \; \EV{ \sup_{t \in [0,1]} \frac{\beta_1}{|\beta_1|} \left(\hat{F}^{k}(t) - \tilde{F}^{k}(t) \right) }\\
  & = \frac{|\beta_1|}{K} \sum_{k=1}^{K} \EV{ \sup_{t \in [0,1]} \left(\hat{F}^{k}(t) - \tilde{F}^{k}(t) \right) } 
    \tag*{\footnotesize (by Lemma~\ref{lem:sup-symmetry})} \\
  & = \frac{|\beta_1|}{K} \sum_{k=1}^{K} c(n) = |\beta_1| c(n).
\end{align*}
We report below Lemma~\ref{lem:sup-symmetry}, which holds here because $\tilde{F}^k$ is continuous for any $k \in [K]$ (Assumption \ref{assumption:score-cdf-continuous}). The result follows by noting that $\beta_0 + | \beta_1 | = \frac{1+\epsilon}{1-\epsilon}$.
\end{proof}

\begin{lemma}
\label{lem:sup-symmetry}
Let $Z_1, \dots, Z_n$ be i.i.d.\ random variables with continuous distribution function $G$,
and let $\hat{G}_n(t) := n^{-1} \sum_{i=1}^n \I{Z_i \leq t}$ denote the corresponding empirical
distribution function. Then
\begin{align}
\label{eq:sup-symmetry}
    \sup_{t \in \mathbb{R}} \left( \hat{G}_n(t) - G(t) \right)
    \; \overset{d}{=} \;
    \sup_{t \in \mathbb{R}} \left( G(t) - \hat{G}_n(t) \right).
\end{align}
Consequently, for any $a \in \mathbb{R}$,
\begin{align}
\label{eq:sup-scalar}
    \EV{ \sup_{t \in \mathbb{R}} \; a \left( \hat{G}_n(t) - G(t) \right) } = |a| \; c(n),
\end{align}
where $c(n)$ is the quantity defined in~\eqref{eq:c-definition}.
\end{lemma}

\begin{proof}[Proof of Lemma~\ref{lem:sup-symmetry}]
Let $U_1, \dots, U_n$ be i.i.d.\ uniform random variables on $[0,1]$ and let
$U_{(1)} \leq \dots \leq U_{(n)}$ be their order statistics. Since $G$ is continuous, the
probability integral transform gives $G(Z_i) \overset{d}{=} U_i$, and hence
(see e.g.~\citealp{durbin1973distribution}, eq.~(2.1.5))
\begin{align}
\label{eq:sup-pit}
    \sup_{t \in \mathbb{R}} \left( \hat{G}_n(t) - G(t) \right)
    \overset{d}{=} \sup_{i \in [n]} \left\{ \frac{i}{n} - U_{(i)} \right\},
    \qquad
    \sup_{t \in \mathbb{R}} \left( G(t) - \hat{G}_n(t) \right)
    \overset{d}{=} \sup_{i \in [n]} \left\{ U_{(i)} - \frac{i-1}{n} \right\}.
\end{align}
It therefore suffices to show that the two suprema on the right-hand sides
of~\eqref{eq:sup-pit} are equal in distribution.
Since $1 - U_1, \dots, 1 - U_n$ are again i.i.d.\ uniform random variables on $[0,1]$, and since
reversing the order of a sample reverses the order of its order statistics, we have the
distributional identity
\begin{align*}
    \big( 1 - U_{(n)},\, 1 - U_{(n-1)},\, \dots,\, 1 - U_{(1)} \big)
    \; \overset{d}{=} \;
    \big( U_{(1)},\, U_{(2)},\, \dots,\, U_{(n)} \big).
\end{align*}
Define $V_j := 1 - U_{(n+1-j)}$ for $j \in [n]$, so that
$(V_1, \dots, V_n) \overset{d}{=} (U_{(1)}, \dots, U_{(n)})$. Substituting $j = n+1-i$,
\begin{align*}
    \frac{i}{n} - U_{(i)}
    = \frac{n+1-j}{n} - U_{(n+1-j)}
    = \big( 1 - U_{(n+1-j)} \big) - \frac{j-1}{n}
    = V_j - \frac{j-1}{n},
\end{align*}
and therefore, since $i \mapsto n+1-i$ is a bijection of $[n]$,
\begin{align*}
    \sup_{i \in [n]} \left\{ \frac{i}{n} - U_{(i)} \right\}
    = \sup_{j \in [n]} \left\{ V_j - \frac{j-1}{n} \right\}
    \; \overset{d}{=} \;
    \sup_{i \in [n]} \left\{ U_{(i)} - \frac{i-1}{n} \right\},
\end{align*}
where the last step uses $(V_1, \dots, V_n) \overset{d}{=} (U_{(1)}, \dots, U_{(n)})$.
Combining this with~\eqref{eq:sup-pit} proves~\eqref{eq:sup-symmetry}.

It remains to establish~\eqref{eq:sup-scalar}. The case $a = 0$ is trivial. For $a \neq 0$,
pulling the positive constant $|a|$ out of the supremum gives
\begin{align*}
    \EV{ \sup_{t \in \mathbb{R}} \; a \left( \hat{G}_n(t) - G(t) \right) }
    = |a| \; \EV{ \sup_{t \in \mathbb{R}} \; \frac{a}{|a|} \left( \hat{G}_n(t) - G(t) \right) }.
\end{align*}
If $a > 0$ the right-hand side is $|a| \, \EV{ \sup_t ( \hat{G}_n - G ) }$, while if $a < 0$
it is $|a| \, \EV{ \sup_t ( G - \hat{G}_n ) }$; by~\eqref{eq:sup-symmetry} these two
expectations coincide. Finally, the first identity in~\eqref{eq:sup-pit} together with the
definition of $c(n)$ in~\eqref{eq:c-definition} gives
$\EV{ \sup_t ( \hat{G}_n - G ) } = \EV{ \sup_{i \in [n]} \{ i/n - U_{(i)} \} } = c(n)$,
which completes the proof.
\end{proof}

\begin{proof}[Proof of Theorem~\ref{thm:finite-sample-factor}]
Let us define the empirical process $\hat{\psi}(t)$ as follows:
\begin{align*}
  \hat{\psi}(t) 
  & := \sum_{k=1}^{K} \sum_{l=1}^{K} W_{kl} \left[ \hat{\rho}_l \hat{F}_l^k(t) - \tilde{\rho}_l \tilde{F}_l^{k}(t) \right].
\end{align*}
Our goal is equivalent to that of proving
\begin{align*}
  \EV{ \sup_{t \in [0,1]} \hat{\psi}(t)  } & \leq \delta^{\mathrm{FS}}(n).
\end{align*}

Recall that the matrix $\bar{W}$ is defined such that $\bar{W}_{kl} = \beta_0 \I{k=l} + \beta_k/K$, for some $\beta_0, (\beta_k)_{k}$. The intuition behind the proof is as follows. We will derive a bound as a function of the parameters $\beta$; then, the result can be obtained by taking the minimum over $\beta$.

Note that
\begin{align*}
  \hat{\psi}(t) 
  & = \frac{1}{n} \sum_{i=1}^{n} \sum_{k=1}^{K} \sum_{l=1}^{K} W_{kl} \left[  \I{\hat{s}(X_i,k) \leq t, \tilde{Y}_i = l} - \tilde{\rho}_l \tilde{F}_l^{k}(t) \right]  \\
  & = \frac{1}{n} \sum_{i=1}^{n} \sum_{k=1}^{K} \sum_{l=1}^{K} \left( W_{kl} - \bar{W}_{kl} \right) \left[  \I{\hat{s}(X_i,k) \leq t, \tilde{Y}_i = l} - \tilde{\rho}_l \tilde{F}_l^{k}(t) \right]  \\
  & \qquad + \frac{1}{n} \sum_{i=1}^{n} \sum_{k=1}^{K} \sum_{l=1}^{K} \left( \beta_0 \I{k=l} + \beta_k/K \right) \left[  \I{\hat{s}(X_i,k) \leq t, \tilde{Y}_i = l} - \tilde{\rho}_l \tilde{F}_l^{k}(t) \right]  \\
  & = \frac{1}{n} \sum_{i=1}^{n} \sum_{k=1}^{K} \sum_{l=1}^{K} \left( W_{kl} - \bar{W}_{kl} \right) \left[  \I{\hat{s}(X_i,k) \leq t, \tilde{Y}_i = l} - \tilde{\rho}_l \tilde{F}_l^{k}(t) \right]  \\
  & \qquad + \beta_0 \frac{1}{n} \sum_{i=1}^{n} \sum_{k=1}^{K} \left[  \I{\hat{s}(X_i,k) \leq t, \tilde{Y}_i = k} - \tilde{\rho}_k \tilde{F}_k^{k}(t) \right]  \\
  & \qquad +  \frac{1}{n} \sum_{i=1}^{n} \sum_{k=1}^{K} (\beta_k/K) \sum_{l=1}^{K} \left[  \I{\hat{s}(X_i,k) \leq t, \tilde{Y}_i = l} - \tilde{\rho}_l \tilde{F}_l^{k}(t) \right]  \\
  & = \frac{1}{n} \sum_{i=1}^{n} \sum_{k=1}^{K} \sum_{l=1}^{K} \left( W_{kl} - \bar{W}_{kl} \right) \left[  \I{\hat{s}(X_i,k) \leq t, \tilde{Y}_i = l} - \tilde{\rho}_l \tilde{F}_l^{k}(t) \right]  \\
  & \qquad + \beta_0 \frac{1}{n} \sum_{i=1}^{n} \left[  \I{\hat{s}(X_i,\tilde{Y}_i) \leq t} - \tilde{F}(t) \right]  \\
  & \qquad + \frac{1}{n} \sum_{i=1}^{n} \sum_{k=1}^{K} (\beta_k/K) \left[  \I{\hat{s}(X_i,k) \leq t} - \tilde{F}^{k}(t) \right]  \\
  & = \hat{\psi}_1(t)  + \hat{\psi}_2(t)  + \hat{\psi}_3(t),
\end{align*}
where
\begin{align*}
  \hat{\psi}_1(t)
  & := \beta_0 \frac{1}{n} \sum_{i=1}^{n} \left[  \I{\hat{s}(X_i,\tilde{Y}_i) \leq t} - \tilde{F}(t) \right],  \\
  \hat{\psi}_2(t) 
  & := \frac{1}{n} \sum_{i=1}^{n} \sum_{k=1}^{K} (\beta_k/K) \left[  \I{\hat{s}(X_i,k) \leq t} - \tilde{F}^{k}(t) \right] \\
  \hat{\psi}_3(t) 
  & := \frac{1}{n} \sum_{i=1}^{n} \sum_{k=1}^{K} \sum_{l=1}^{K} (W_{kl}-\bar{W}_{kl}) \left[  \I{\hat{s}(X_i,k) \leq t, \tilde{Y}_i = l} - \tilde{\rho}_l \tilde{F}_l^{k}(t) \right], \\
  \tilde{F}^{k}(t)
  & := \P{\hat{s}(X_i,k) \leq t}, \\
  \tilde{F}(t)
  & := \P{\hat{s}(X_i,\tilde{Y}_i) \leq t}.
\end{align*}

We now bound the suprema of these three processes separately, since
\begin{align*}
  \EV{ \sup_{t \in [0,1]} \hat{\psi}(t)  } & \leq \EV{ \sup_{t \in [0,1]} \hat{\psi}_1(t)  } + \EV{ \sup_{t \in [0,1]} \hat{\psi}_2(t)  } + \EV{ \sup_{t \in [0,1]} \hat{\psi}_3(t)  }.
\end{align*}

For $\hat{\psi}_1(t)$, we apply Lemma~\ref{lem:sup-symmetry}:
\begin{equation*}
  \EV{\sup_{t \in [0,1]} \hat{\psi}_1(t)} = \EV{ \sup_{t \in [0,1]}  \beta_0 \left( \hat{F}(t) - \tilde{F}(t) \right) } = |\beta_0| \; c(n).
\end{equation*}
We proceed similarly for $\hat{\psi}_2(t)$, also using Lemma~\ref{lem:sup-symmetry}. Let $\sigma_k:=\mathrm{sign}(\beta_k)$, with the convention $\mathrm{sign}(0) = +1$, so that $\beta_k = \sigma_k |\beta_k|$. Then,
\begin{align*}
  \EV{ \sup_{t \in [0,1]} \hat{\psi}_2(t)  }
  & = \EV{ \sup_{t \in [0,1]} \sum_{k=1}^{K} \frac{\beta_k}{K}  \left( \hat{F}^{k}(t) - \tilde{F}^{k}(t) \right) }  \\
  & = \EV{ \sup_{t \in [0,1]} \sum_{k=1}^{K} \frac{\sigma_k |\beta_k|}{K}  \left( \hat{F}^{k}(t) - \tilde{F}^{k}(t) \right) }  \\
  &\leq \EV{\sum_{k=1}^{K} \frac{|\beta_k|}{K} \sup_{t \in [0,1]} \sigma_k \left( \hat{F}^{k}(t) - \tilde{F}^{k}(t) \right) }  \\
  &= \sum_{k=1}^{K} \frac{|\beta_k|}{K} \EV{ \sup_{t \in [0,1]} \sigma_k \left( \hat{F}^{k}(t) - \tilde{F}^{k}(t) \right) }  \\
  &= \sum_{k=1}^{K} \frac{|\beta_k|}{K} \EV{ \sup_{t \in [0,1]} \left( \hat{F}^{k}(t) - \tilde{F}^{k}(t) \right) }  \tag*{\footnotesize (by Lemma~\ref{lem:sup-symmetry})} \\
  & = \frac{\sum_{k=1}^{K} |\beta_k|}{K} c(n).
\end{align*}

What remains to be bound is $\hat{\psi}_3(t)$.
For simplicity, define $\Omega_{kl} := W_{kl}-\bar{W}_{kl}$, so that
\begin{align*}
  \hat{\psi}_3(t) 
  & := \frac{1}{n} \sum_{i=1}^{n} \sum_{k=1}^{K} \sum_{l=1}^{K} \Omega_{kl} \left[  \I{\hat{s}(X_i,k) \leq t, \tilde{Y}_i = l} - \tilde{\rho}_l \tilde{F}_l^{k}(t) \right].
\end{align*}
The bound on $\hat{\psi}_3(t)$ is found following two alternative approaches leading to two different bounds, which we combine by taking the minimum of the two. Both approaches focus on bounding the supremum of $|\hat{\psi}_3(t)|$. They both rely on symmetrization and on bounding of the Rademacher complexity of a specific set of functions, and differ in the way the bound on the Rademacher dimension is found. One approach employs Massart's inequality, the other employs the chaining technique.
The result is gathered in Lemma~\ref{lemma:bound-on-psi3} and reads
\begin{align*}
  \EV{ \sup_{t \in [0,1]} \hat{\psi}_{3}(t)  } \leq \EV{ \sup_{t \in [0,1]} \left|\hat{\psi}_{3}(t) \right| }
  \leq \frac{1}{\sqrt{n}} B(K,n, \beta),
\end{align*}
where $B(K,n, \beta)$ is defined in~\eqref{eq:B-definition}.

Putting everything together, we find that, for any $\beta \in \mathbb{R}^{K+1}$,
\begin{align*}
  \EV{ \sup_{t \in [0,1]} \hat{\psi}(t)  } 
  & \leq c(n) \left(|\beta_0|+\frac{\sum_{k=1}^{K} |\beta_k|}{K} \right) + \frac{1}{\sqrt{n}} B(K,n, \beta).
\end{align*}
Therefore,
\begin{align*}
  \EV{ \sup_{t \in [0,1]} \hat{\psi}(t)  } 
  & \leq \inf_{\beta} \left\{ c(n) \left(|\beta_0|+\frac{\sum_{k=1}^{K} |\beta_k|}{K} \right) + \frac{1}{\sqrt{n}} B(K,n,\beta) \right\}
  =: \delta^{\mathrm{FS}}(n).
\end{align*}

\end{proof}

\begin{lemma}
\label{lemma:bound-on-psi3}
    Let $(X_i, \tY_i)_{i=1}^n \sim P$ be an independent and identically distributed random sample. For all $i$, let $A_i(t) = \sum_{k=1}^{K} \sum_{l=1}^K \Omega_{kl} \I{\hat{s}(X_i,k)\leq t, \tY_i=l}$. 
Then,
    \begin{align} \label{eq:bound-on-psi3}  
\begin{split}
        & \EV{\sup_{t \in [0,1]} \left|\frac{1}{n} \sum_{i=1}^n A_i(t) - \EV{A(t)}\right|} \\
        & \qquad \leq \frac{2 \|\Omega\|_1 }{\sqrt{n}} \min \left\{ \sqrt{ 2 \log(2K n + 2)},
          24 \sqrt{\log (K + 2)} + \frac{24}{\sqrt{\log (K + 2)}} \right\},
      \end{split}
    \end{align}
where $\|\Omega\|_1 := \max_{l \in [K]} \sum_{k=1}^{K} |\Omega_{kl}|$. 
\end{lemma}

\begin{proof}[Proof of Lemma~\ref{lemma:bound-on-psi3}]
For ease of notation, let $Z_i=(X_i,\tY_i)$ and $Z_{1}^n=(Z_1,\ldots,Z_n)$. In the following, we denote by $\EV{\cdot}$ the expected value of a random quantity taken with respect to the joint distribution of $Z_1,\ldots,Z_n$.

We study the quantity
\begin{equation}
    \EV{\sup_{t \in [0,1]} \left| \frac{1}{n} \sum_{i=1}^n A_{i}(t) - \EV{A(t)} \right|}
    \label{eq:sup_psi1}
\end{equation}
by using a symmetrization argument which links~\eqref{eq:sup_psi1} with the Rademacher complexity of a properly defined set of vectors. We then get the result by finding a bound for the Rademacher complexity.

First, we define $\mathcal{H}$ as the following set of functions, indexed by $t \in [0,1]$:
\begin{align*}
    \mathcal{H} := \left\{ h_t : \mathcal{Z} \to \mathbb{R},\; h_t(z) = \sum_{k=1}^K \sum_{l=1}^K \Omega_{kl} \I{y=l} \I{\hat{s}(x,k)\leq t}, \; z = (x,y) \in \mathcal{Z} \; : \; t \in [0,1] \right\}.
\end{align*}
For any $h_t \in \mathcal{H}$, we write $h_t(Z)$, $h_t(Z_i)$, or $h_t(z)$ to denote the function $h_t$ evaluated at the random variable $Z$, at the $i$-th data point $Z_i$, or at a fixed realization $z \in \mathcal{Z}$, respectively. In particular, $h_t(Z)$ is itself a scalar random variable. In the following, we sometimes omit the subscript $t$ for ease of notation, writing $h$ for $h_t$, when no ambiguity arises.

Note that
\begin{equation}
    \sup_{t \in [0,1]} \left| \frac{1}{n} \sum_{i=1}^n A_{i}(t) - \EV{A(t)} \right| = \sup_{h \in \mathcal{H}} \left|\frac{1}{n} \sum_{i=1}^n h(Z_i) - \EV{h(Z)} \right|.
    \label{eq:correspondence_of_sup}
\end{equation}
We denote by $\mathcal{H}(Z_1^n)$ the set of vectors in $\mathbb{R}^n$ that can be realized by applying any function $h \in \mathcal{H}$ to the collection $Z_1^n$, i.e., $\left\{ ( h(Z_1),\ldots,h(Z_n))\right\}_{h \in \mathcal{H}}$. 
In addition, we denote by $\mathcal{R}(\mathcal{H}, Z_1^n)$ the Rademacher complexity of $\mathcal{H}(Z_1^n)$. That is,
\begin{equation*}
    \mathcal{R}(\mathcal{H}, Z_1^n) = \mathbb{E}_{\eta_1^n}\left[\sup_{h\in \mathcal{H}} \left|\frac{1}{n} \sum_{i=1}^n \eta_i h(Z_i) \right|\right],
\end{equation*}
where $\eta_1^n = (\eta_1,\ldots,\eta_n)$ is a sequence of independent Rademacher random variables.

Following from the symmetrization lemma (e.g., \cite{wainwright2019high}, p.~106), which we recall in Lemma~\ref{lemma:symmetrization}, it holds that
\begin{equation}
    \EV{\sup_{h \in \mathcal{H}} \left|\frac{1}{n} \sum_{i=1}^n h(Z_i) - \EV{h(Z)} \right|}
    \leq
    \mathbb{E} \left[2 \mathcal{R}(\mathcal{H}, Z_1^n)\right].
    \label{eq:symmetrization}
\end{equation}

We now follow two alternative approaches to bound the Rademacher complexity of $\mathcal{H}(Z_1^n)$. In one approach, the straightforward application of Massart's lemma leads to the first bound. In the second approach, a chaining technique is employed to find a convenient covering number for $\mathcal{H}(Z_1^n)$, which is then used to bound the Rademacher complexity of $\mathcal{H}(Z_1^n)$ via Dudley's entropy integral. The final bound for~\eqref{eq:sup_psi1} is obtained by taking the minimum of the two bounds.

\noindent \textit{Bound via direct use of Massart's lemma.}\\
First, we show that $\mathcal{H}(Z_1^n)$ is bounded and with finite cardinality.

Concerning cardinality, one easily checks that $|\mathcal{H}(Z_1^n)|\leq Kn + 1$. Indeed, it is sufficient to note that a realization $Z_1^n$ is associated to $Kn$ scores $\hat{s}(X_i,k)$, with $i=1,\ldots,n$ and $k=1,\ldots,K$. The ordered scores $\hat{S}_{(1)},\ldots,\hat{S}_{(Kn)}$ partition the real line in $Kn + 1$ intervals. The result on cardinality follows by noting that a component of the vectors in $\mathcal{H}(Z_1^n)$ changes value whenever $t$ shifts between two intervals.

Concerning boundedness, it suffices to observe that
\begin{align*}
    \max_{a \in \mathcal{H}(Z_1^n)} ||a|| 
  &= \max_{j \in [Kn+1]} ||a_j|| 
    = \max_{j \in [Kn+1]} \sqrt{\sum_{i=1}^n a_{ji}^2}\\
    &= \max_{s \in \{-\infty, \hat{S}_{(1)},\dots, \hat{S}_{(Kn)}\}} \sqrt{\sum_{i=1}^n \left( \sum_{k,l} \Omega_{kl} \I{\tilde{Y_i}=l} \I{\hat{s}(X_i, k) \leq s} \right)^2}\\
    &= \max_{s \in \{-\infty, \hat{S}_{(1)},\dots, \hat{S}_{(Kn)}\}} \sqrt{\sum_{i=1}^n \left( \sum_{k} \Omega_{k \tY_i} \I{\hat{s}(X_i, k) \leq s} \right)^2}\\
    &\leq \max_{s \in \{-\infty, \hat{S}_{(1)},\dots, \hat{S}_{(Kn)}\}} \sqrt{\sum_{i=1}^n \left( \sum_{k} \left|\Omega_{k \tY_i} \I{\hat{s}(X_i, k) \leq s} \right| \right)^2}\\
    &\leq \sqrt{\sum_{i=1}^n \left( \sum_{k} \left|\Omega_{k \tY_i} \right| \right)^2} \leq \sqrt{\sum_{i=1}^n \left( \max_{l \in [K]} \sum_k |\Omega_{kl} |\right)^2}\\
    &=  \sqrt{n} \, \|\Omega\|_1.
\end{align*}

Recall that Massart's lemma (\cite{massart2000some}, Lemma 5.2) applies to expected values in the form $\mathbb{E}_{\eta_1^n}\left[\sup_{a \in \mathcal{A}} \sum_{i=1}^n \eta_i a_i  \right]$, for some finite set $\mathcal{A}$, while our definition of $\mathcal{R}(\mathcal{H}, Z_1^n)$ has an absolute value inside the supremum.
To fill this gap and make Massart's lemma applicable in our case, note that
\begin{align*}
    \mathcal{R}(\mathcal{H}, Z_1^n) 
    & = \mathbb{E}_{\eta_1^n}\left[\sup_{h\in \mathcal{H}} \left|\frac{1}{n} \sum_{i=1}^n \eta_i h(Z_i) \right|\right] \\
    & = \mathbb{E}_{\eta_1^n}\left[\sup_{a \in \mathcal{H}(Z_1^n) \cup - \mathcal{H}(Z_1^n) } \left( \frac{1}{n} \sum_{i=1}^n \eta_i a_i \right) \right],
\end{align*}
where $-\mathcal{H}(Z_1^n)$ is the finite subset of $\mathbb{R}^{n}$ obtained by flipping the signs of all elements of $\mathcal{H}(Z_1^n)$.
Since $|\mathcal{H}(Z_1^n) \cup - \mathcal{H}(Z_1^n)| \leq 2 |\mathcal{H}(Z_1^n)|$, and $\max_{a \in \mathcal{H}(Z_1^n) \cup - \mathcal{H}(Z_1^n)} ||a|| = \max_{a \in \mathcal{H}(Z_1^n)} ||a|| $, Massart's lemma gives:
\begin{align} \label{eq:Massart's}
\begin{split}
    \mathcal{R}(\mathcal{H},Z_{1}^{n}) &\leq  \frac{1}{n} \sqrt{n} \, \|\Omega\|_1 \sqrt{2 \log [2(Kn + 1)]}
    = \|\Omega\|_1 \sqrt{\frac{2 \log (2Kn + 2)}{n}}.
  \end{split}
\end{align}
By combining~\eqref{eq:Massart's} with~\eqref{eq:symmetrization} and~\eqref{eq:correspondence_of_sup}, we obtain
\begin{equation}
    \EV{\sup_{t \in [0,1]} \left| \frac{1}{n} \sum_{i=1}^n A_{i}(t) - \EV{A(t)} \right| } \leq 2 \|\Omega\|_1 \sqrt{\frac{2 \log(2Kn + 2)}{n}}.
    \label{eq:Mass-bound-on-psi1}
\end{equation}

\noindent \textit{Bound via chaining.}\\
The bound above depends on $n$ through $\sqrt{\log(Kn)}$. We now derive an alternative
bound, free of $n$, by covering $\mathcal{H}$ at every scale and applying Dudley's entropy
integral conditionally on $Z_1^n$. Note first that $h_t \in \mathcal{H}$ may be expressed as
\begin{equation*}
    h_t(Z) = \sum_{k=1}^K \Omega_{k\tY} \I{\hat{s}(X,k)\leq t}.
\end{equation*}

For any $h,g : \mathcal{Z} \to \mathbb{R}$, define the empirical metric
\begin{equation*}
    d_n(h,g) := \sqrt{\frac{1}{n} \sum_{i=1}^n \left( h(Z_i) - g(Z_i) \right)^2 },
\end{equation*}
and recall that an $\eta$-cover of $\mathcal{H}$ with respect to $d_n$ is a set of functions
$\mathcal{G}$ such that for all $h \in \mathcal{H}$ there exists $g \in \mathcal{G}$ with
$d_n(h,g) \leq \eta$. The $\eta$-covering number
$N(\eta, \mathcal{H}, L_2(P_n))$ is the cardinality of the smallest such cover
(e.g., \cite{wainwright2019high}, Definition 5.1). Note that $d_n$, and hence
$N(\eta, \mathcal{H}, L_2(P_n))$, is a function of the realization $Z_1^n$.
We now compute $N(\eta, \mathcal{H}, L_2(P_n))$.

The class $\mathcal{H}$ is indexed by the single scalar parameter $t \in [0,1]$, and this is
what we discretize. Define the pooled empirical distribution
function of the $Kn$ scores,
\begin{equation*}
    \hat{F}_{\mathrm{pool}}(t) := \frac{1}{Kn} \sum_{i=1}^n \sum_{k=1}^K \I{\hat{s}(X_i,k) \leq t},
    \qquad t \in [0,1].
\end{equation*}
For a fixed integer $m \geq 1$, set $t_0 := 0$ and
\begin{equation*}
    t_j := \inf \left\{ t \in [0,1] : \hat{F}_{\mathrm{pool}}(t) \geq \frac{j}{m} \right\},
    \qquad j = 1,\ldots,m,
\end{equation*}
and consider the finite set of functions
\begin{equation*}
    \mathcal{G}(m) := \left\{ h_{t_0}, h_{t_1}, \ldots, h_{t_m} \right\} \subset \mathcal{H},
    \qquad |\mathcal{G}(m)| \leq m+1 .
\end{equation*}

Now, we show that $\mathcal{G}(m)$ is an $\eta$-cover of $\mathcal{H}$ with respect to $d_n$,
with $\eta = \sqrt{\|\Omega\|_1 \|\Omega\|_{\mathrm{max}} K / m}$ and $\|\Omega\|_{\mathrm{max}} := \max_{k,l \in [K]} |\Omega_{kl}|$. 
Fix any $t \in [0,1]$, let
$j := \max \{ j' : t_{j'} \leq t \}$, and write
$\alpha_{ik} := \I{t_j < \hat{s}(X_i,k) \leq t} \in \{0,1\}$. 
For every $i \in [n]$,
\begin{equation*}
    \left| h_t(Z_i) - h_{t_j}(Z_i) \right|
    = \left| \sum_{k=1}^K \Omega_{k \tY_i} \alpha_{ik} \right|
    \leq \sum_{k=1}^K \left| \Omega_{k \tY_i} \right| \alpha_{ik}
    \leq \sum_{k=1}^K \left| \Omega_{k \tY_i} \right|
    \leq \|\Omega\|_1 .
\end{equation*}
Therefore,
\begin{align} \label{eq:dn}
\begin{split}
    d_n(h_t, h_{t_j})^2
    &= \frac{1}{n} \sum_{i=1}^n \left( h_t(Z_i) - h_{t_j}(Z_i) \right)^2 \\
    &\leq \|\Omega\|_1 \cdot \frac{1}{n} \sum_{i=1}^n \sum_{k=1}^K
        \left| \Omega_{k \tY_i} \right| \alpha_{ik} \\
    &\leq \|\Omega\|_1 \|\Omega\|_{\mathrm{max}} \cdot \frac{1}{n} \sum_{i=1}^n \sum_{k=1}^K \alpha_{ik}.
  \end{split}
\end{align}
By construction of the grid,
\begin{align} \label{eq:grid-mass}
\begin{split}
  \frac{1}{n} \sum_{i=1}^n \sum_{k=1}^K \alpha_{ik}
  & = \frac{1}{n} \sum_{i=1}^n \sum_{k=1}^K \left( \I{\hat{s}(X_i,k) \leq t} - \I{\hat{s}(X_i,k) \leq t_j}\right)  \\
  &  = K \left[ \hat{F}_{\mathrm{pool}}(t) - \hat{F}_{\mathrm{pool}}(t_j) \right] 
   \leq K \left[ \frac{j+1}{m} - \frac{j}{m} \right] 
  = \frac{K}{m},
\end{split}
\end{align}    
where the inequality holds because $\hat{F}_{\mathrm{pool}}(t) < (j+1)/m$ by maximality of $j$. 

Combining~\eqref{eq:dn} and~\eqref{eq:grid-mass} gives
\begin{align*}
    d_n(h_t, h_{t_j})^2
    & \leq \frac{\|\Omega\|_1 \|\Omega\|_{\mathrm{max}} K}{m} = \eta^2.
\end{align*}
Note that
all quantities above are evaluated at the realized sample, so this establishes an $\eta$-cover
in the empirical metric $d_n$, as required.
Writing $C := \sqrt{\|\Omega\|_1 \|\Omega\|_{\mathrm{max}} K}$ and choosing $m = \lceil C^2 / \eta^2 \rceil$,
it follows that, for every $\eta > 0$,
\begin{equation}
    N(\eta, \mathcal{H}, L_2(P_n)) \leq |\mathcal{G}(m)| \leq m + 1 \leq \frac{C^2}{\eta^2} + 2 .
    \label{eq:covering-number}
\end{equation}

Next, we apply Dudley's entropy integral (e.g., \cite{wainwright2019high}, Eq. (5.48)), conditionally on $Z_1^n$.
To determine the upper bound of integration, we need to upper bound the diameter of $\mathcal{H}$ in the empirical norm:
\begin{equation*}
  D_n(\mathcal{H}) := \sup_{h,g \in \mathcal{H}} d_n(h,g)
  = \sup_{t,t' \in [0,1]} \sqrt{\frac{1}{n} \sum_{i=1}^n \left( h_t(Z_i) - h_{t'}(Z_i) \right)^2 }.
\end{equation*}
For any $t, t' \in [0,1]$ with $t' \leq t$, and any
$i \in [n]$,
\begin{equation*}
  \left| h_t(Z_i) - h_{t'}(Z_i) \right|
  = \left| \sum_{k=1}^K \Omega_{k \tY_i}
      \left( \I{\hat{s}(X_i,k) \leq t} - \I{\hat{s}(X_i,k) \leq t'} \right) \right|
  \leq \sum_{k=1}^K \left| \Omega_{k \tY_i} \right|
  \leq \|\Omega\|_1 ,
\end{equation*}
since each difference of indicators lies in $\{0,1\}$. As the same bound holds when
$t \leq t'$, by symmetry, we obtain
\begin{equation}
  D_n(\mathcal{H})
  = \sup_{t,t' \in [0,1]} \sqrt{\frac{1}{n} \sum_{i=1}^n
    \left( h_t(Z_i) - h_{t'}(Z_i) \right)^2 }
  \leq \|\Omega\|_1.
  \label{eq:diameter}
\end{equation}

Therefore, Dudley's entropy integral (\cite{wainwright2019high}, Eq. (5.48)) gives:
\begin{align*}
    \mathcal{R}(\mathcal{H}, Z_1^n)
    &\leq \frac{24}{\sqrt{n}} \int_{0}^{D_n(\mathcal{H})}
        \sqrt{\log N(\eta, \mathcal{H}, L_2(P_n))} \, d\eta \\
    &\leq \frac{24}{\sqrt{n}} \int_{0}^{\|\Omega\|_1} \sqrt{\log \left( \frac{C^2}{\eta^2} + 2\right)} \, d\eta.
\end{align*}
By solving the integral (as shown below), one finds that
\begin{equation}
    \mathcal{R}(\mathcal{H}, Z_1^n) \leq \frac{24}{\sqrt{n}} \, \|\Omega\|_1
    \left[ \sqrt{\log (K + 2)} + \frac{1}{\sqrt{\log (K + 2)}} \, \right],
    \label{eq:Chaining-bound-on-Rademacher}
\end{equation}
uniformly in $Z_1^n$, from which it directly follows that
\begin{equation}
    \EV{\sup_{t \in [0,1]} \left| \frac{1}{n} \sum_{i=1}^n A_{i}(t) - \EV{A(t)} \right| }
    \leq \frac{48}{\sqrt{n}} \, \|\Omega\|_1
    \left[ \sqrt{\log (K + 2)} + \frac{1}{\sqrt{\log (K + 2)}} \, \right],
    \label{eq:Chaining-bound-on-psi1}
\end{equation}
where the factor $2$ comes from the symmetrization
bound~\eqref{eq:symmetrization} and~\eqref{eq:correspondence_of_sup}.

For the sake of completeness, we here report the calculations that lead
to~\eqref{eq:Chaining-bound-on-Rademacher}. Since $\eta \leq \|\Omega\|_1$ on the domain of
integration, and recalling $C^2 = \|\Omega\|_1 \|\Omega\|_{\mathrm{max}} K$,
\begin{equation*}
    \frac{C^2}{\eta^2} + 2
    = \frac{C^2}{\eta^2} \left( 1 + \frac{2 \eta^2}{C^2} \right)
    \leq \frac{C^2}{\eta^2} \left( 1 + \frac{2 \|\Omega\|_1^2}{C^2} \right)
    = \frac{B^2}{\eta^2},
    \qquad B := C \sqrt{1 + \frac{2 \|\Omega\|_1}{K \|\Omega\|_{\mathrm{max}}}}.
\end{equation*}
Therefore, defining $u_0 := \log ( B / \|\Omega\|_1 )$,
\begin{align*}
    \int_{0}^{\|\Omega\|_1} \sqrt{\log \left( \frac{C^2}{\eta^2} + 2 \right)} \, d\eta
    &\leq \int_{0}^{\|\Omega\|_1} \sqrt{2 \log \left( \frac{B}{\eta} \right)} \, d\eta \\
    &= \sqrt{2} \, B \int_{-\infty}^{-u_0} \sqrt{-u} \; e^{u} \, du \\
    &= \sqrt{2} \, B \int_{u_0}^{\infty} \sqrt{v} \, e^{-v} \, dv \\
    &= \sqrt{2} \, B \, \Gamma\!\left(\frac{3}{2}, u_0\right) \\
    &\leq \sqrt{2} \, B \cdot \sqrt{u_0} \, e^{-u_0} \left( 1 + \frac{1}{2 u_0} \right) \\
    &= \sqrt{2} \, \|\Omega\|_1 \sqrt{u_0} \left( 1 + \frac{1}{2 u_0} \right)
    = \|\Omega\|_1 \left[ \sqrt{2 u_0} + \frac{1}{\sqrt{2 u_0}} \right] ,
\end{align*}
where the first equality follows from the change of variables $u = \log(\eta/B)$, the second
from $v = -u$, the inequality from a known upper bound for the incomplete Gamma function
\citep{natalini2000inequalities}, and the last line from $e^{-u_0} = \|\Omega\|_1/B$. 

It remains to bound $u_0$. By definition of $B$ and $C$,
\begin{equation*}
    2 u_0 = \log \left( \frac{B^2}{\|\Omega\|_1^2} \right)
    = \log \left( \frac{K \|\Omega\|_{\mathrm{max}}}{\|\Omega\|_1} + 2 \right)
    \in \left[ \log 3, \; \log (K+2) \right],
\end{equation*}
where the two bounds follow from $\|\Omega\|_{\mathrm{max}} \leq \|\Omega\|_1 \leq K \|\Omega\|_{\mathrm{max}}$. In
particular $2 u_0 \geq \log 3 > 1$, so that the above bound on $\Gamma(3/2, u_0)$ is applied
away from its singularity at the origin, and the map $x \mapsto \sqrt{x} + 1/\sqrt{x}$, which
is increasing for $x \geq 1$, may be evaluated at the upper end $2u_0 \leq \log (K+2)$.
Therefore,
\begin{align*}
    \int_{0}^{\|\Omega\|_1} \sqrt{\log \left( \frac{C^2}{\eta^2} + 2 \right)} \, d\eta
    &\leq \|\Omega\|_1 \left[ \sqrt{\log (K+2)} + \frac{1}{\sqrt{\log (K+2)}} \right].
\end{align*}

\end{proof}

\begin{lemma}\label{lemma:symmetrization}
Let $Z_1^n=(Z_1,\ldots,Z_n)$ be a collection of independent and identically distributed random samples. Then,
\begin{equation*}
    \mathbb{E}_{Z_1^n}\left[\sup_{h \in \mathcal{H}} \left|\frac{1}{n} \sum_{i=1}^n h(Z_i) - \EV{h(Z)} \right|\right] \leq \mathbb{E}_{Z_1^n} \left[2 \mathcal{R}(\mathcal{H}, Z_1^n)\right],
\end{equation*}
where $\mathcal{H}$ is the set defined in the proof of Lemma~\ref{lemma:bound-on-psi3} and $\mathcal{R}(\mathcal{H},Z_1^n)$ denotes the Rademacher complexity of $\mathcal{H}(Z_1^n)$.
\end{lemma}

\begin{proof}[Proof of Lemma~\ref{lemma:symmetrization}]

Let $Z'_1,\ldots,Z'_n$ be a second i.i.d. sequence, distributed as $Z_1,\ldots,Z_n$ and independent of $Z_1,\ldots,Z_n$.
Let $\eta_1^n = (\eta_1, \ldots, \eta_n)$ be an i.i.d sequence of Rademacher variables, independent of $Z_1,\ldots,Z_n$ and $Z'_1,\ldots,Z'_n$.  This implies that the variables $h(Z_i) - h(Z'_i)$ are independent, symmetric and distributed as $\eta_i(h(Z_i) - h(Z'_i))$.
Then,
\begin{align*}
    \mathbb{E}_{Z_1^n} \left[ \sup_{h \in \mathcal{H}} \left| \frac{1}{n} \sum_{i=1}^{n} h(Z_i) - \mathbb{E}_{Z}[h(Z)]\right| \right] &= \mathbb{E}_{Z_1^n} \left[ \sup_{h \in \mathcal{H}} \left| \frac{1}{n} \sum_{i=1}^{n} h(Z_i) - \mathbb{E}_{Z_1}[h(Z')]\right| \right]\\
    &= \mathbb{E}_{Z_1^n} \left[ \sup_{h \in \mathcal{H}} \left| \mathbb{E}_{Z'^n_1}\!\left[\frac{1}{n} \sum_{i=1}^{n} \left(h(Z_i) - h(Z'_i)\right)\right]\right| \right]\\
    &\leq \mathbb{E}_{Z_1^n, Z'^n_1} \left[ \sup_{h \in \mathcal{H}} \left| \frac{1}{n} \sum_{i=1}^{n} (h(Z_i) - h(Z'_i))\right| \right]\\
    &= \mathbb{E}_{Z_1^n, Z'^n_1, \eta_1^n} \left[ \sup_{h \in \mathcal{H}} \left| \frac{1}{n} \sum_{i=1}^{n} \eta_i \left(h(Z_i) - h(Z'_i)\right)\right| \right]\\
    &\leq 2 \mathbb{E}_{Z_1^n, \eta_1^n} \left[ \sup_{h \in \mathcal{H}} \left| \frac{1}{n} \sum_{i=1}^{n} \eta_i h(Z_i)\right| \right]\\
    &= \mathbb{E}_{Z_1^n}\left[2 \mathbb{E}_{\eta_1^n} \left[ \sup_{h \in \mathcal{H}} \left| \frac{1}{n} \sum_{i=1}^{n} \eta_i h(Z_i)\right| \right]\right]\\
    &= \mathbb{E}_{Z_1^n}\left[2 \mathcal{R}(\mathcal{H}, Z_1^n)\right].
\end{align*}
Above, the first inequality follows from the application of Jensen's inequality.
\end{proof}

\begin{lemma}\label{lemma:bound-on-c}
Let $U_1,\ldots, U_n$ be i.i.d uniform random variables on $[0,1]$, and denote their order statistics as $U_{(1)},\ldots, U_{(n)}$. Then
\begin{equation}
    \EV{\sup_{i \in [n]} \left\{ \frac{i}{n} - U_{(i)}\right\}} \leq \sqrt{\frac{\pi}{8n}}.
    \label{eq:bound-on-c}
\end{equation}
\end{lemma}

\begin{proof}[Proof of Lemma~\ref{lemma:bound-on-c}]

Let $\hat{F}(t)$ denote the empirical distribution function based on\\
$U_1,\dots,U_n$. Then, the following pathwise identity holds
\begin{equation*}
    \underset{i \in [n]}{\sup} \left\{ \frac{i}{n} - U_{(i)} \right\} {=} \underset{t \in [0,1]}{\sup} \left( \hat{F}(t) - t \right).
\end{equation*}
This implies that
\begin{align*}
  \EV{\sup_{i \in [n]} \left\{ \frac{i}{n} - U_{(i)}\right\}}
  & = \EV{ \sup_{t \in [0,1]} \left( \hat{F}(t) - t \right) }  \\
  & = \int_{0}^{\infty} \P{ \sup_{t \in [0,1]} \left( \hat{F}(t) - t \right) > z} dz \\
  & \leq \int_{0}^{\infty} e^{ -2 n z^2 }  dz \\
  & = \sqrt{\frac{\pi}{8n}}.
\end{align*}
Above, the inequality follows from the DKW inequality with Massart's sharp constant.
\end{proof}

\subsection{Asymptotic Approximation of the Correction Factor}

\begin{proof}[Proof of Theorem~\ref{thm:asymptotic-factor}]
Consider the definition of the empirical process $\hat{\psi}(t)$:
\begin{align*}
    \hat{\psi}(t) &= \sum_{k=1}^{K} \sum_{l=1}^{K} W_{kl} \left[ \hat{\rho}_l \hat{F}_l^{k}(t) - \tilde{\rho}_l \tilde{F}_l^{k}(t) \right]\\
    &=  \frac{1}{n} \sum_{i=1}^{n} \sum_{k=1}^{K} \sum_{l=1}^{K} W_{kl} \left[ \I{\hat{s}(X_i, k) \leq t, \tY_i = l} - \tilde{\rho}_l \tilde{F}_l^{k}(t) \right].
\end{align*}
Note that $\hat{\psi}(t)$ is an empirical process with mean 0. The convergence of $\hat{\psi}$ to a Gaussian limit can be established via Donsker's theorem; e.g., see \citet[Ch.~19]{van2000asymptotic}.

Let $\mathcal{F}$ be the following class of functions of $Z=(X,\tilde{Y})$, indexed by $t \in [0,1]$,
\begin{align}
    \mathcal{F} = \left\{ \sum_{k=1}^K \sum_{l=1}^K W_{kl} \I{\hat{s}(X, k) \leq t, \tY = l} : t \in [0,1] \right\}.
\end{align}
Let $f_t$ be the element of $\mathcal{F}$ corresponding to $t$, and note that it may be expressed as
\begin{align*}
    f_t(Z) = \sum_{k=1}^K W_{k\tY} \I{\hat{s}(X,k) \leq t}.
\end{align*}
Now let $Z_i=(X_i, \tY_i)$, $i=1,\dots,n$, be independent and identically distributed realizations from some unknown distribution $P$. Now denote
\begin{align*}
    & \mathbb{P}_n f_t = \frac{1}{n} \sum_{i=1}^n f_t(Z_i),
    & P f_t = \EV{\mathbb{P}_n f_t}.
\end{align*}
Note that $\hat{\psi}(t) = \mathbb{P}_n f_t - P f_t$.

Let us denote $L_2(P)$ the set of measurable functions whose second powers are $P$-integrable, i.e., $\EV{\frac{1}{n} \sum_{i=1}^n g(Z_i)^2} < \infty.$
Given two functions $l$ and $u$, the bracket $[l, u]$ is the set of all functions $g$ with  $l \leq g \leq u$. An $\eta$-bracket in $L_2(P)$ is a bracket $[l, u]$ with
\begin{align*}
    \EV{\frac{1}{n} \sum_{i=1}^n \left(u(Z_i)-l(Z_i)\right)^2} &< \eta^2.
\end{align*}

The bracketing number $N_{[\,]}(\eta, \mathcal{F}, L_2(P))$ is the minimum number of $\eta$-brackets needed to cover $\mathcal{F}$; e.g., see \citet[p.~270]{van2000asymptotic}. The bracketing integral \citep[p.~270]{van2000asymptotic} is defined starting from the bracketing number as
\begin{align*}
    J(\delta, \mathcal{F}, L_2(P)) := \int_{0}^{\delta} \sqrt{\log N_{[\,]}(\eta, \mathcal{F}, L_2(P)) } d \eta.
\end{align*}
We now show that this integral is finite.

For $k,l \in [K]$, let $W^+_{kl}, W^-_{kl} \geq 0$ denoting the positive and negative parts of $W_{kl}$, respectively, so that $W_{kl} = W^+_{kl} - W^-_{kl}$ and $|W_{kl}| = W^+_{kl} + W^-_{kl}$.
Then, for any $f_t \in \mathcal{F}$,
\begin{align*}
  f_t(Z) 
  &= \sum_{k=1}^K W_{k \tY} \I{\hat{s}(X, k) \leq t}\\
  &= \sum_{k=1}^K W^+_{k \tY} \I{\hat{s}(X, k) \leq t} - \sum_{k=1}^K W^-_{k \tY} \I{\hat{s}(X, k) \leq t},
\end{align*}
For $\epsilon > 0$, define the following set of functions
\begin{align*}
    \mathcal{M}(\epsilon) = \bigg\{ & m(Z) = \sum_{k=1}^K W^+_{k \tY} \mathbb{I}[\hat{s}(X,k) \leq c_{k j_k}] - \sum_{k=1}^K W^-_{k \tY} \mathbb{I}[\hat{s}(X,k) \leq c_{k j'_k}], \\
  & \; j_k,j'_k = 0,1,\dots,d, \; d:= \left\lceil K^2/\epsilon^2 \right\rceil \bigg\},
\end{align*}
where $c_{k j_k} := \left(\tF^k\right)^{-1}(\frac{j_k}{d})$ and $c_{k j'_k} := \left(\tF^k\right)^{-1}(\frac{j'_k}{d})$ for all $k \in [K]$. 
These quantile functions are well defined; in fact, $\tilde{F}^k$ is continuous, as discussed in the proof of Theorem~\ref{thm:finite-sample-factor-simplified}, and strictly increasing by
Assumption~\ref{ass:score-cdf-strictly-increasing}, hence it is a bijection from $[0,1]$ onto
$[0,1]$. In particular $\left(\tilde{F}^k\right)^{-1}(0) = 0$ and
$\left(\tilde{F}^k\right)^{-1}(1) = 1$, since $\hat{s}$ takes values in $[0,1]$, and
\begin{equation}\label{eq:quantile-grid}
    \tilde{F}^k\!\left(c_{k j_k}\right) = \frac{j_k}{d}, \qquad j_k = 0, 1, \ldots, d, \quad k \in [K].
\end{equation}

We now show that, for each element $f_t \in \mathcal{F}$, there exist $u_t, l_t \in \mathcal{M}(\epsilon)$ such that $[l_t,u_t]$ is a $\eta$-bracket containing $f_t$, for a suitable value of $\eta$.

Specifically, consider the two functions $l_t, u_t \in \mathcal{M}(\epsilon)$ defined as
\begin{align*}
    l_t(Z) &= \sum_{k=1}^K W^+_{k\tY} \I{\hat{s}(X, k) \leq c_{k (j_k-1)}} - \sum_{k=1}^K W^-_{k\tY} \I{\hat{s}(X, k) \leq c_{k j_k}},\\
    u_t(Z) & = \sum_{k=1}^K W^+_{k\tY} \I{\hat{s}(X, k) \leq c_{k j_k}} - \sum_{k=1}^K W^-_{k\tY} \I{\hat{s}(X, k) \leq c_{k (j_k-1)}},
\end{align*}
where $j_k := \min\{j \geq 1: t \leq c_{kj}\}$ for each $k \in [K]$.
It is clear that $l_t \leq f_t \leq u_t$.
Next, we now show that $[l_t,u_t]$ is an $\eta$-bracket in $L_2(P)$, with $\eta = \epsilon \max_{k,l} |W_{kl}|$.

Indeed,
\begin{align*}
  u_t(Z)-l_t(Z)
  & = \sum_{k=1}^K |W_{k\tY}| \left( \I{\hat{s}(X, k) \leq c_{k j_k}} - \I{\hat{s}(X, k) \leq c_{k (j_k-1)}} \right) \\
  & = \sum_{k=1}^K |W_{k\tY}| H_k(X),
\end{align*}
where $H_k(X) := \I{ c_{k (j_k-1)} < \hat{s}(X, k) \leq c_{k j_k}}$.
This implies
\begin{align*}
  \left(u_t(Z)-l_t(Z)\right)^2
    & = \sum_{k,k'=1}^K |W_{k\tY}||W_{k'\tY}| H_k(X) H_{k'}(X) \\
    & \leq \max_{k,l \in [K]} |W_{kl}|^2 \sum_{k,k'=1}^K H_k(X) H_{k'}(X)  \\
    & \leq \max_{k,l \in [K]} |W_{kl}|^2 \sum_{k,k'=1}^K H_k(X).
\end{align*}
Therefore,
\begin{align*}
    \EV{ \frac{1}{n} \sum_{i=1}^n \left(u_t(Z_i)-l_t(Z_i)\right)^2 }
    & \leq \max_{k,l \in [K]} |W_{kl}|^2 K \sum_{k=1}^K \left[ \tF^k(c_{k j_k}) - \tF^k(c_{k (j_k - 1)}) \right] \\
    & = \max_{k,l \in [K]} |W_{kl}|^2  \frac{K^2}{d} \\
    & \leq \max_{k,l \in [K]}|W_{kl}|^2 \epsilon^2 = \eta^2.
\end{align*}
Note that the second equality follows from~\eqref{eq:quantile-grid}, which gives $\tilde{F}^k(c_{k_{j_k}}) - \tilde{F}^k(c_{k_{(j_k-1)}}) = 1/d$ for every $k \in [K]$.

Then, $N_{[\,]}(\eta, \mathcal{F}, L_2(P)) \leq |\mathcal{M}(\epsilon)|$.
Note that each function $m$ in $\mathcal{M}(\epsilon)$ is identified by a pair of $K$-dimensional vectors each in the form $c = (c_{1 j_1},\ldots, c_{K j_K})$. Since each of their components can take $d+1$ values, $|\mathcal{M}(\epsilon)| = (d+1)^{2K} = \left( \lceil K^2 / \epsilon^2 \rceil + 1 \right)^{2K} \leq \left( K/\epsilon + 2 \right)^{4K}$.

It is then possible to find an upper bound for the bracketing number of $\mathcal{F}$ with $\delta=1$, which reads
\begin{align*}
  J(1, \mathcal{F}, L_2(P)) 
  & = \int_{0}^{1} \sqrt{\log N_{[\,]}(\eta, \mathcal{F}, L_2(P)) } d \eta \\
  & \leq \int_{0}^{1} \sqrt{4K \log \left( K \max_{k,l} |W_{kl}| /\eta + 2 \right) }  d \eta \\
  & = \max_{k,l} |W_{kl}| \int_{0}^{1/\max_{k,l} |W_{kl}|} \sqrt{4K \log \left( K/\epsilon + 2 \right) }  d \epsilon \\
  & \leq \max_{k,l} |W_{kl}| \int_{0}^{1} \sqrt{4K \log \left( K/\epsilon + 2 \right) }  d \epsilon,
\end{align*}
where the last inequality follows from $1/\max_{k,l} |W_{kl}| \leq 1$, since $W$ is the inverse of a stochastic transition matrix.
The fact that this integral is finite can be proven by repeating similar calculations as those shown at the end of the proof of Lemma~\ref{lemma:bound-on-psi3}.

As $J(1, \mathcal{F}, L_2(P))$ is finite, it follows from Donsker's theorem (e.g., \citet{van2000asymptotic}, Ch. 19, Theorem 19.4) that $\mathcal{F}$ is a $P$-Donsker class of functions. Consequently, $\sqrt{n}\left(\mathbb{P}_n f_t - P f_t \right)$ converges in distribution to a centered Gaussian process with covariance function

\begin{equation*}
G(t_1,t_2) = \EV{f_{t_1} f_{t_2}} - \EV{f_{t_1}}\EV{f_{t_2}}, \; \forall t_1,t_2 \in [0,1].
\end{equation*}
Let $\mathrm{GBB}$ denote the limiting Gaussian process. Then, for $n$ that tends to infinity,
\begin{align*}
    \sqrt{n} \cdot \hat{\psi} \rightsquigarrow \mathrm{GBB},
\end{align*}
which also implies that, for all $t \in [0,1]$
\begin{align*}
    \sqrt{n} \cdot \hat{\psi}(t) \overset{d}{\to} \mathrm{GBB}(t).
\end{align*}

\end{proof}

\subsection{Theoretical Upper Bound on Coverage}

\begin{lemma}\label{lemma:mcdiarmid-psi}
Under the assumptions of Theorem~\ref{thm:algorithm-lower-bound}, let $\hat{\psi}$ be the empirical process in~\eqref{eq:hat-psi}, $\delta^{**}(n) := \EV{\sup_{t \in [0,1]} |\hat{\psi}(t)|}$, and $\|W\|_1 := \max_{l \in [K]} \sum_{k=1}^{K} |W_{kl}|$.
Then, for any $u \geq 0$,
\begin{align*}
    \P{ \sup_{t \in [0,1]} \left| \hat{\psi}(t) \right| \geq \delta^{**}(n) + u } \leq \exp\left( - \frac{n u^2}{2 \|W\|_1^2} \right).
\end{align*}
\end{lemma}

\begin{proof}[Proof of Lemma~\ref{lemma:mcdiarmid-psi}]
Write $Z_i = (X_i, \tY_i)$ and, for $t \in [0,1]$,
\[
  f_t(Z) := \sum_{k=1}^{K} W_{k \tY} \I{\hat{s}(X,k) \leq t},
\]
so that
$\hat{\psi}(t) = \frac{1}{n} \sum_{i=1}^{n} f_t(Z_i) - \EV{f_t(Z)}$, as in the proof of Theorem~\ref{thm:asymptotic-factor}.

Since $|f_t(Z)| \leq \sum_{k=1}^{K} |W_{k \tY}| \leq \|W\|_1$ for all $t \in [0,1]$ and all $Z$, replacing a single $Z_i$ by $Z_i'$ changes $\hat{\psi}(t)$ by at most $2\|W\|_1/n$, uniformly in $t$, and hence changes $g(Z_1, \ldots, Z_n) := \sup_{t \in [0,1]} |\hat{\psi}(t)|$ by at most $2\|W\|_1/n$.


Note also that the supremum is measurable. Indeed, $\hat{\psi}$ is the difference between a right-continuous step function, with at most $Kn$ jumps, and $F$, which is also right-continuous. Its supremum over $[0,1]$ therefore coincides with its supremum over a countable dense subset of $[0,1]$
containing the endpoint $1$, and is measurable as a countable supremum of
measurable functions.

Then, the one-sided version of the bounded differences inequality \citep[Corollary 2.21]{wainwright2019high} with $c_i = 2\|W\|_1/n$ gives
\[
  \P{g - \EV{g} \geq u} \leq \exp\left(-\frac{2u^2}{\sum_{i=1}^{n} c_i^2} \right) = \exp\left(-\frac{n u^2}{2\|W\|_1^2}\right),
\]
and $\EV{g} = \delta^{**}(n)$ by definition.
\end{proof}

\begin{proof}[Proof of Theorem~\ref{thm:algorithm-upper-bound}]
Define the events $\mathcal{A}_1=\{\hat{\mathcal{I}}=\emptyset \}$ and $\mathcal{A}_2=\{\hat{i}=1\}$.
Then,
\begin{align}
  \begin{split} 	\label{eq:upper_bound_events-marg}
	\P{Y_{n+1} \in \hat{C}(X_{n+1})}
    & \leq \EV{\P{Y_{n+1} \in \hat{C}(X_{n+1}) \mid \cD}\I{\cA_1}}\\
    & \qquad + \EV{\P{Y_{n+1} \in \hat{C}(X_{n+1}) \mid \cD}\I{\cA_2}} \\
	& \qquad + \EV{\P{Y_{n+1} \in \hat{C}(X_{n+1}) \mid \cD}\I{\mathcal{A}_1^c\cap \mathcal{A}_2^c}}\\
	& = \P{\cA_1} + \EV{\P{Y_{n+1} \in \hat{C}(X_{n+1}) \mid \cD}\I{\cA_2}} \\
	& \qquad + \EV{\P{Y_{n+1} \in \hat{C}(X_{n+1}) \mid \cD}\I{\mathcal{A}_1^c\cap \mathcal{A}_2^c}}.
      \end{split}
\end{align}

We will now separately bound the three terms on the right-hand-side of~\eqref{eq:upper_bound_events-marg}.
The following notation will be useful for this purpose.
For all $i \in [n]$, let $U_i \sim \text{Uniform}(0,1)$ be independent and identically distributed uniform random variables, and
denote their order statistics as $U_{(1)} < U_{(2)} < \ldots < U_{(n)}$.

Note that, by Assumption~\ref{assumption:regularity-dist}, $\tilde{F}$ is continuous and strictly increasing, so we may take $U_{(i)} := \tilde{F}(S_{(i)})$. Moreover, since $F$ and $\tF$ are Lipschitz with constants $f_{\max}$ and $\tf_{\max}$, and $\tilde{F}^{-1}$ is Lipschitz with constant $1/\tf_{\min}$, the map $\Delta \circ \tilde{F}^{-1}$ is Lipschitz on $[0,1]$ with constant $L = (f_{\max} + \tf_{\max})/\tf_{\min}$. Since $\Delta(0) = \Delta(1) = 0$ (the scores have continuous distributions on $[0,1]$) and $\tilde{F}^{-1}(0)=0$, $\tilde{F}^{-1}(1)=1$, it follows that, for all $i \in [n]$,
\begin{align}\label{eq:delta-lipschitz-endpoints}
  - L \left( 1 - U_{(i)} \right) \leq \Delta(S_{(i)}) \leq L \, U_{(i)}.
\end{align}

\begin{itemize}
\item We show below that the probability of the event $\cA_1$ can be bounded from above as:
  \begin{align} \label{eq:upper_bound_event1-marg}
    \P{\hat{\mathcal{I}} = \emptyset} \leq \frac{3}{n}.
  \end{align}

  First, note that
  \begin{align}
      \P{\hat{\cI} \neq \emptyset} \geq \P{n \in \hat{\cI}} = \P{1 - \alpha - \hDelta(S_{(n)}) + \delta(n) \leq 1}.
  \end{align}
  We show now that
  \begin{align}
     \P{1 - \alpha - \hDelta(S_{(n)}) + \delta(n) \leq 1} \geq 1 - \frac{3}{n}.
  \end{align}
  Indeed, for any $i \in [n]$,
  \begin{align*}
    & 1 - \alpha - \hDelta(S_{(i)}) + \delta(n) \\
    & \qquad = 1 - \alpha - \Delta(S_{(i)}) + \delta(n) + \Delta(S_{(i)}) - \hDelta(S_{(i)}) \\
    & \qquad = 1 - \alpha - \Delta(S_{(i)}) + \delta(n) \\
    & \qquad + \sum_{k=1}^K \sum_{l=1}^K W_{kl} \left( \tilde{\rho}_l \tilde{F}_l^{k}(S_{(i)}) - \hat{\rho}_l \hat{F}_l^{k}(S_{(i)}) \right) - \left(\tilde{F}(S_{(i)}) - \hat{F}(S_{(i)})\right)\\
    & \qquad \leq 1 - \alpha - \Delta(S_{(i)}) + \delta(n) \\
    & \qquad + \underset{t \in [0,1]}{\sup} \left|\sum_{k=1}^K \sum_{l=1}^K W_{kl} \left( \hat{\rho}_l \hat{F}_l^{k}(t) - \tilde{\rho}_l \tilde{F}_l^{k}(t)\right)\right| +
    \underset{t \in [0,1]}{\sup} \left| \hat{F}(t) - \tilde{F}(t)\right|.
  \end{align*}

By applying the Dvoretzky–Kiefer–Wolfowitz inequality on the last term above, we have that, with probability at most $1/n$,
\begin{align*}
    \underset{t \in [0,1]}{\sup} \left| \hat{F}(t) - \tilde{F}(t)\right| > \sqrt{\frac{\log (2 n)}{2 n}}.
\end{align*}
Further, by Lemma~\ref{lemma:mcdiarmid-psi} applied with $u = \|W\|_1 \sqrt{2 \log(n)/n}$, 
\begin{align*}
    \underset{t \in [0,1]}{\sup} \left|\sum_{k=1}^K \sum_{l=1}^K W_{kl} \left( \hat{\rho}_l \hat{F}_l^{k}(t) - \tilde{\rho}_l \tilde{F}_l^{k}(t)\right)\right| > \delta^{**}(n) + \|W\|_1 \sqrt{\frac{2 \log n}{n}},
\end{align*}
with probability at most $1/n$.
Finally, focusing on $i=n$, by~\eqref{eq:delta-lipschitz-endpoints}, $-\Delta(S_{(n)}) \leq L (1 - U_{(n)})$, and since $\P{1 - U_{(n)} > \log(n)/n} = (1 - \log(n)/n)^n \leq 1/n$, with probability at most $1/n$, $-\Delta(S_{(n)}) > L \log(n)/n$.

Combining these three results, we have that, with probability at least $1 - \frac{3}{n}$,
  \begin{align*}
     & 1 - \alpha - \hDelta(S_{(n)}) + \delta(n) \\
    & \qquad \leq 1 - \alpha + \delta(n) + \sqrt{\frac{\log (2 n)}{2 n}} + \delta^{**}(n) + \|W\|_1 \sqrt{\frac{2 \log n}{n}} + L \frac{\log n}{n}.
  \end{align*}

From Theorem~\ref{thm:upper-bound-on-abs-psi} and the definition of $d(n)$ in Assumption~\ref{assumption:regularity-dist-delta-marg}, we have that
  \begin{align*}
     1 - \alpha - \hDelta(S_{(n)}) + \delta(n)
    \leq 1 - \alpha + \delta(n) + \sqrt{\frac{\log (2 n)}{2 n}} + d(n) + L \frac{\log n}{n}.
  \end{align*}

  Then, it follows from Assumption~\ref{assumption:regularity-dist-delta-marg} that, with probability at least $1 - \frac{3}{n}$,
    \begin{align*}
    & 1 - \alpha - \hDelta(S_{(n)}) + \delta(n) \leq 1.
  \end{align*}

\item Under $\cA_2$, we have $\htau = S_{(1)}$, and hence, by~\eqref{eq:delta-lipschitz-endpoints},
  \begin{align*}
    \P{Y_{n+1} \in \hat{C}(X_{n+1}) \mid \cD}
    = F(S_{(1)})
    = \tilde{F}(S_{(1)}) + \Delta(S_{(1)})
    \leq (1 + L) \, U_{(1)}.
  \end{align*}
  Therefore,
  \begin{align}
    \EV{\P{Y_{n+1} \in \hat{C}(X_{n+1}) \mid \cD}\I{\cA_2}}
    \leq (1+L) \, \EV{U_{(1)}} = \frac{1+L}{n+1}.
    \label{eq:upper_bound_event2-marg}
  \end{align}

\item Under $\cA_1^c\cap \cA_2^c$, by definition of $\hat{\mathcal{I}}$, for any $i \leq \hat{i}  - 1$,
  \begin{align*}
    \frac{i}{n} < 1 - \left( \alpha + \hDelta(S_{(i)}) - \delta(n) \right).
  \end{align*}
  Therefore, choosing $i = \hat{i}-1$, we get:
 \begin{align*}
   \frac{\hat{i}}{n} < 1 - \left( \alpha + \hDelta(S_{(\hat{i}-1)}) - \delta(n) \right) + \frac{1}{n}.
 \end{align*}

As in the proof of Theorem~\ref{thm:algorithm-lower-bound}, the probability of coverage conditional on the labeled data in $\mathcal{D}$ can be written as:

  \begin{align*}
    & \P{Y_{n+1} \in \hat{C}(X_{n+1}) \mid \mathcal{D}} \\
    & \qquad = \tilde{F}(S_{(\hat{i})}) + \Delta(S_{(\hat{i} )}) \\
    & \qquad = \hat{F}(S_{(\hat{i})}) + \hDelta(S_{(\hat{i} )}) + \sum_{k=1}^K \sum_{l=1}^K W_{kl} \left( \tilde{\rho}_l \tilde{F}_l^k(S_{(\hat{i} )}) - \hat{\rho}_l\hat{F}_l^k (S_{(\hat{i} )}) \right) \\
    & \qquad = \frac{\hat{i} }{n} + \hDelta(S_{(\hat{i} )}) + \sum_{k=1}^K \sum_{l=1}^K W_{kl} \left( \tilde{\rho}_l \tilde{F}_l^k(S_{(\hat{i} )}) - \hat{\rho}_l\hat{F}_l^k (S_{(\hat{i} )}) \right)\\
    & \qquad < 1 - \left( \alpha + \hDelta(S_{(\hat{i} -1)}) - \delta (n) \right) + \frac{1}{n} \\
    & \qquad \qquad + \hDelta(S_{(\hat{i} )}) + \sum_{k=1}^K \sum_{l=1}^K W_{kl} \left( \tilde{\rho}_l \tilde{F}_l^k(S_{(\hat{i} )}) - \hat{\rho}_l\hat{F}_l^k (S_{(\hat{i} )}) \right)\\
    & \qquad = 1 - \alpha + \delta (n) + \frac{1}{n} \\
    & \qquad \qquad + \left( \hDelta(S_{(\hat{i} )}) - \hDelta(S_{(\hat{i} -1)}) \right) + \sum_{k=1}^K \sum_{l=1}^K W_{kl} \left( \tilde{\rho}_l \tilde{F}_l^k(S_{(\hat{i} )}) - \hat{\rho}_l\hat{F}_l^k (S_{(\hat{i} )}) \right) \\
    & \qquad \leq 1 - \alpha + \delta (n) + \frac{1}{n} \\
    & \qquad \qquad + \left( \hDelta(S_{(\hat{i} )}) - \hDelta(S_{(\hat{i} -1)}) \right) + \underset{t \in [0,1]}{\sup} \left|\sum_{k=1}^K \sum_{l=1}^K W_{kl} \left( \hat{\rho}_l\hat{F}_l^k (t) - \tilde{\rho}_l \tilde{F}_l^k(t) \right) \right|.
  \end{align*}

  The expected value of the last term above is $\delta^{**}(n)$.

  The term $\hDelta(S_{(\hat{i} )}) - \hDelta(S_{(\hat{i} -1)})$ above is given by
  \begin{align*}
    & \hDelta(S_{(\hat{i} )}) - \hDelta(S_{(\hat{i} -1)}) \\
     & \qquad = \left[ \hDelta(S_{(\hat{i} )}) - \Delta(S_{(\hat{i} )}) \right] + \left[ \Delta(S_{(\hat{i} )}) - \Delta(S_{(\hat{i}-1 )}) \right] + \left[ \Delta(S_{(\hat{i}-1 )}) - \hDelta(S_{(\hat{i}-1 )}) \right] \\
     & \qquad \leq 2 \sup_{t \in [0,1]} \left| \hDelta(t) - \Delta(t) \right| + \left| \Delta(S_{(\hat{i} )}) - \Delta(S_{(\hat{i}-1 )}) \right|.
    \end{align*}
    We already know how to bound the expected value of the first term on the right-hand-side above. Indeed,
    \begin{align*}
        &\sup_{t \in [0,1]} \left| \hDelta(t) - \Delta(t) \right|\\
        & \qquad \leq \sup_{t \in [0,1]} \left| \sum_{kl} W_{kl} \left( \hat{\rho}_l\hat{F}_l^k (t) - \tilde{\rho}_l \tilde{F}_l^k(t) \right) \right| + \sup_{t \in [0,1]}\left| \hat{F}(t) - \tF(t) \right|,
    \end{align*}
    so that by making use of the DKW inequality we get
    \begin{align*}
        & 2 \EV{\sup_{t \in [0,1]} \left| \hDelta(t) - \Delta(t) \right|} \leq 2 \delta^{**}(n) + \sqrt{\frac{2 \pi}{n}}.
    \end{align*}
    
    Concerning the second term, since $\Delta(\tilde{F}^{-1}(\cdot))$ is $L$-Lipschitz,
    \begin{align*}
        \left| \Delta( \tilde{F}^{-1}(U_{(\hat{i} )})) - \Delta( \tilde{F}^{-1}(U_{(\hat{i}-1 )})) \right| & \leq L \cdot \left(U_{(\hat{i})} - U_{(\hat{i}-1 )} \right)\\
        & \leq L \cdot\underset{2 \leq i \leq n}{\max} \left(U_{(i)} - U_{(i-1 )} \right)\\
        & \leq L \cdot \underset{1 \leq i \leq n+1}{\max}D_{i},
    \end{align*}
    where $D_{1} = U_{(1)}$, $D_i = U_{(i)} - U_{(i-1)}$ for $i=2,\dots,n$, and $D_{n+1} = 1 - U_{(n)}$. By a standard result on maximum uniform spacing,
    \begin{align*}
        \EV{\underset{1 \leq i \leq n+1}{\max}D_{i}} = \frac{1}{n+1} \sum_{j=1}^{n+1} \frac{1}{j} 
      \leq \frac{1 + \log(n+1)}{n+1}.
    \end{align*}

At this point, we have proven that
\begin{align}
\begin{split}
    & \EV{ \P{Y_{n+1} \in \hat{C} (X_{n+1}) \mid \cD}\I{\cA_1^c\cap \cA_2^c}} \\
    & \qquad \leq 1 - \alpha +  \delta (n) + \frac{1}{n} + 3 \delta^{**} (n) + \sqrt{\frac{2 \pi}{n}} + L \cdot \frac{1 + \log(n+1)}{n+1}.
\end{split} \label{eq:upper_bound_event3-marg}
\end{align}
\end{itemize}

Finally, combining~\eqref{eq:upper_bound_events-marg} with \eqref{eq:upper_bound_event1-marg}, \eqref{eq:upper_bound_event2-marg}, and \eqref{eq:upper_bound_event3-marg} leads to the desired result:
\begin{align*}
   & \P{Y_{n+1} \in \hat{C} (X_{n+1})}  \\
   & \qquad \leq \P{\cA_1} + \EV{\P{Y_{n+1} \in \hat{C} (X_{n+1}) \mid \cD}\I{\cA_2}} \\
   & \qquad \qquad + \EV{\P{Y_{n+1} \in \hat{C} (X_{n+1}) \mid \cD}\I{\cA_1^c\cap \cA_2^c}} \\
    & \qquad \leq 1 - \alpha + \delta (n) + 3 \delta^{**} (n) + \sqrt{\frac{2 \pi}{n}} + \frac{4}{n} + \frac{1+L}{n+1} + L \frac{1 + \log(n+1)}{n+1} \\
  & \qquad = 1 - \alpha + \delta(n) + \varphi(n).
\end{align*}

\end{proof}

\begin{theorem} \label{thm:upper-bound-on-abs-psi}
    Under the assumptions of Theorem~\ref{thm:algorithm-lower-bound}, consider the empirical process $\hat{\psi}(t)$ defined in~\eqref{eq:hat-psi}. Then,
    \begin{align*}
        \delta^{**}(n) & \leq \inf_{\beta} \left\{ \frac{1}{\sqrt{n}} \sqrt{\frac{\pi}{2}} \left(|\beta_0|+\frac{\sum_{k=1}^{K} |\beta_k|}{K} \right) + \frac{1}{\sqrt{n}} B(K,n,\beta) \right\},
    \end{align*}
    where $B(K,n, \beta)$ is defined in~\eqref{eq:B-definition}.
\end{theorem}

\begin{proof}[Proof of Theorem~\ref{thm:upper-bound-on-abs-psi}]
Our goal is that of proving
\begin{align*}
    \EV{ \sup_{t \in [0,1]} \left|\hat{\psi}(t) \right| } & \leq \inf_{\beta} \left\{ \frac{1}{\sqrt{n}} \sqrt{\frac{\pi}{2}} \left(|\beta_0|+\frac{\sum_{k=1}^{K} |\beta_k|}{K} \right) + \frac{1}{\sqrt{n}} B(K,n,\beta) \right\}.
\end{align*}

We proceed similarly to the proof of Theorem~\ref{thm:finite-sample-factor}.
Recall that the matrix $\bar{W}$ is defined such that $\bar{W}_{kl} = \beta_0 \I{k=l} + \beta_k/K$, for some $\beta_0, (\beta_k)_{k}$. As in the proof of Theorem~\ref{thm:finite-sample-factor}, we will derive a bound as a function of the parameters $\beta$; then, the result can be obtained by taking the minimum over $\beta$.

Recall that $\hat{\psi}(t) = \hat{\psi}_1(t)  + \hat{\psi}_2(t)  + \hat{\psi}_3(t)$, where
\begin{align*}
  \hat{\psi}_1(t)
  & := \beta_0 \frac{1}{n} \sum_{i=1}^{n} \left[  \I{\hat{s}(X_i,\tilde{Y}_i) \leq t} - \tilde{F}(t) \right],  \\
  \hat{\psi}_2(t) 
  & := \frac{1}{n} \sum_{i=1}^{n} \sum_{k=1}^{K} (\beta_k/K) \left[  \I{\hat{s}(X_i,k) \leq t} - \tilde{F}^{k}(t) \right] \\
  \hat{\psi}_3(t) 
  & := \frac{1}{n} \sum_{i=1}^{n} \sum_{k=1}^{K} \sum_{l=1}^{K} (W_{kl}-\bar{W}_{kl}) \left[  \I{\hat{s}(X_i,k) \leq t, \tilde{Y}_i = l} - \tilde{\rho}_l \tilde{F}_l^{k}(t) \right], \\
  \tilde{F}^{k}(t)
  & := \P{\hat{s}(X_i,k) \leq t}, \\
  \tilde{F}(t)
  & := \P{\hat{s}(X_i,\tilde{Y}_i) \leq t}.
\end{align*}

Then, it follows that
\begin{align*}
    \left| \hat{\psi}(t) \right| \leq \left| \hat{\psi}_1(t) \right| + \left| \hat{\psi}_2(t) \right| + \left| \hat{\psi}_3(t) \right|.
\end{align*}

We now bound the suprema of these three processes separately, since
\begin{align*}
  \EV{ \sup_{t \in [0,1]} \left| \hat{\psi}(t) \right| } & \leq \EV{ \sup_{t \in [0,1]} \left| \hat{\psi}_1(t) \right|  } + \EV{ \sup_{t \in [0,1]} \left| \hat{\psi}_2(t)\right|  } + \EV{ \sup_{t \in [0,1]} \left| \hat{\psi}_3(t) \right|  }.
\end{align*}

Let's start from $\hat{\psi}_1(t)$. First, note that
\begin{equation*}
  \EV{\sup_{t \in [0,1]} \left| \hat{\psi}_1(t) \right|} \leq  \left|\beta_0\right| \EV{ \sup_{t \in [0,1]} \left| \hat{F}(t) - \tilde{F}(t) \right| }.
\end{equation*}

Then, it follows from the DKW inequality that, for any $\eta > 0$
\begin{align*}
    \P{\sup_{t \in [0,1]} \left| \hat{F}(t) - \tilde{F}(t) \right| > \eta} \leq 2 \exp \left\{ -2n \eta^2 \right\}.
\end{align*}
This directly implies that
\begin{equation*}
  \EV{\sup_{t \in [0,1]} \left| \hat{\psi}_1(t) \right|} \leq \left| \beta_0 \right| \sqrt{\frac{\pi}{2 n}}.
\end{equation*}

Similarly, we can deal with $\hat{\psi}_2(t)$:
\begin{align*}
  \EV{ \sup_{t \in [0,1]} \left| \hat{\psi}_2(t) \right| }
  & = \EV{ \sup_{t \in [0,1]} \left| \sum_{k=1}^{K} \frac{\beta_k}{K}  \left( \hat{F}^{k}(t) - \tilde{F}^{k}(t) \right) \right|}  \\
  & \leq \sum_{k=1}^{K} \frac{\left|\beta_k \right|}{K} \EV{ \sup_{t \in [0,1]} \left| \hat{F}^{k}(t) - \tilde{F}^{k}(t) \right| }  \\
  & \leq \frac{\sum_{k=1}^{K} \left| \beta_k \right|}{K} \sqrt{\frac{\pi}{2 n}}.
\end{align*}

A bound for $\hat{\psi}_3(t)$ is identified by defining $\Omega_{kl} := W_{kl}-\bar{W}_{kl}$, so that
\begin{align*}
  \hat{\psi}_3(t) 
  & := \frac{1}{n} \sum_{i=1}^{n} \sum_{k=1}^{K} \sum_{l=1}^{K} \Omega_{kl} \left[  \I{\hat{s}(X_i,k) \leq t, \tilde{Y}_i = l} - \tilde{\rho}_l \tilde{F}_l^{k}(t) \right].
\end{align*}
Then, it follows from Lemma~\ref{lemma:bound-on-psi3} that
\begin{align*}
  \EV{ \sup_{t \in [0,1]} \left|\hat{\psi}_{3}(t) \right| }
  &\leq \frac{1}{\sqrt{n}} B(K,n, \beta),
\end{align*}
where $B(K,n, \beta)$ is defined in~\eqref{eq:B-definition}.

Putting everything together, we find that, for any $\beta \in \mathbb{R}^{K+1}$,
\begin{align*}
  \EV{ \sup_{t \in [0,1]} \left| \hat{\psi}(t) \right| } 
  & \leq \frac{1}{\sqrt{n}} \sqrt{\frac{\pi}{2}} \left(\left| \beta_0 \right| +\frac{\sum_{k=1}^{K} \left| \beta_k \right| }{K} \right) + \frac{1}{\sqrt{n}} B(K,n, \beta).
\end{align*}
Therefore,
\begin{align*}
  \EV{ \sup_{t \in [0,1]} \left| \hat{\psi}(t) \right| } 
  & \leq \inf_{\beta} \left\{ \frac{1}{\sqrt{n}} \sqrt{\frac{\pi}{2}} \left(|\beta_0|+\frac{\sum_{k=1}^{K} |\beta_k|}{K} \right) + \frac{1}{\sqrt{n}} B(K,n,\beta) \right\}.
\end{align*}

\end{proof}

\clearpage

\section{Additional Numerical Results} \label{appendix:additional-numerical-results}

\subsection{The Impact of the Label Contamination Strength}

\begin{figure}[!htb]
    \centering
    \includegraphics[width=0.8\linewidth]{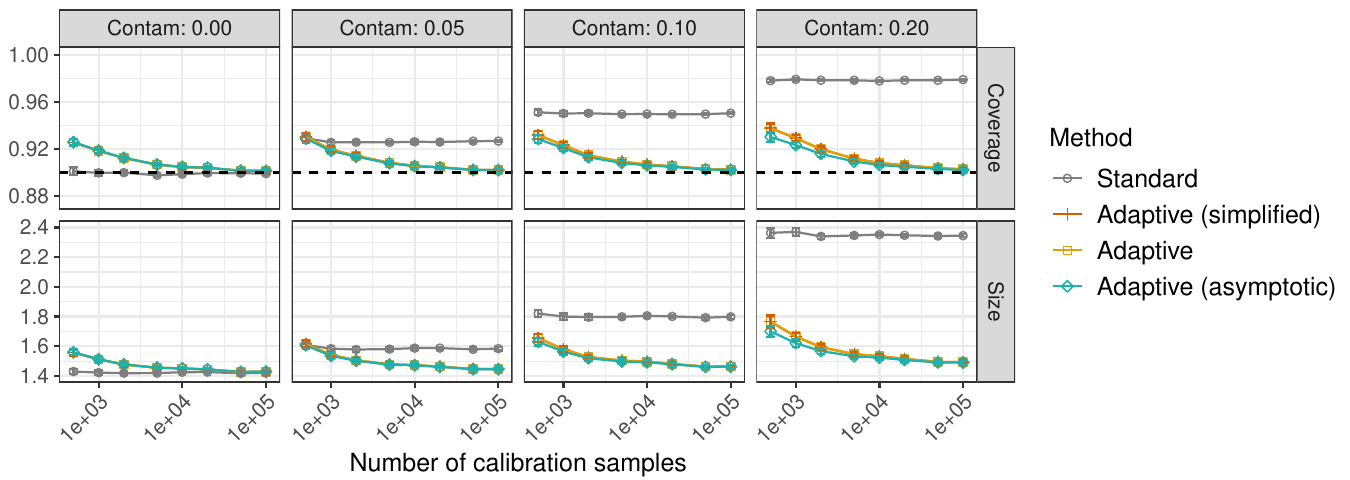}
    \caption{Performances of different conformal methods on simulated data with labels contaminated by a randomized response model, as function of the number of calibration samples. The empirical coverage and average size of the prediction sets are stratified based on the strength $\epsilon$ of the label contamination process. The number of classes is $K=4$.}
    \label{fig:exp1_synthetic1_ntrain10000_K4_nu0.2_marginal_uniform_optimisticFALSE}
\end{figure}

\begin{figure}[!htb]
    \centering
    \includegraphics[width=0.8\linewidth]{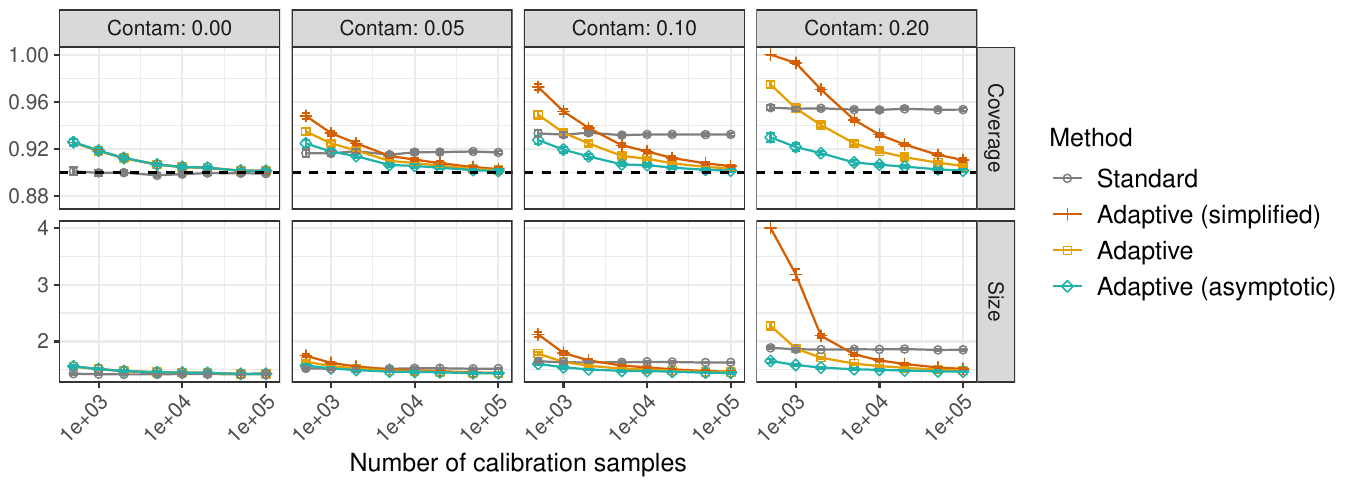}
    \caption{Performances of different conformal methods on simulated data with labels contaminated by a block randomized response model, as function of the number of calibration samples. The empirical coverage and average size of the prediction sets are stratified based on the strength $\epsilon$ of the label contamination process. The number of classes is $K=4$.}
    \label{fig:exp1_synthetic1_ntrain10000_K4_nu0.2_marginal_block_optimisticFALSE}
\end{figure}

\FloatBarrier
\clearpage

\subsection{The Impact of the Label Contamination Model}
\label{appendix:additional-numerical-results-contamination-model}
\begin{figure}[!htb]
\centering
\includegraphics[width=0.8\linewidth]{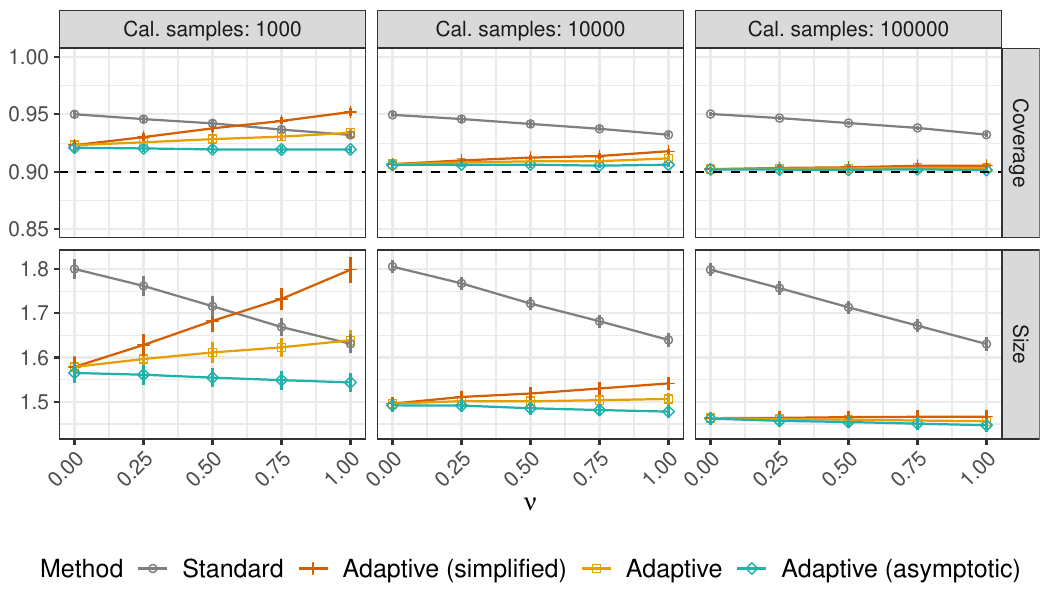}
\caption{Performances of different conformal methods on simulated data with labels contaminated by a two-level randomized response model, as function of $\nu$. The reported empirical coverage and average size of the prediction sets are stratified based on the number of calibration samples. The strength parameter of the label contamination process is $\epsilon=0.1$. The number of labels is $K=4$.}
\label{fig:Aexp2_synthetic1_ntrain10000_K4_eps0.1_marginal_RRB_optimisticFALSE}
\end{figure}

\FloatBarrier

\subsection{The Impact of the Number of Classes}
\label{appendix:impact-number-classes}
\begin{figure}[!htb]
    \centering
    \includegraphics[width=0.8\linewidth]{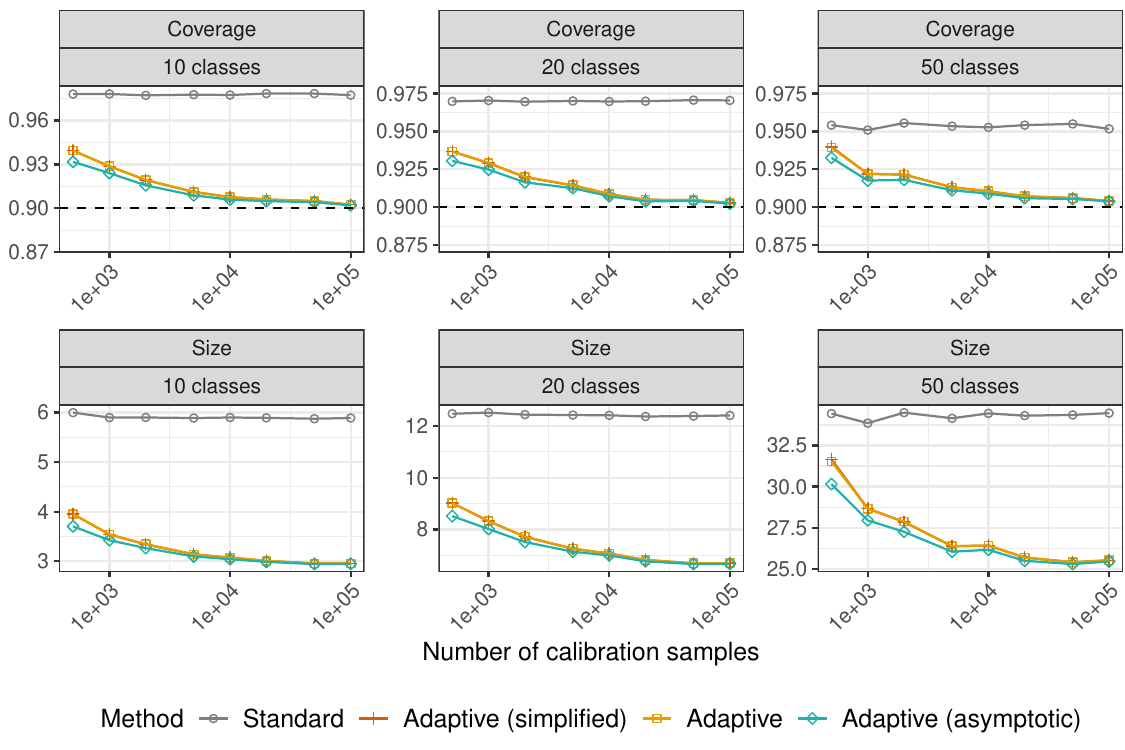}
    \caption{Performances of different conformal prediction methods on simulated data with contaminated labels. The label contamination process is the randomized response model with $\epsilon=0.2$. The reported empirical coverage and average size of the prediction sets are stratified based on the number of labels. The dashed horizontal line indicates the 90\% nominal marginal coverage.}
    \label{fig:exp3_synthetic1_ntrain10000_eps0.200000_nu0.8_marginal_uniform_optimisticFALSE}
\end{figure}

\begin{figure}[!htb]
\centering
\includegraphics[width=0.8\linewidth]{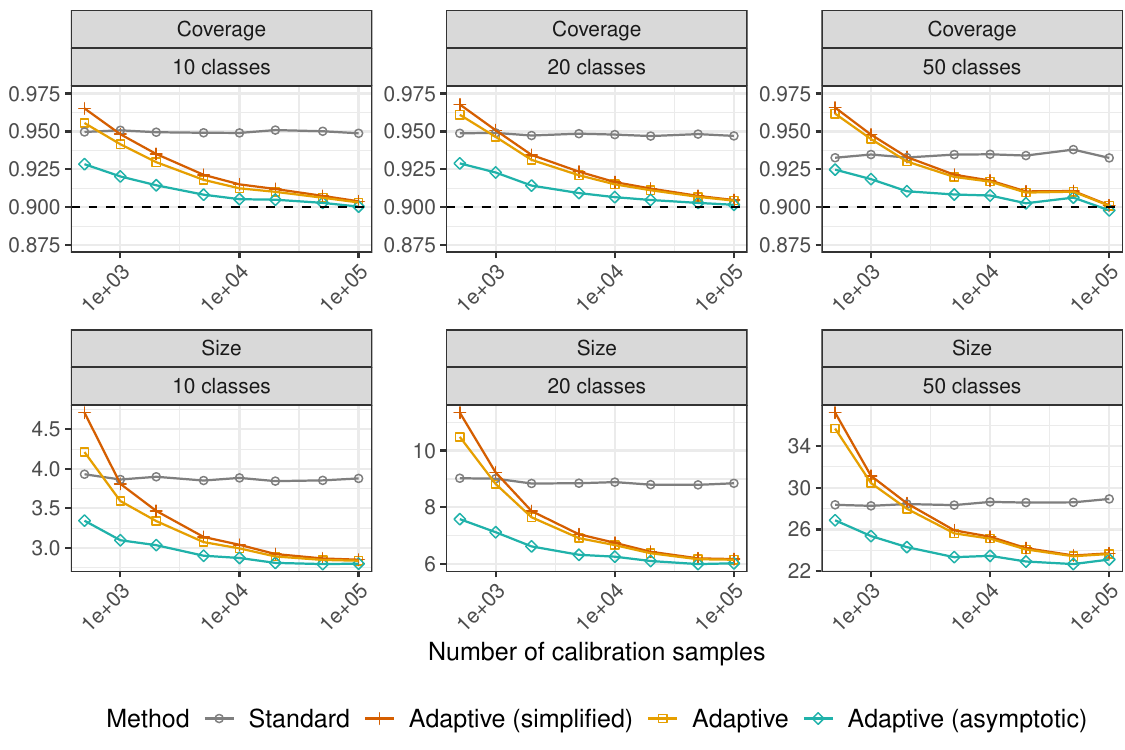}
\caption{Performances of different conformal prediction methods on simulated data with contaminated labels. The label contamination process is the two-level randomized response model described in Appendix~\ref{appendix:two-level-rmm}, with $\epsilon=0.1$ and $\nu = 0.8$. Other details are as in Figure~\ref{fig:exp3_synthetic1_ntrain10000_eps0.200000_nu0.8_marginal_uniform_optimisticFALSE}.}
\label{fig:exp3_synthetic1_ntrain10000_eps0.100000_nu0.8_marginal_RRB_optimisticFALSE}
\end{figure}

\begin{figure}[!htb]
\centering
\includegraphics[width=0.8\linewidth]{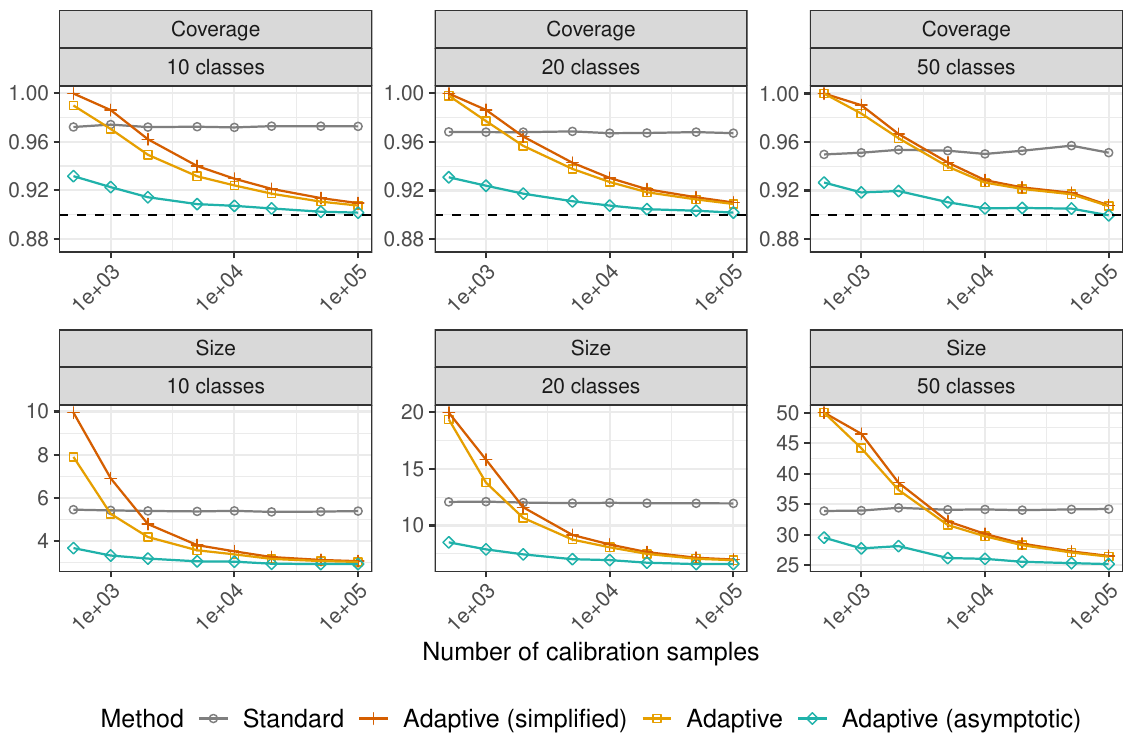}
\caption{Performances of different conformal prediction methods on simulated data with contaminated labels. The label contamination process is the two-level randomized response model described in Appendix~\ref{appendix:two-level-rmm}, with $\epsilon=0.2$ and $\nu = 0.8$. Other details are as in Figure~\ref{fig:exp3_synthetic1_ntrain10000_eps0.200000_nu0.8_marginal_uniform_optimisticFALSE}.}
\label{fig:exp3_synthetic1_ntrain10000_eps0.200000_nu0.8_marginal_RRB_optimisticFALSE}
\end{figure}

\begin{figure}[!htb]
    \centering
    \includegraphics[width=0.8\linewidth]{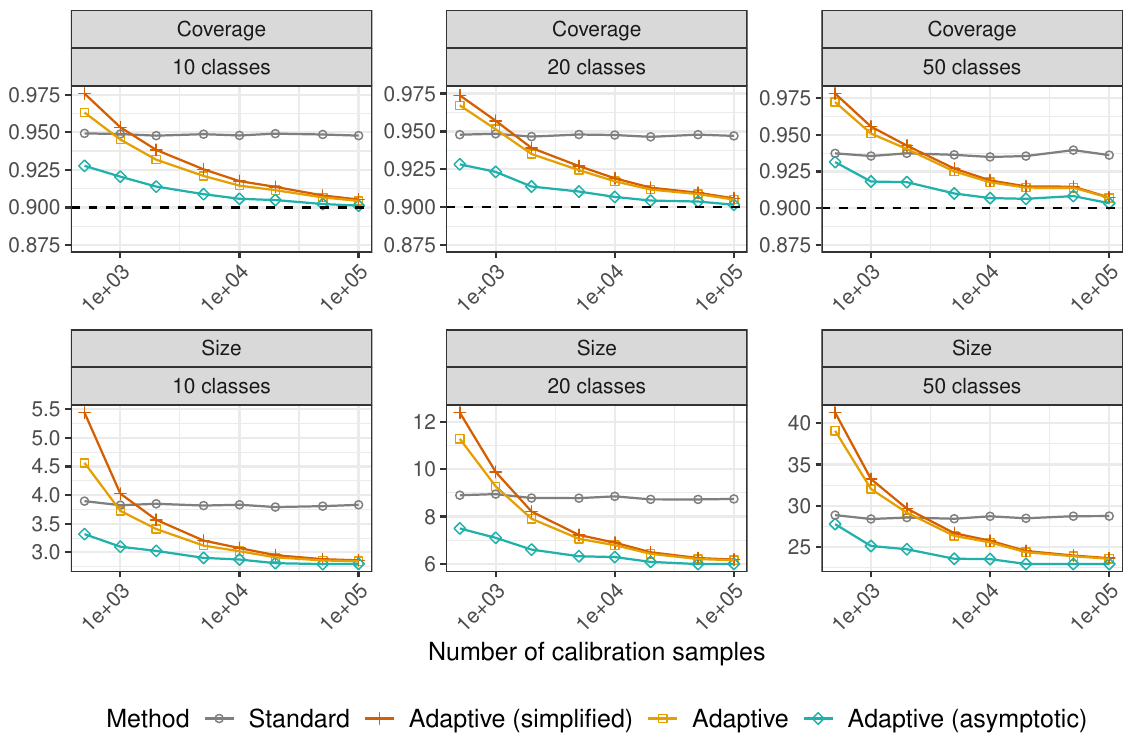}
    \caption{Performances of different conformal prediction methods on simulated data with contaminated labels. The label contamination process is the block-randomized response model, with two blocks, with $\epsilon=0.1$. Other details are as in Figure~\ref{fig:exp3_synthetic1_ntrain10000_eps0.200000_nu0.8_marginal_uniform_optimisticFALSE}.}
    \label{fig:exp3_synthetic1_ntrain10000_eps0.100000_nu0.8_marginal_block_optimisticFALSE}
\end{figure}

\begin{figure}[!htb]
    \centering
    \includegraphics[width=0.8\linewidth]{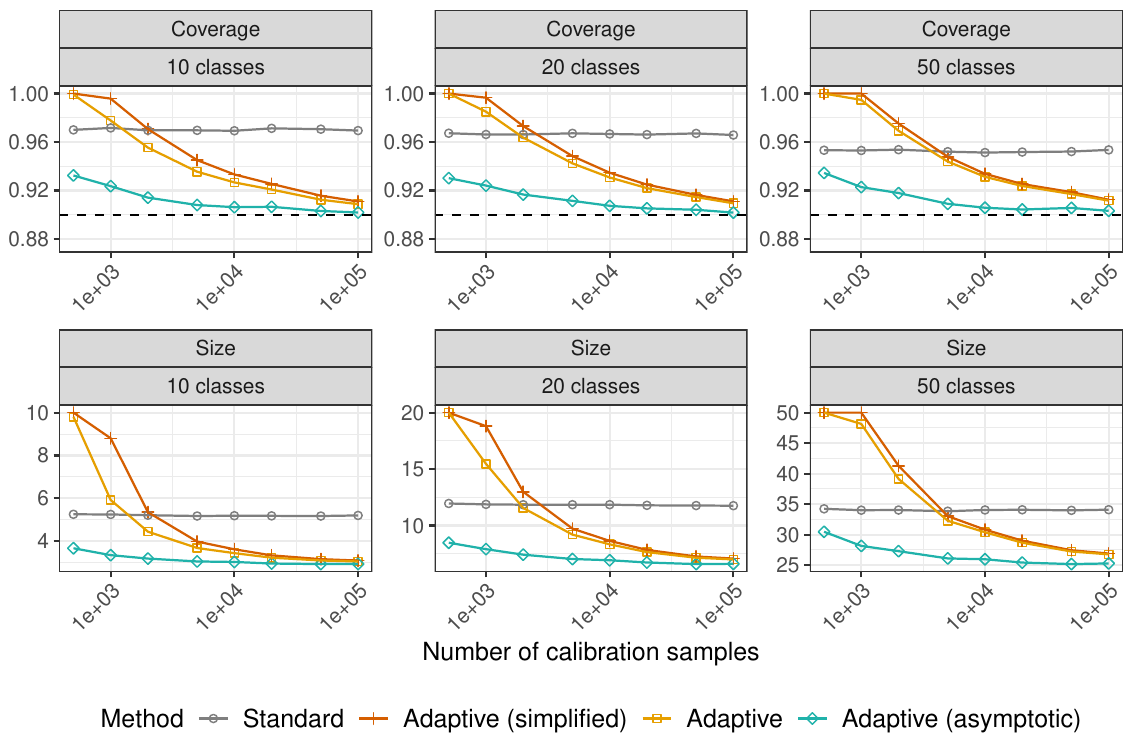}
    \caption{Performances of different conformal prediction methods on simulated data with contaminated labels. The label contamination process is the block-randomized response model, with two blocks, with $\epsilon=0.2$. Other details are as in Figure~\ref{fig:exp3_synthetic1_ntrain10000_eps0.200000_nu0.8_marginal_uniform_optimisticFALSE}.}
    \label{fig:exp3_synthetic1_ntrain10000_eps0.200000_nu0.8_marginal_block_optimisticFALSE}
\end{figure}

Figure~\ref{fig:exp3_synthetic1_ntrain10000_eps0.200000_nu0.8_marginal_uniform_optimisticFALSE} suggests that all implementations of our adaptive method remain similarly effective even when the number $K$ of possible labels becomes very large. However, while this is often true, it is not always the case, depending on the structure of the contamination model and on the label contamination strength $\epsilon$. 
The experiments in Figure~\ref{fig:exp3_synthetic1_ntrain10000_eps0.200000_nu0.8_marginal_uniform_optimisticFALSE} use a randomized response model, namely the setting for which the finite-sample implementations are optimized.
In contrast, Figure~\ref{fig:exp3_synthetic1_ntrain10000_eps0.100000_nu0.8_marginal_RRB_optimisticFALSE} examines the case where the contamination model is a two-level randomized response model with $\nu = 0.8$ and $\epsilon = 0.1$, corresponding to a significant deviation from the simple randomized response model.
Under this scenario, the finite-sample evaluation of the correction term becomes increasingly conservative relative to the more efficient {\em Asymptotic} implementation as $K$ grows. Figure~\ref{fig:exp3_synthetic1_ntrain10000_eps0.200000_nu0.8_marginal_RRB_optimisticFALSE} shows that this effect gets more pronounced as $\epsilon$ increases. Analogous results are obtained for the block-randomized response model with two blocks, with $\epsilon = 0.1$ and $\epsilon = 0.2$, as shown in Figures~\ref{fig:exp3_synthetic1_ntrain10000_eps0.100000_nu0.8_marginal_block_optimisticFALSE} and~\ref{fig:exp3_synthetic1_ntrain10000_eps0.200000_nu0.8_marginal_block_optimisticFALSE}.\\

\clearpage

\noindent \textbf{Randomized response model.}

\FloatBarrier

\noindent \textbf{Two-level randomized response model with $\nu=0.8$.}

\FloatBarrier

\noindent \textbf{Block-randomized response model.}

\FloatBarrier

\subsection{The Advantage of Targeting Marginal Coverage}
\label{appendix:comparison-with-label-cond}
\begin{figure}[!htb]
    \centering
    \includegraphics[width=0.8\linewidth]{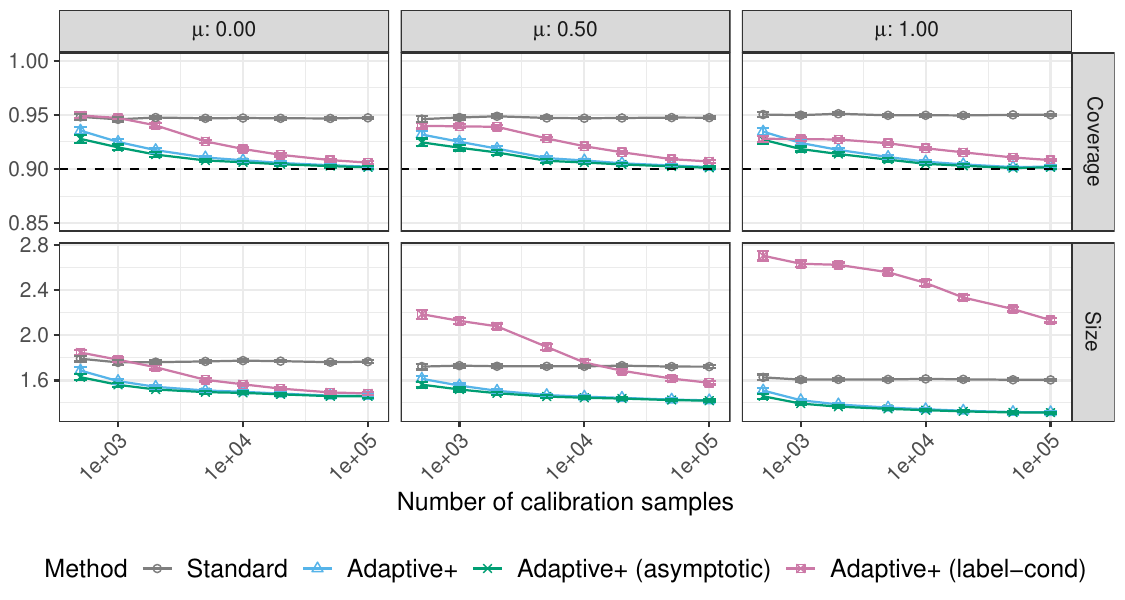}
    \caption{Performances of different conformal methods on simulated data with labels contaminated by a two-level randomized response, as function of the number of samples. The reported empirical coverage and average size of the prediction sets are stratified based on the class imbalance, controlled by the parameter $\mu$. The number of classes is $K=4$, the strength parameter of the label contamination process is $\epsilon=0.1$, and $\nu=0.2$. The dashed horizontal line indicates the 90\% nominal marginal coverage.}
    \label{fig:exp201_synthetic4_ntrain10000_K4_eps0.1_nu0.2_marginal_RRB_optimisticTRUE}
\end{figure}

In these experiments, class imbalance is introduced and controlled through an imbalance indicator $\mu$. The imbalance indicator adjusts the class distribution using an exponential decay function, where higher values of $\mu$ result in greater disparity among class proportions. These ratios are applied as a post-sampling step, where the sampled data is re-weighted or re-sampled to align with the desired class balance. This ensures precise control over the degree of imbalance across different experimental settings. Figure~\ref{fig:exp201_synthetic4_ntrain10000_K4_eps0.1_nu0.2_marginal_RRB_optimisticTRUE} compares the performances of our optimistic adaptive methods for marginal coverage with the optimistic adaptive method for label-conditional coverage of~\citet{sesia2025adaptive}. 
The comparison is made in terms of marginal coverage and average size of the prediction sets output by the methods, and the results are stratified based on $\mu$. As expected, the results show that the label-conditional method produces increasingly large prediction sets, as the class imbalance gets more severe. The performance of the method for marginal coverage, on the other hand, is not impacted by class imbalance, and the prediction sets output by the method remain informative also for high values of $\mu$.

\FloatBarrier

\clearpage

\subsection{The Impact of Estimating the Contamination Model}
\label{appendix:additional_results_impact_contamination_estimation}

\noindent \textbf{Impact of the number of noisy training samples}

\begin{figure}[!htb]
\centering
\includegraphics[width=\linewidth]{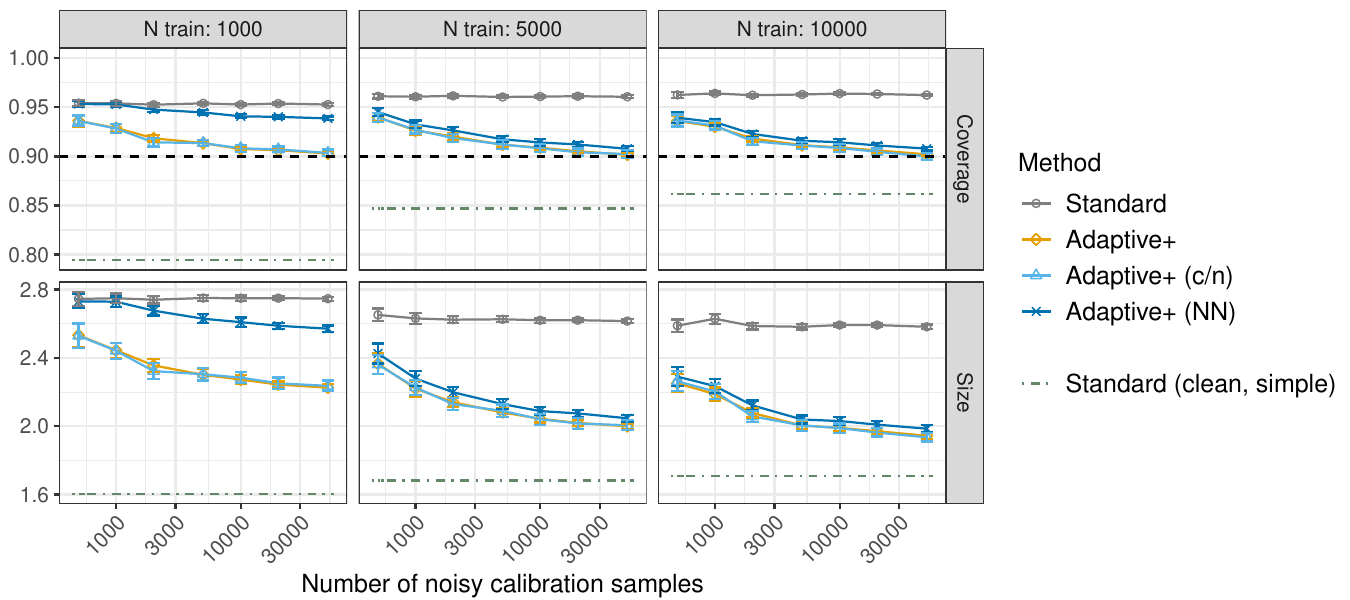}
\caption{Performances of different conformal methods on simulated data with labels contaminated by a randomized response model, as in Figure~\ref{fig:exp401_synthetic6_K4_nt5000_marginal_uniform_expeasyFALSE}. The number of clean observations used by {\em Adaptive+ (c/n)} and {\em Adaptive+ (NN)} to estimate the label contamination model is $n_{\mathrm{cl}}=500$. The number of classes is $K=4$. The data generating scheme follows a multinomial logistic regression and the contamination process is a randomized response model with $\epsilon=0.2$. In these experiments, the contamination layer of the neural network is parametrized as a randomized response model, with a single parameter $\epsilon \in [0,1)$.}
\label{fig:exp403_synthetic6_K4_ncl500_marginal_uniform_expeasyFALSE}
\end{figure}

\begin{figure}[!htb]
\centering
\includegraphics[width=\linewidth]{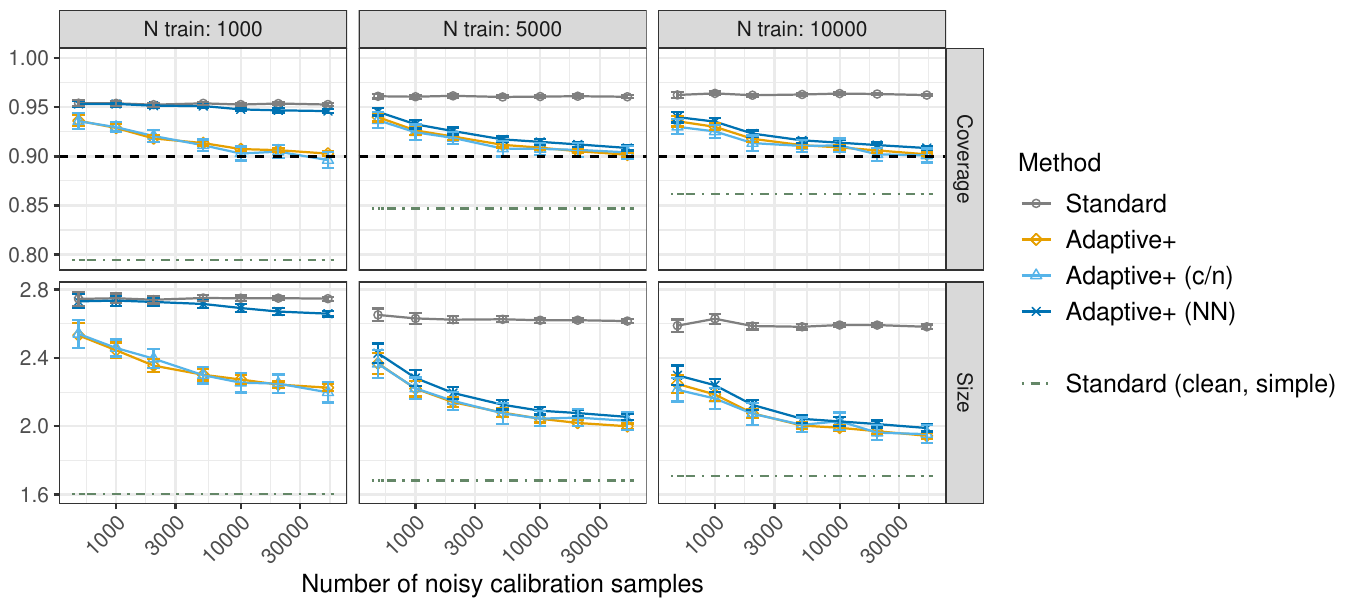}
\caption{Performances of different conformal methods on simulated data with labels contaminated by a randomized response model.
The number of clean observations used by {\em Adaptive+ (c/n)} and {\em Adaptive+ (NN)} to estimate the label contamination model is $n_{\mathrm{cl}}=100$. Other details are as in Figure~\ref{fig:exp403_synthetic6_K4_ncl500_marginal_uniform_expeasyFALSE}.}
\label{fig:exp403_synthetic6_K4_ncl100_marginal_uniform_expeasyFALSE}
\end{figure}

\FloatBarrier

\noindent \textbf{Impact of the contamination process}

We now simulate synthetic data with noisy labels obtained from contamination processes beyond the randomized response model.
The three contamination processes considered are: the block-randomized response model with two blocks and $\epsilon=0.2$ {\em block}; the two-level randomized response model with $\epsilon=0.2$ and $\nu=0.2$ {\em two-level}; and a nearly diagonal contamination matrix {\em near-diag}, having the form
    \begin{align*}
        T =
        \begin{bmatrix}
            0.80 & 0.10 & 0.05 & 0.05 \\
            0.10 & 0.80 & 0.05 & 0.05 \\
            0.05 & 0.05 & 0.80 & 0.10 \\
            0.05 & 0.05 & 0.10 & 0.80
        \end{bmatrix}.
    \end{align*}

\begin{figure}[!htb]
\centering
\includegraphics[width=\linewidth]{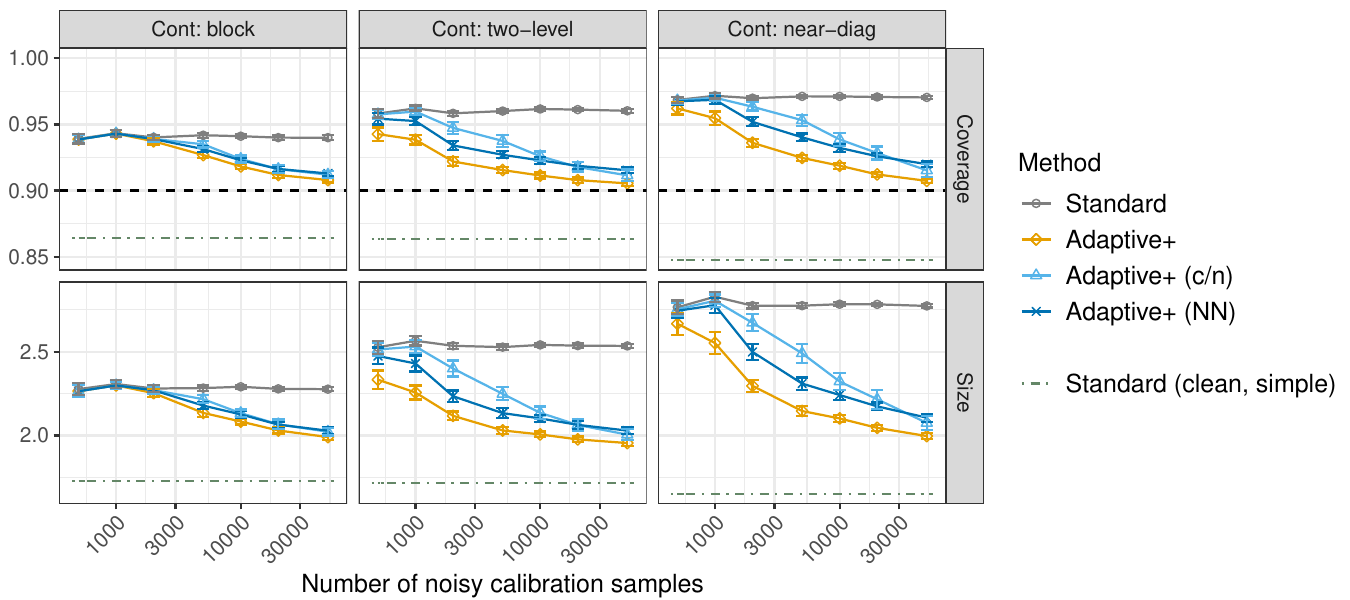}
\caption{Performances of different conformal methods on simulated data with contaminated labels, as function of the number of calibration samples. The empirical coverage and average size of the prediction sets are stratified based on the synthetic contamination process. The number of clean, simple observations is $n_{\mathrm{cl}}=500$ and the size of the noisy training set is $n_{\mathrm{train}}=10000$. The number of classes is $K=4$. The data generating scheme follows a multinomial logistic regression. In these experiments, the contamination layer of the neural network is parametrized as a general contamination with $K^2$ parameters.}
\label{fig:exp406_synthetic6_K4_nt10000_ncl500_marginal_expeasyFALSE}
\end{figure}

\FloatBarrier

\subsection{Estimation Accuracy for Contamination Model}
\label{appendix:effectiveness_of_contamination_estimation}

In this appendix, we evaluate empirically the accuracy of the neural network approach from Section~\ref{sec:contamination_estimation-unpaired} for estimating the label contamination model.

We consider two variants of the network: {\em NN}, whose backbone is an MLP with two hidden layers of 16 and 8 units, respectively; and {\em NNs}, a simplified variant, in which the backbone reduces to a single linear layer feeding directly into the softmax classifier and the contamination layer. Both variants are trained using the same alternating optimization procedure, for two sequential steps of 100 epochs each with batch size 128, using AdamW with learning rate $5 \times 10^{-2}$ in the first step and $10^{-3}$ in the second. The contamination layer is initialized so that the initial contamination matrix $T_0$ is a randomized response model with $\epsilon=10^{-6}$.

We evaluate performance based on two metrics: the accuracy of the classifier resulting from each method's backbone on the clean labels of a held-out test set of 2,000 observations; and the network's ability to recover the true contamination matrix $T$, measured either as the (signed) real-valued residual $\hat\epsilon - \epsilon$ (for experiments considering the randomized response model) or as the Frobenius norm of the difference between the estimated and true contamination matrices (for experiments considering a general contamination model).

\subsubsection{Impact of the Size of Clean Observations}

\begin{figure}[!htb]
\centering
\includegraphics[width=\linewidth]{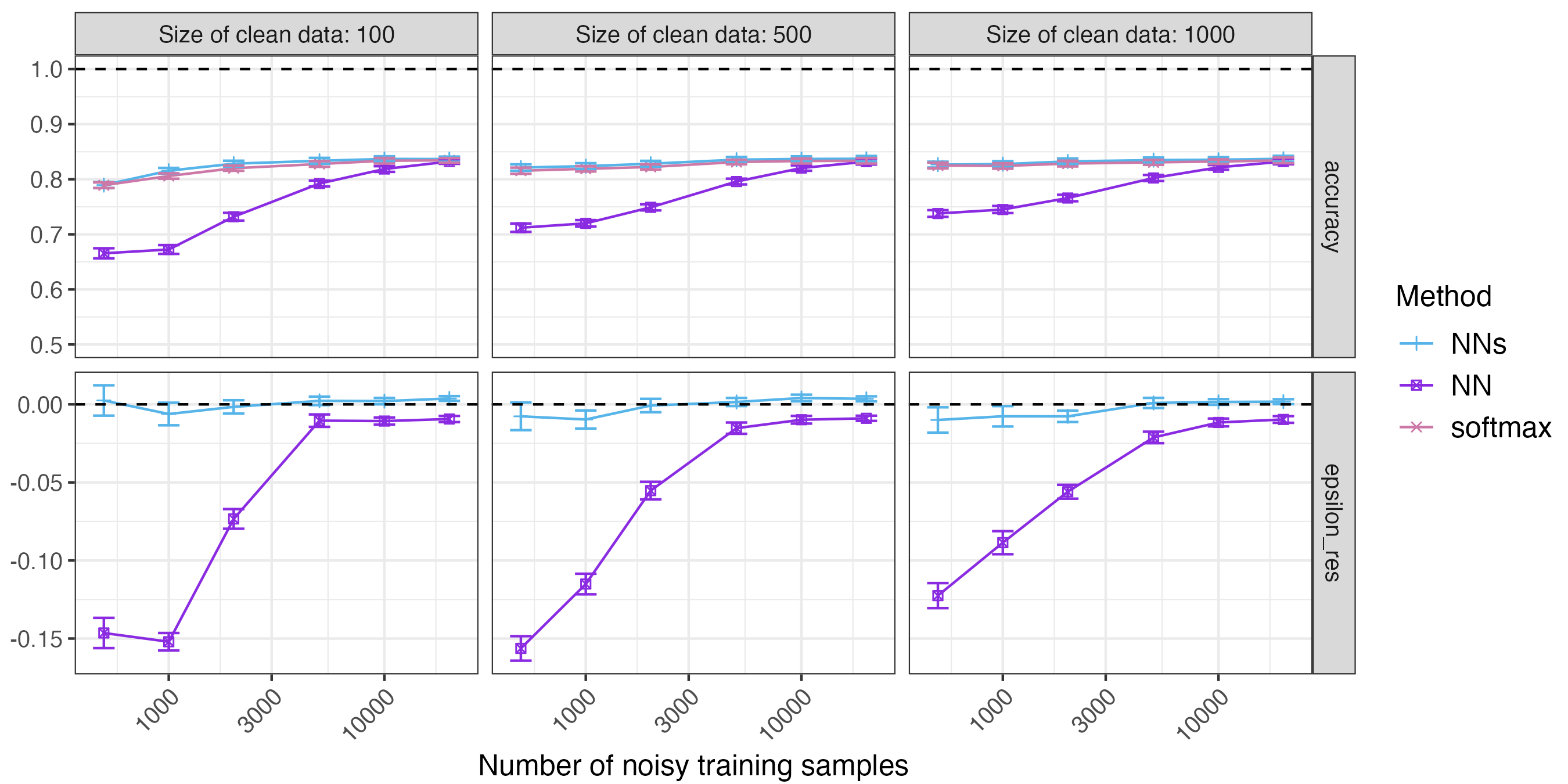}
\caption{Performance of the methods for $T$ estimation as function of the training set size. The results are stratified based on the number of clean observations available. The contamination process is a randomized response model with contamination strength $\epsilon = 0.2$. The number of possible labels is $K=4$. The data generating scheme follows a multinomial logistic regression, for which the {\em NNs} backbone is well-specified and the {\em NN} backbone is over-parametrized. The contamination layer of the neural networks parametrizes a randomized response model with a single parameter $\epsilon$.}
\label{fig:exp301_synthetic6_K4_uniform_FALSE}
\end{figure}

Figure~\ref{fig:exp301_synthetic6_K4_uniform_FALSE} shows the performance of the two network variants, as a function of the number of noisy training observations, stratified by the number of clean observations ($n_\mathrm{clean} = 100, 500, 1000$). Data are generated according to a multinomial logistic regression with $K=4$ labels, and contaminated with a randomized response model with $\epsilon=0.2$. The contamination layer of both networks parametrizes a randomized response model with a single parameter $\epsilon$. As a benchmark, we also report the accuracy of a multinomial softmax classifier fit directly on the same data available to the two networks, which allows us to assess whether the backbone of each network is learning the structure underlying the clean data.

With enough noisy training data, both network variants recover the contamination process accurately, even when only $n_\mathrm{clean}=100$ clean observations are available. Having fewer parameters to train, {\em NNs} reaches an accuracy close to that of the softmax benchmark, and an accurate estimate of $\epsilon$, already for small values of $n_\mathrm{clean}$ and of the noisy training sample size. The MLP network ({\em NN}) instead converges more slowly to the true contamination level. A larger clean sample size improves its starting point and speeds up this convergence. Notably, {\em NN} approaches the nominal value of $\epsilon$ from below throughout this process. This is a desirable property, since underestimating the contamination level translates into a conservative correction in our noise-adaptive methodology, avoiding the risk of inducing under-coverage by over-correcting for contamination that is not actually present.

\FloatBarrier

\subsubsection{Impact of the Data Simulation Scheme}

\begin{figure}[!htb]
\centering
\includegraphics[width=\linewidth]{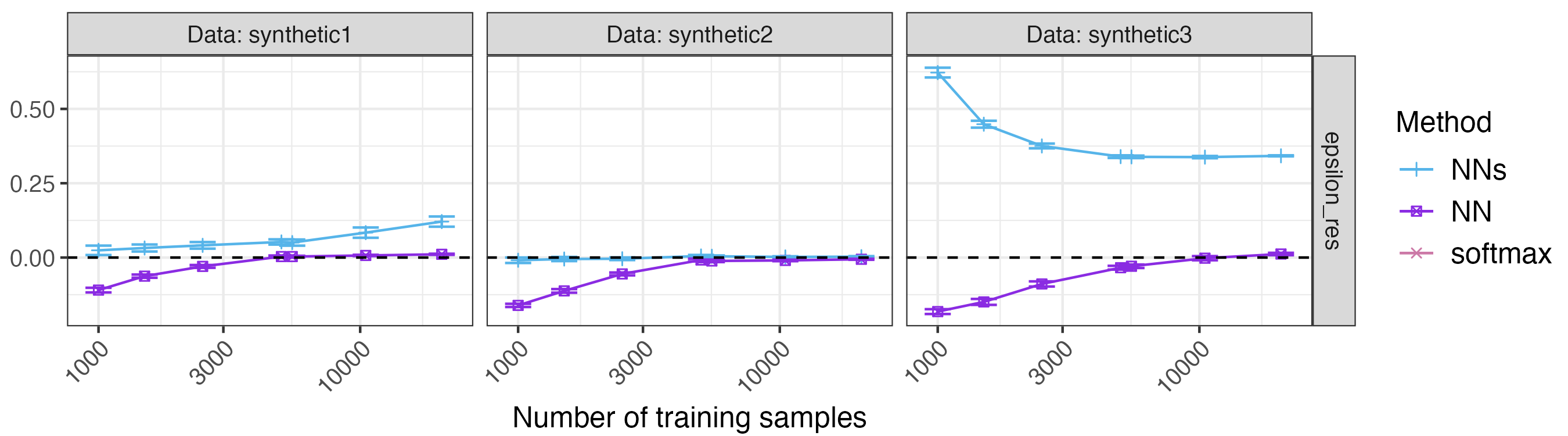}
\caption{Performance of the methods for $T$ estimation as function of the training set size. The results are stratified based on data simulation scheme. The {\em NNs} backbone is well-specified for the second simulation scheme (center) and biased for the other two. The {\em NN} backbone is sufficiently flexible for all three scenarios.
The contamination process is a randomized response model with contamination strength $\epsilon = 0.2$. The number of clean observations used to estimate the label contamination model is $n_{\mathrm{cl}}=500$. The number of possible labels is $K=4$. The contamination layer of the neural networks parametrizes a randomized response model with a single parameter $\epsilon$.}
\label{fig:exp304_eps0.2_ncl500_K4_uniform}
\end{figure}

Figure~\ref{fig:exp304_eps0.2_ncl500_K4_uniform} shows the residual $\hat\epsilon-\epsilon$ for the two network variants, as a function of the number of noisy training observations, for three different data-generating schemes, with $n_\mathrm{clean}=500$ fixed and contamination given by a randomized response model with $\epsilon=0.2$. The contamination layer of both networks parametrizes a randomized response model with a single parameter $\epsilon$. The three schemes are: {\em synthetic1}, a generator producing Gaussian clusters separated according to a class-separation parameter; {\em synthetic2}, a multinomial logistic regression model; {\em synthetic3}, a decision-tree-based generator introducing heteroscedasticity in the class-conditional distributions.

The performance of {\em NNs} depends on how well the data-generating scheme matches its single linear layer. Under {\em synthetic2}, which matches this linear structure, {\em NNs} remains close to unbiased throughout. Under {\em synthetic1} and, especially, {\em synthetic3}, which deviate from a linear class-conditional structure, {\em NNs} instead displays a substantial, persistent positive bias that does not vanish as the noisy training sample size grows, and is most severe for {\em synthetic3}. The {\em NN} starts with a negative bias across all three schemes, but this bias consistently shrinks toward zero as the noisy training sample size increases, regardless of the data-generating scheme. As in the previous experiment, {\em NN} approaches the true $\epsilon$ from below, which is again a desirable, conservative behavior for our noise-adaptive methodology. These results indicate that the added flexibility of the MLP backbone is necessary to reliably recover the contamination process when the classifier's decision boundary is not well approximated by a linear model, while the simpler {\em NNs} architecture is only reliable when this linear approximation holds.

\FloatBarrier

\subsubsection{Impact of the Contamination Process}
\label{appendix:impact_of_contamination_process_on_estimation}

\begin{figure}[!htb]
\centering
\includegraphics[width=\linewidth]{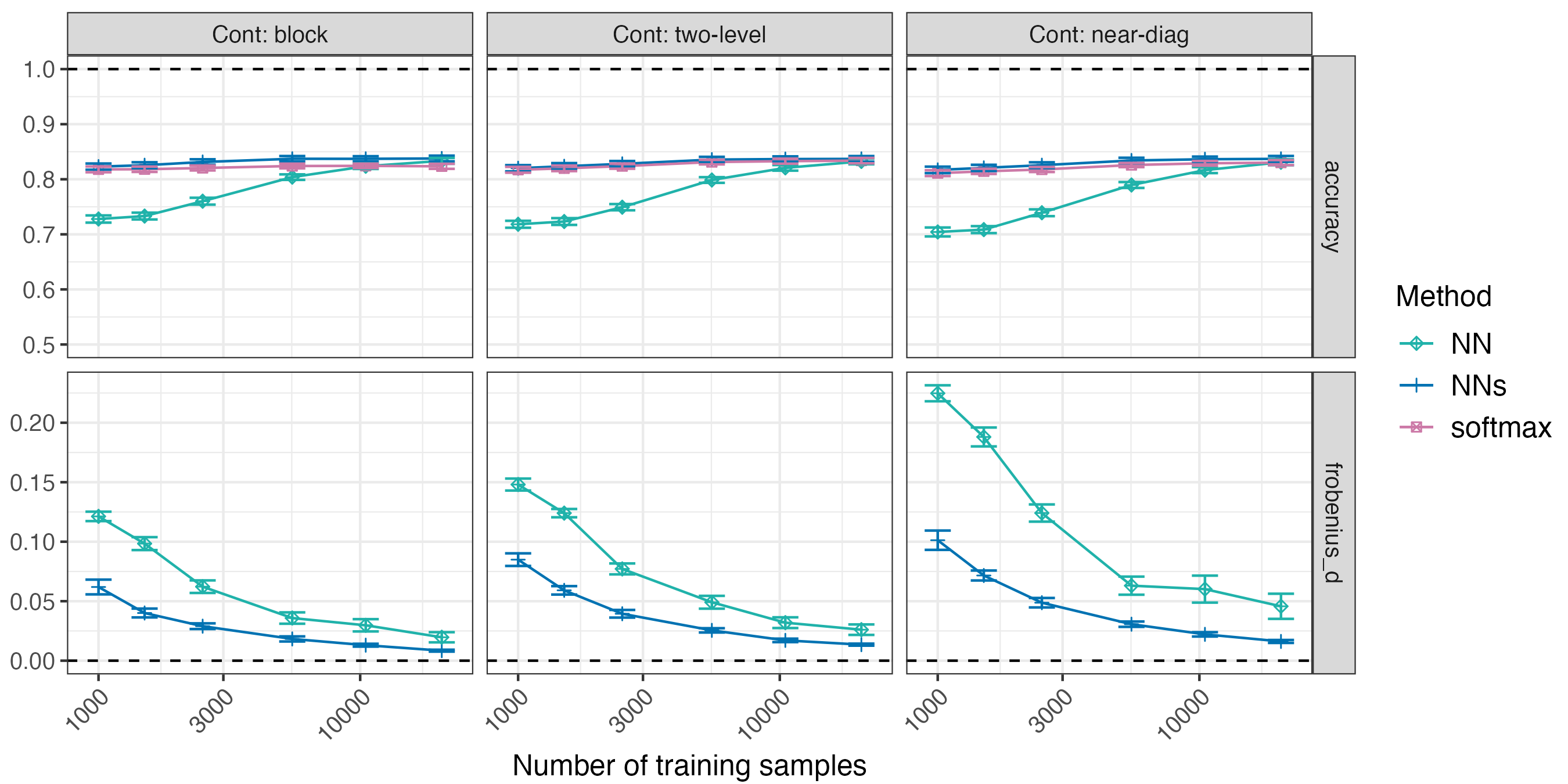}
\caption{Performance of the methods for $T$ estimation as function of the training set size. The results are stratified based on contamination process. The number of clean and simple observations is $n_{\mathrm{cl}}=500$. The number of possible labels is $K=4$. The data generating scheme follows a multinomial logistic regression. The contamination layer of the neural networks parametrizes a generic contamination model with $K^2$ parameters.}
\label{fig:exp305_synthetic6_eps0.2_ncl500_K4}
\end{figure}

Figure~\ref{fig:exp305_synthetic6_eps0.2_ncl500_K4} shows the accuracy and the Frobenius-norm estimation error of the two network variants, as a function of the number of noisy training observations, for three different contamination processes, with $n_\mathrm{clean}=500$ fixed and data generated according to a multinomial logistic regression with $K=4$ labels. Here, the contamination layer of both networks parametrizes a general contamination matrix with $K^2$ parameters, rather than the single-parameter randomized response model used in the previous experiments. As before, we also report the accuracy of a multinomial softmax classifier fit directly on the same data available to the two networks. The three contamination processes considered are: the block-randomized response model with two blocks and $\epsilon=0.2$ {\em block}; the two-level randomized response model with $\epsilon=0.2$ and $\nu=0.2$ {\em two-level}; and a nearly diagonal contamination matrix {\em near-diag}, having the form
    \begin{align*}
        T =
        \begin{bmatrix}
            0.80 & 0.10 & 0.05 & 0.05 \\
            0.10 & 0.80 & 0.05 & 0.05 \\
            0.05 & 0.05 & 0.80 & 0.10 \\
            0.05 & 0.05 & 0.10 & 0.80
        \end{bmatrix}.
    \end{align*}

Both {\em NNs} and {\em NN} are able to recover a general, $K^2$-parameter contamination matrix accurately, across all three contamination processes considered. As expected, {\em NNs} is faster to converge, given its smaller number of parameters. It is nonetheless important that the more flexible {\em NN} architecture is also able to recover the contamination process well, since, as shown in the previous experiment, this flexibility becomes convenient when the data-generating scheme does not match the simple linear structure assumed by {\em NNs}.

\FloatBarrier

\clearpage

\subsection{Analysis of CIFAR-10H Data}
\label{appendix:cifar}

This appendix reports a second application to real data, complementing the BigEarthNet analysis of Section~\ref{sec:case_study} and following the same general strategy described there.

The CIFAR-10H data set \citep{peterson2019human} contains 10,000 real-world images in 32x32 resolution, each annotated with both ``ground-truth'' labels, from the original CIFAR-10 data set, and imperfect labels provided by approximately 50 human annotators via Amazon Mechanical Turk. Variability in annotator opinions introduces label noise, which we model as a contamination process corrupting the true labels.
To simplify the analysis, we assign to each image a single corrupted label $\tilde{Y}$, drawn at random from a multinomial distribution based on the relative frequency of the annotator-provided labels. In this setup, the true and noisy labels differ with frequency 0.046.

Non-conformity scores are computed using a pre-trained ResNet-18 convolutional neural network \citep{sesia2025adaptive}.
The transition matrix is estimated as described in Section~\ref{sec:case_study}, identifying $n_{\mathrm{cl}}=500$ clean, simple images among $2{,}000$ images, with the remaining $n_{\mathrm{train}}=1,500$ treated as noisy training samples; the contamination network is trained on features extracted from the penultimate layer of a ResNet-18 pre-trained on ImageNet. No regularization of the loss was necessary in this case.

\begin{figure}[!htb]
\centering
\includegraphics[width=0.9\linewidth]{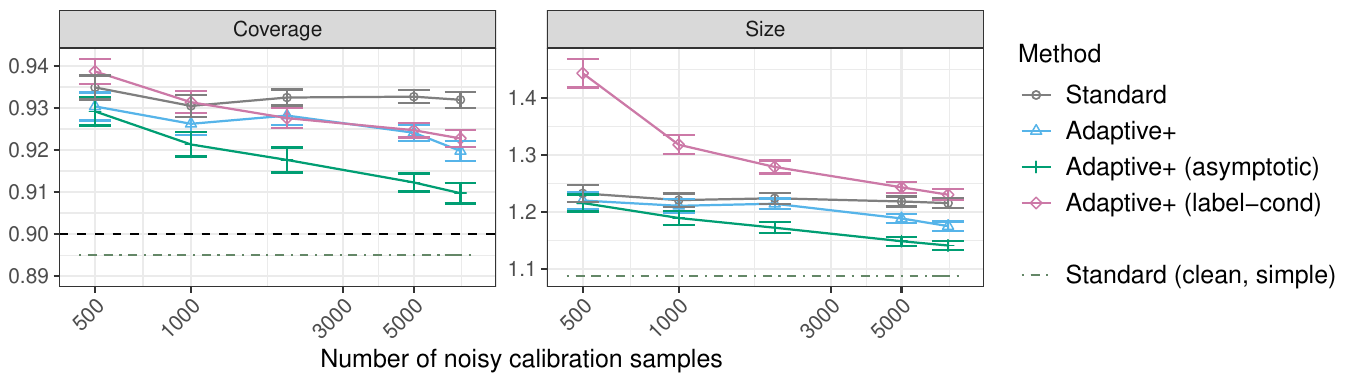}
\caption{Performance of conformal prediction methods on the CIFAR-10H data set with contaminated labels, as a function of the number of calibration samples, with $K=10$. The dashed line indicates the nominal 90\% marginal coverage level.}
\label{fig:cifar10_marginal_optimisticTRUE}
\end{figure}

Figure~\ref{fig:cifar10_marginal_optimisticTRUE} presents results for a random test set of 500 images, with calibration sample sizes $n$ ranging from 500 to 7,000, comparing the same methods as in Section~\ref{sec:case_study}. Each experiment is repeated 100 times using different random splits of CIFAR-10H into training, calibration, and test sets, re-estimating the contamination matrix every time.
Both the standard method and the label-conditional adaptive method are overly conservative, whereas our marginal adaptive methods---especially the one based on the asymptotic approximation of the correction factor---achieve valid coverage with smaller prediction sets, even for small calibration sets. Once again, targeting marginal coverage yields considerably more informative prediction sets than targeting label-conditional coverage.

\end{document}